\documentclass[prc,showpacs,preprintnumbers,amsmath,amssymb,superscriptaddress,groupedaddress,nofootinbib,tightenlines,twocolumn]{revtex4}

\usepackage{graphicx}
\usepackage{amsmath}
\usepackage{bm}
\usepackage{bbm}
\usepackage{color}
\usepackage{hyperref}
\usepackage{cleveref}

\def\beqn{\begin{eqnarray}}
\def\eeqn{\end{eqnarray}}
\def\barr{\begin{array}}
\def\earr{\end{array}}
\def\btab{\begin{tabular}}
\def\etab{\end{tabular}}
\def\bite{\begin{itemize}}
\def\eite{\end{itemize}}
\def\bcen{\begin{center}}
\def\ecen{\end{center}}

\def\magenta{\color{magenta}}
\def\black{\color{black}}
\def\eq{\begin{equation}}
\def\ee{\end{equation}}
\def\nn{\nonumber}

\def\pdagger{p\hspace{-0.17cm}/}

\def\qdagger{q\hspace{-0.18cm}/}
\def\keldagger{k\hspace{-0.2cm}/}

\def\q2dagger{q_2\hspace{-0.35cm}/\;}

\newcommand{\be}{\begin{equation}}
\newcommand{\bea}{\begin{eqnarray}}
\newcommand{\eea}{\end{eqnarray}}

\begin{document}
\preprint{MITP/18-096}
\preprint{ACFI-T18-19}
\title{Dispersive Evaluation of the Inner Radiative Correction\\ in Neutron and Nuclear $\beta$-decay}

\author{Chien-Yeah Seng} 
\email{cseng@hiskp.uni-bonn.de}
\affiliation{
INPAC, Shanghai Key Laboratory for Particle Physics and Cosmology,\\
MOE Key Laboratory for Particle Physics, Astrophysics and Cosmology,\\
School of Physics and Astronomy, Shanghai Jiao-Tong University, Shanghai 200240, China}
\affiliation{Helmholtz-Institut f\"{u}r Strahlen- und Kernphysik and Bethe Center for Theoretical Physics,\\
	Universit\"{a}t Bonn, 53115 Bonn, Germany}

\author{Mikhail Gorchtein} 
\email{gorshtey@uni-mainz.de}
\affiliation{Institut f\"ur Kernphysik, PRISMA Cluster of Excellence\\
Johannes Gutenberg-Universit\"at, Mainz, Germany}


\author{Michael J. Ramsey-Musolf}
\email{mjrm@physics.umass.edu}
\affiliation{Amherst Center for Fundamental Interactions, Department of Physics, University of Massachusetts, Amherst, MA 01003}
\affiliation{Kellogg Radiation Laboratory, California Institute of Technology, Pasadena, CA 91125 USA}

\begin{abstract}
We propose a novel dispersive treatment of the so-called inner radiative correction to the neutron and nuclear $\beta$-decay. We show that it requires knowledge of the parity-violating structure function $F_3^{(0)}$ that arises from the interference of the axial vector charged current and the isoscalar part of the electromagnetic current. By isospin symmetry, we relate this structure function to the charged current inelastic scattering of neutrinos and antineutrinos. Applying this new data-driven analysis we obtain a new, more precise evaluation for the universal radiative correction $\Delta_{R}^{V,\,new}=0.02467(22)$ that supersedes the previous estimate by Marciano and Sirlin, $\Delta_R^V=0.02361(38)$. 
The substantial shift in the central value of 
$\Delta_R^V$ reflects in a respective shift of $V_{ud}$ and a considerable tension in the unitarity constraint on the first row of the CKM matrix which is used as one of the most stringent constraints on New Physics contributions in the charged current sector.
We also point out that dispersion relations offer a unifying tool for treating hadronic and nuclear corrections within the same framework. 
We explore the potential of the dispersion relations for addressing the nuclear structure corrections absorbed in the ${\cal F}t$ values, a crucial ingredient alongside $\Delta_R^V$ in extracting $V_{ud}$ from superallowed nuclear decays. In particular, we estimate the quenching of the free neutron Born contribution in the nuclear environment, corresponding to a quasielastic single-nucleon knockout, and find a significantly stronger quenching effect as compared to currently used estimates based on the quenching of spin operators in nuclear transitions. This observation suggests that the currently used theoretical uncertainties of ${\cal F}t$ values might be underestimated and require a renewed scrutiny, while emphasizing the importance of new, more precise measurements of the free neutron decay where nuclear corrections are absent.
\end{abstract}
\date{\today}
\maketitle

\section{Introduction}
\label{sec:intro}
Precise studies of neutron and nuclear $\beta$-decays provide stringent tests of the Standard Model (SM) of fundamental interactions and probes of possible physics that may lie beyond it. 
In particular, a comparison of the nucleon (bound and free) $\beta$-decay to that of the muon yields a test of the universality of the weak interaction  upon introducing the  mixing amongst the quark weak eigenstates as reflected in the unitary Cabibbo-Kobayashi-Maskawa (CKM) matrix. 
Probes of CKM unitarity provide one of the most stringent tests of the SM,  and any significant deviation from unitarity would unavoidably signal the presence of physics beyond the Standard Model (BSM) ( see Ref. \cite{Gonzalez-Alonso:2018omy} for the most recent review of BSM effects in $\beta$-decays). The sensitivity to BSM physics relies on both a high degree of experimental precision and robust theoretical computations used to extract CKM matrix elements from experimental observables. 

Here, we focus on the hadronic and nuclear theory relevant to tests of the first row CKM unitarity condition : $|V_{ud}|^2+|V_{us}|^2+|V_{ub}|^2=1$. The matrix element $|V_{ud}| =0.97420\pm0.00021$ \cite{PDG2018} is the main contributor to the first row unitarity, and is relevant for charged pion, neutron, and nuclear $\beta$-decay. Currently, the most precise determination of the value of $V_{ud}$ is obtained with the superallowed $0^+$-$\,0^+$ nuclear $\beta$ decays. Since both initial and final nuclei have no spin, only the vector current interaction with the nucleus contributes at leading order. The conservation of the vector current (CVC) protects the vector coupling from being renormalized by the strong interaction and makes $0^+$-$\,0^+$ nuclear $\beta$ decays an especially robust method for determining $V_{ud}$. Precision tests require, apart from the purely experimental accuracy, an accurate computation of SM electroweak radiative corrections (RC). The present day framework for computing these corrections was formulated in the classic paper by Sirlin \cite{Sirlin:1977sv}, and subsequent refinements by Marciano and Sirlin (``MS") have represented the state-of-the-art for this topic (see, {\it e.g.}, Ref.~\cite{Marciano:2005ec}). In this context, the corrections are separated into the ``outer" correction that bears dependence on the electron spectrum, and ``inner" correction that is electron energy-independent. The outer corrections are generally nucleus-dependent, while the inner corrections contain both nucleus-dependent part and a universal, nucleus-independent contribution, $\Delta_R^V$. The latter also enters the rate for the decay of the free neutron.

The extraction of  $V_{ud}$ from superallowed decays relies on several inputs: (i) measurement of the reduced half-life $ft$ consisting of the 
half-life $t$ of the given decay channel and the respective branching ratio, and the statistical rate function $f$ which depends on the available phase-space (or $Q$-value) of a decay; \black (ii) extraction of $\mathcal{F}t$ -- a ``corrected $ft$ value", obtained by factoring out nucleus-dependent parts of the radiative corrections and nuclear structure-dependent parts of the transition matrix elements; (iii) combination of the results from a variety of decays to yield a global $\mathcal{F}t$ value. From the latter one obtains \cite{Hardy:2014qxa}: 
\begin{eqnarray}
\mathrm{Superallowed}\;\beta\;\mathrm{decays}\,:\;\left|V_{ud}\right|^2=\frac{2984.43s}{\mathcal{F}t(1+\Delta_R^V)}\ \ .
\label{eq:superallowed}
\end{eqnarray}
Here,
\begin{equation}
\mathcal{F}t=ft(1+\delta'_R)(1+\delta_{NS}-\delta_C),
\end{equation}
where $\delta'_R$ is the outer correction that depends on the electron energy and the charge $Z$ of the final nucleus and accounts for the Coulomb distortion and other QED effects; \black and $\delta_{NS},\delta_C$ are nuclear structure-dependent corrections that are independent of the electron energy. 

The analogous relationship for the free neutron is given by \cite{Czarnecki:2004cw}:
\begin{eqnarray}
\mathrm{Free}\;\;\mathrm{neutron}\,:\;\left|V_{ud}\right|^2=\frac{5099.34s}{\tau_n(1+3\lambda^2)(1+\Delta_R)}\,.
\label{eq:freen}
\end{eqnarray}
where $\tau_n$ is the neutron lifetime; $\lambda=g_A/g_V$ gives the ratio of the axial and vector nucleon charged current couplings;  and $\Delta_R$ includes both the outer corrections and $\Delta_R^V$. At present, one obtains a more precise value of $V_{ud}$ from superallowed decays via Eq.~(\ref{eq:superallowed}) than from neutron decay, despite the presence of the additional nuclear structure-dependent corrections that one must apply to obtain $\mathcal{F}t$. With the advent of future, more precise measurements of $\tau_n$ and neutron decay correlations that yield $\lambda$, the precision of the $V_{ud}$ determinations from nuclear and neutron decays may become comparable \cite{Kahlenberg:2017vnx,Maisonobe:2015len,Darius:2017arh,Broussard:2017tab}. 

In both cases, the dominant theoretical uncertainty for some time has been the hadronic contribution to the $W\gamma$ box diagram entering $\Delta_R^V$. The associated uncertainty has been obtained by MS in Ref.~\cite{Marciano:2005ec}, corresponding to a 0.018\% uncertainty in the value of $V_{ud}$. For the nuclear structure-dependent corrections, $\delta_{NS}$ and $\delta_C$ the analyses of Towner and Hardy (TH) have provided the canonical inputs used in the determination of the averaged $\mathcal{F}t$ values \cite{Hardy:2014qxa}. Importantly, application of the TH computed corrections to the most precisely measured superallowed transitions yields a nucleus-independent result of $\mathcal{F}t=3072.07(63)$s \cite{Hardy:2018zsb}, in impressive agreement with the CVC property of the SM charged current interaction. The associated nuclear structure uncertainty in $V_{ud}$ as obtained by TH is smaller than the hadronic uncertainty arising from $\Delta_R^V$. 

In what follows, we present new analyses of both sources of theoretical uncertainty using a dispersion relation framework that was recently presented in Ref.~\cite{Seng:2018yzq}. In that work, we applied this framework to the computation of $\Delta_R^V$, obtaining both a new value for this quantity and a significant reduction in the theoretical uncertainty. Below, we provide extensive details  entering that treatment which led to the value of $|V_{ud}|=0.97370(10)_{\mathcal{F}t}(10)_\mathrm{RC}$ quoted in that paper. In addition, we revisit one contribution to the nuclear structure-dependent correction $\delta_{NS}$ that is associated with the Born contribution to the nuclear transition amplitude. We point out that earlier work has omitted a significant contribution from the quasielastic (QE) nuclear response, and provide a first quantification of this contribution within the context of the dispersion relation framework. Combining the new hadronic and $\delta_{NS}$ results, we obtain 
\begin{equation}
|V_{ud}|=0.97395(21)_{\mathcal{F}t}(10)_\mathrm{RC} \ \ \ 
\end{equation}
that is to be compared with the present value quoted in the PDG: $|V_{ud}|=0.97420(10)_{\mathcal{F}t}(18)_{\mathrm{RC}}$. 

Our discussion of these analyses is organized as follows. In Section \ref{sec:basics} we provide a brief overview of the current situation for the experimental and theoretical treatment of both superallowed and free neutron $\beta$-decays. Section \ref{sec:DR} introduces the dispersion relation formalism. In Sections \ref{sec:input}-\ref{sec:resultsn} we present in detail the computation of the $W\gamma$ box contribution to $\Delta_R^V$. Section \ref{sec:nuclear} contains our analysis of the QE contribution to $\delta_{NS}$. In Section \ref{sec:relationtodata} we show how future measurements of parity-violating asymmetries in polarized electron-nucleon scattering may afford additional tests of the hadronic contribution to $\Delta_R^V$. We summarize our work in Section \ref{sec:conclude}. A variety of technical, computational details are given in a series of appendices.

\section{Neutron $\beta$-decay observables at tree- and one-loop level}
\label{sec:basics}
The differential decay rate of an unpolarized neutron is given at tree level by:
\begin{eqnarray}
\frac{d\Gamma}{dE_{e}} & = & \frac{G_{F}^{2}V_{ud}^{2}}{2\pi^3}(1+3\lambda^{2})|\vec{p}_{e}|E_{e}(E_{m}-E_{e})^{2}
\end{eqnarray}
where $E_m=(M_n^2-M_p^2+m_e^2)/2M_n$ is the maximum electron energy. The Fermi constant $G_F$ is that obtained from the muon decay. This definition allows to absorb an entire class of radiative corrections that are common to the muon and the neutron decay processes. Above, we denote $\lambda=g_A/g_V$;  
the vector and axial couplings ${g_V,\,g_A}$, corresponding to the Fermi and Gamow-Teller amplitudes in the neutron beta decay, respectively, are defined through the matrix element of the charged weak current in the non-recoil limit:
\begin{equation}
\left\langle p\right|\bar{u}\gamma^\mu(1-\gamma_5)d\left|n\right\rangle=\bar{u}_p\gamma^\mu(g_V-g_A\gamma_5)u_n
\end{equation}
and $g_V=1$ in the exact isospin limit due to CVC. 
The object of primary interest, the CKM matrix element $V_{ud}$ measures the coupling of the $W$ boson to the quark first generation in units of that to the leptons. 

Higher-order corrections to the tree-level expression, which do not make part of the one-loop result for $G_F$, 
modify the differential decay rate according to,
\begin{eqnarray}
\frac{d\Gamma}{dE_{e}}&=& \frac{G_{F}^{2}V_{ud}^{2}}{2\pi^3}(1+3\lambda^2)|\vec{p}_{e}|E_{e}(E_{m}-E_{e})^{2}F(\beta)\times\nonumber\\
&&\left(1+\frac{\alpha}{2\pi}\bar{g}(E_m)\right)
\left(1+\Delta_R^V\right).\label{eq:dGamma_mod}\end{eqnarray}

Above, the Coulomb interaction leads to the appearance of the Fermi function $F(Z,\beta)\approx1\pm Z\alpha\pi/\beta$ where $+ (-)$ should be taken for the electron (positron) in the final state, respectively. The Fermi function depends on the charged lepton velocity $\beta=|\vec{p}_e|/E_e$ and the atomic number of the final nucleus $Z$. The function $\bar{g}(E_m)$ is the outer correction to the decay rate: it represents the extreme infrared part of the radiative correction and is exactly calculable \cite{Sirlin:1967zza} (we refer the reader to an exhaustive review of outer corrections in Ref.~\cite{Hayen:2017pwg}). Meanwhile, the vector and axial coupling constants $g_V,g_A$ are also modified by radiative corrections,
\begin{eqnarray}
g_{V,A} & \rightarrow & g_{V}\left(1+\frac{1}{2}\Delta_R^{V,A}\right)
,\label{eq:replace}
\end{eqnarray}
with $\Delta_R^{V,A}$ the inner corrections to the neutron beta decay; they are constant numbers that depend on the details of hadronic structure but are independent of the electron spectrum, and are regular in the limit $m_e,E_e\rightarrow 0$. 
The already familiar term $\Delta_R^V$ introduced in Eq.~(\ref{eq:superallowed}) corrects the squared Fermi matrix element and makes part of the full radiative correction to the free neutron decay introduced in Eq.~(\ref{eq:freen}), 
\beqn
\Delta_R=(\alpha/2\pi)\bar{g}(E_m)+\Delta_R^V.
\eeqn
The analogous correction to the pure Gamow-Teller rate $\Delta_R^A$ is absorbed in the definition of $\lambda$ in Eq.~\eqref{eq:dGamma_mod} via 
\beqn
\lambda\rightarrow\lambda\left(1+\frac{1}{2}\Delta_R^A-\frac{1}{2}\Delta_R^V\right).
\eeqn 
If $\lambda$ is taken from an experimental measurement of the angular correlations, there is no need in evaluating $\Delta_R^A$ separately. If $g_A$ is computed instead, {\it e.g.} in lattice QCD, one has to include both inner corrections in a comparison of the theoretical and experimental values.
\black 
The analyses of Refs. \cite{Marciano:2005ec,Czarnecki:2004cw} give for the radiative correction to the free neutron decay, 
	\begin{eqnarray}
	\Delta_R&=&\frac{\alpha}{2\pi}
	\Big[\bar g(E_m)+3\ln\frac{M_Z}{M_p}+\ln\frac{M_Z}{\Lambda}+A_g\nn\\
	&&\quad\quad+2C_B+2C_{\rm INT}\Big]+0.0013.\label{eq:deltaRdef}
	\end{eqnarray}
The terms in the square bracket in Eq. \eqref{eq:deltaRdef} are the one-loop electroweak RC as presented in Ref. \cite{Marciano:2005ec}, whereas the +0.0013 originates from the resummation of leading-log corrections of the form $\alpha^n\ln^n(M_Z/M_p)$, $\alpha^n\ln^n(M_p/2E_m)$ as well as the inclusion of some important $\mathcal{O}(\alpha^2)$ effects \cite{Czarnecki:2004cw}. Among the one-loop radiative corrections, the first two terms in the square bracket originate from loop corrections and bremsstrahlung involving the electromagnetic and weak vector interactions, while the last four terms are due to a combination of the $\gamma W$ and $WZ$ boxes  in which the axial current interferes with the electromagnetic or neutral weak vector current. Out of these, the large logarithm and the perturbative QCD correction thereto $A_g$ originate from short distance contributions above some hadronic scale $\Lambda\sim1.5$ GeV, and are independent of the details of the hadronic structure. The appearance of a scale $\Lambda$ in the argument of the logarithm signals some residual sensitivity to the hadronic structure. This sensitivity is absorbed into the last two terms:
the term $C_B$ captures the long distance contributions; finally, $C_{\rm INT}$ contains an implicit dependence on the scale $\Lambda$ and was introduced in Ref. \cite{Marciano:2005ec} to interpolate between short and long-distance regimes. It is these two latter terms that dominate the uncertainty of $\Delta_R^V$. We postpone the detailed discussion of each term to the next Section, and note here that the splitting of the loop integral into short, long and intermediate distance contributions is rather arbitrary, and so are the uncertainties assigned to each contribution. It is our motivation to independently reassess the model dependent part of $\Delta_R$ and the uncertainty thereof in a data-driven dispersive approach. \black

\section{Dispersion representation of the ``inner" $\gamma W$-box correction to $g_V$.}
\label{sec:DR}
\begin{figure}[h]
	\begin{center}
		\includegraphics[width=6.5cm]{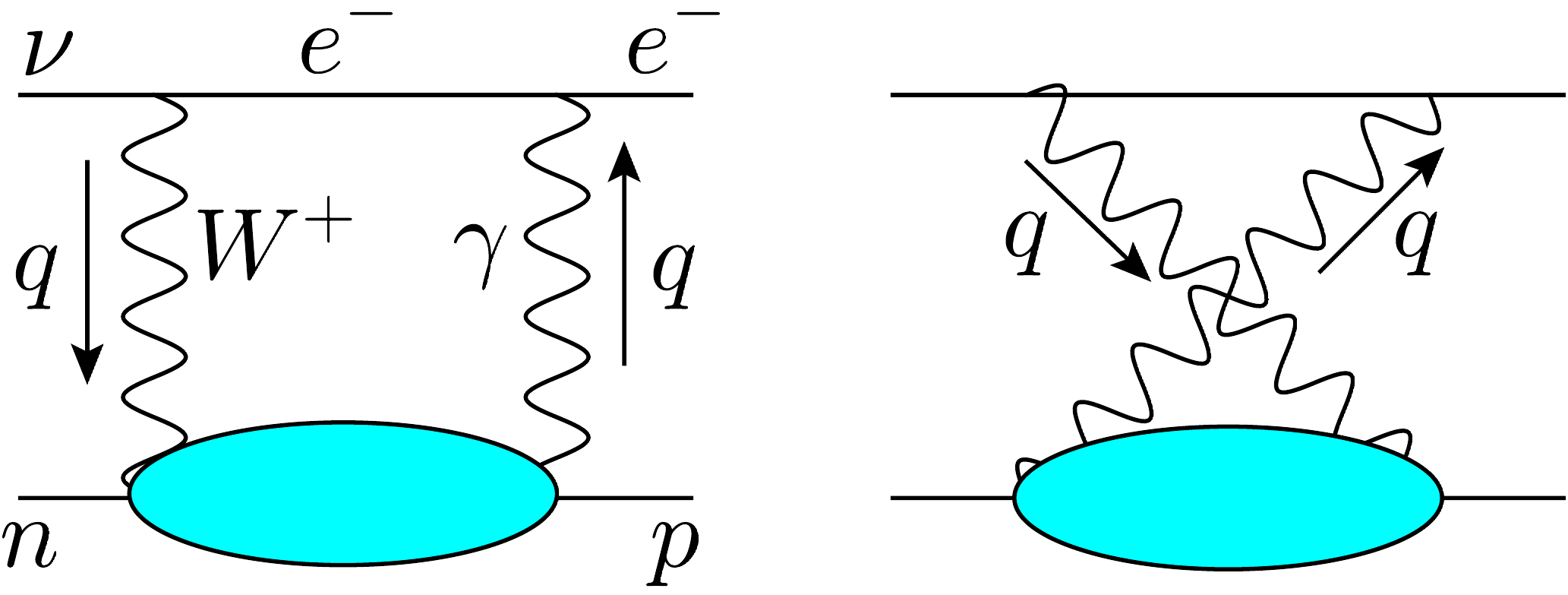}
		\caption{The $\gamma W$-box diagram relevant for the neutron decay. The blob represents the generalized forward Compton tensor.}\label{fig:Boxdiag}
	\end{center}
\end{figure}

The $\gamma W$-box correction is shown in Fig. \ref{fig:Boxdiag}, and is defined as
\beqn
&&T_{\gamma W}=-\sqrt{2}e^2G_FV_{ud}\label{eq:gaWloop}\\
&&\times\int\frac{d^4q}{(2\pi)^4}\frac{\bar u_e\gamma^\mu(\keldagger-\qdagger+m_e)\gamma^\nu(1-\gamma_5)v_\nu\,T_{\mu\nu}^{\gamma W}}{q^2[(k-q)^2-m_e^2][1-q^2/M_W^2]},\nn
\eeqn
where $k$ is the outgoing momentum of the electron. 
The forward generalized Compton tensor for the $\beta^-$ decay process $ W^+n\rightarrow\gamma p$ ($ W^-p\rightarrow\gamma n$ for the $\beta^+$ process relevant for nuclei) represented by the lower blob in Fig. \ref{fig:Boxdiag} is given by 
\beqn
T^{\mu\nu}_{\gamma W}=\frac{1}{2}\int dxe^{iq\cdot x}\langle p|T[J^\mu_{em}(x)J^{\nu}_W(0)]|n\rangle\label{eq:Compton}
\eeqn
with the following definitions of the electromagnetic and charged weak current:
\begin{eqnarray}
J_{em}^\mu&=&\frac{2}{3}\bar{u}\gamma^\mu u-\frac{1}{3}\bar{d}\gamma^\mu d\nonumber\\
J_W^\mu&=&\bar{u}\gamma^\mu(1-\gamma_5) d.
\end{eqnarray}
Notice that the definition of $T_{\gamma W}^{\mu\nu}$ above follows that in {the seminal paper by Sirlin ~\cite{Sirlin:1977sv}. The apparent extra factor of 1/2 is due to the difference in the normalization of the charged weak current: Sirlin defined $J_w^\mu=\bar{u}_L\gamma^\mu d_L$ (in the $V_{ud}=1$ limit) whereas our definition is two times larger, as the later is a more common definition in modern theory and experimental papers.

As the box diagram contains only one heavy boson propagator, it receives contribution from the loop momentum $q$ at all scales, ranging from infrared (i.e. $q\sim m_e$) to ultraviolet. The infrared-singular piece in $T_{\gamma W}$, together with the electron and proton wavefunction renormalization, as well as the real-photon bremsstrahlung diagrams, give rise to the Fermi function $F(\beta)$ and the outer-correction $\bar g(E_m)$ which are known analytically. In the meantime, most parts of the inner corrections from $T_{\gamma W}$ to $g_V$ are either exactly known due to current algebra, or depend only on physics at high scale and are calculable perturbatively. The only piece that depends on the physics at the hadron scale involves the vector-axial vector correlator in $T_{\gamma W}^{\mu\nu}$. Following a notation similar to that in Ref.~\cite{Marciano:2005ec}, we define its correction to the tree-level $W$ exchange Fermi amplitude as 
\beqn
T_{W}+T_{\gamma W}^{VA}=-{\sqrt2}{G_FV_{ud}}\left(1+\Box^{VA}_{\gamma W}\right)\bar u_e\pdagger(1-\gamma_5) v_\nu,\nn\\
\label{eq:BoxgaW}
\eeqn
so that it is straightforwardly connected to the universal radiative correction $\Delta_R^V$ via
\begin{equation}\Box^{VA}_{\gamma W}=\frac{1}{2}\left(\Delta_R^V\right)_{\gamma W}^{VA}.
\end{equation}
\black

The explicit expression of $\Box^{VA}_{\gamma W}$ is given by:
\begin{equation}
\Box^{VA}_{\gamma W}=4\pi\alpha\mathrm{Re}\int\frac{d^4q}{(2\pi)^4}\frac{M_W^2}{M_W^2+Q^2}\frac{Q^2+\nu^2}{Q^4}\frac{T_3(\nu,Q^2)}{M\nu}\label{eq:boxexplicit}
\end{equation}
where $Q^2=-q^2$, $\nu=p\cdot q/M$ with $M$ the average nucleon mass, and $T_3(\nu,Q^2)$ the parity-odd spin-independent invariant amplitude of the forward Compton tensor $T^{\mu\nu}_{\gamma W}$ defined through:
\beqn
T^{\mu\nu}_{\gamma W}=\left[-g^{\mu\nu}+\frac{q^\mu q^\nu}{q^2}\right]T_1
+\frac{\hat p^\mu\hat p^\nu}{(p\cdot q)} T_2
+\frac{i\epsilon^{\mu\nu\alpha\beta}p_\alpha q_\beta}{2(p\cdot q)}T_3,\nn\\
\label{eq:Wmunu}
\eeqn
with $\hat p^\mu=p^\mu-q^\mu (p\cdot q)/q^2$. 
Notice that since $\Box^{VA}_{\gamma W}$ is insensitive to physics at the scale $q\sim m_e$, we have set $m_e,k\rightarrow 0$ as well as $m_n=m_p=M$ to arrive Eq.~\eqref{eq:boxexplicit}. Furthermore, the fact that the electromagnetic current comes as a mixture of an isoscalar and isovector permits a decomposition of the forward amplitude in two isospin channels,
\beqn
T_3=T_3^{(0)}+T_3^{(3)}.
\eeqn

We apply Cauchy's theorem to the definite isospin amplitudes $T_3^{(I)}(\nu,Q^2)$ ($I=0,3$) accounting for their singularities in the complex $\nu$ plane. These lie on the real axis: poles due to a single nucleon intermediate state in the $s-$ and $u$-channels at $\nu=\pm\nu_B=\pm\frac{Q^2}{2M}$, respectively, and unitarity cuts at $\nu\geq \nu_\pi$ and $\nu\leq-\nu_\pi$ where $\nu_\pi=(2Mm_\pi+m_\pi^2+Q^2)/(2M)$, $m_\pi$ being the pion mass. The contour is constructed such as to go around all these singularities, and is closed at infinity, see Fig. \ref{fig:contour}. 
\begin{figure}
	\begin{center}
		\includegraphics[width=6.5cm]{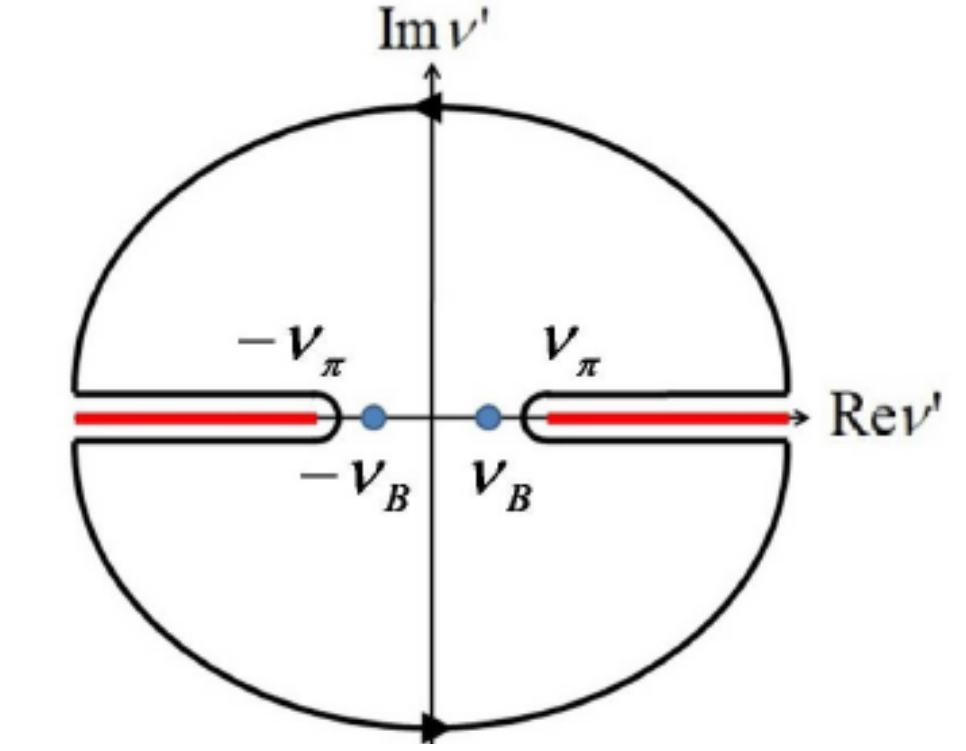}
		\caption{The contour in the complex $\nu$ plane.}\label{fig:contour}
	\end{center}
\end{figure}
The discontinuity of the forward amplitude in the physical region (i.e. $\nu>0$) is given by the generalization of the DIS structure functions to the $\gamma W$-interference in the standard normalization, 
\beqn
{\mathrm {Dis}}\,T_3^{(I)}(\nu,Q^2)&=&T_3^{(I)}(\nu+i\epsilon,Q^2)-T_3^{(I)}(\nu-i\epsilon,Q^2)\nn\\
&=&4\pi F^{(I)}_3(\nu,Q^2)
\eeqn
where
\begin{widetext}
\begin{eqnarray}
W_{\gamma W}^{(I)\mu\nu}=\frac{1}{8\pi}\sum_X(2\pi)^4\delta^4(p+q-p_X)\left\langle p\right|J_{em}^{(I)\mu}\left|X\right\rangle\left\langle X\right|J_W^\nu\left|n\right\rangle
=\left[-g^{\mu\nu}+\frac{q^\mu q^\nu}{q^2}\right]F_1^{(I)}
+\frac{\hat p^\mu\hat p^\nu}{(p\cdot q)} F_2^{(I)}+\frac{i\epsilon^{\mu\nu\alpha\beta}p_\alpha q_\beta}{2(p\cdot q)}F_3^{(I)},\nn\\
\end{eqnarray}
\end{widetext}
(we define $W_{\gamma W}^{\mu\nu}$ with a coefficient of $(8\pi)^{-1}$ instead of the more common $(4\pi)^{-1}$ to keep in sync with our definition of $T_{\gamma W}^{\mu\nu}$ that contains a factor 1/2) and for the sake of a unified description, within $F_i^{(I)}$ we keep both the $\delta$-functions at the nucleon poles, and the discontinuities along the multi-particle cuts. The full function $T_3^{(I)}(\nu,Q^2)$ is reconstructed from a fixed-$Q^2$ dispersion relation
\beqn
T_3^{(I)}(\nu,Q^2)=\frac{2}{i}\int\limits_0^\infty d\nu'\left[\frac{1}{\nu'-\nu}+\frac{\xi^I}{\nu'+\nu}\right]F_3^{(I)}(\nu',Q^2),\nn\\
\label{eq:DR}
\eeqn
modulo possible subtractions which are needed to make the dispersion integral convergent. 
The form of the dispersion relation depends on the crossing behavior, the relative sign $\xi^I$ between the contributions along the positive and negative real $\nu$ axis. It can be shown that the isoscalar amplitude is an odd function of $\nu$, hence $\xi^0=-1$, while the isovector amplitude is even (see Appendix \ref{app:crossing}). Correspondingly, the isoscalar requires no subtractions, while the isovector one may have to be subtracted one time.

Putting together Eqs. (\ref{eq:boxexplicit},\ref{eq:DR}) and performing the loop integral via Wick rotation we arrive at 
\beqn
\Box^{VA\,{(0)}}_{\gamma W}&=&\frac{\alpha}{\pi M}\int\limits_0^\infty \frac{dQ^2 M_W^2}{M_W^2+Q^2} \int\limits_0^\infty d\nu \frac{(\nu+2q)}{\nu(\nu+q)^2}F_3^{(0)}(\nu,Q^2),\nn\\
\Box^{VA\,{(3)}}_{\gamma W}&=&0,\label{eq:BoxDR}
\eeqn
where we introduced the virtual photon three-momentum $q=\sqrt{\nu^2+Q^2}$.
The vanishing of the isovector contribution is the consequence of the crossing symmetry, as has already been noticed by Sirlin \cite{Sirlin:1967zza}. Thus from now onward we shall represent $\Box_{\gamma W}^{VA,(0)}$ simply by $\Box_{\gamma W}^{VA}$ without causing any confusion. Changing the variables $\nu\to Q^2/(2Mx)$ we notice that the $x$ integral is, up to a factor, precisely the first Nachtmann moment of the structure function $F_3^{(0)}$,
\beqn
\int\limits_0^\infty d\nu \frac{(\nu+2q)}{M\nu(\nu+q)^2}F_3^{(0)}(\nu,Q^2)=\frac{3}{2Q^2}M_3^{(0)}(1,Q^2).
\eeqn
The definition of the Nachtmann moments of $F_3$ reads \cite{Nachtmann:1973mr,Nachtmann:1974aj}
\beqn
M_3^{(0)}(N,Q^2)=\frac{N+1}{N+2}\int\limits_0^1\frac{dx\xi^{N}}{x^2}\left[2x-\frac{N\xi}{N+1}\right]F_3^{(0)}
,\nn\\
\eeqn
where we introduced the Nachtmann variable $\xi=2x/(1+\sqrt{1+4M^2x^2/Q^2})$. This gives our master formula
\beqn
\Box^{VA}_{\gamma W}&=&\frac{3\alpha}{2\pi}\int_0^\infty \frac{dQ^2 M_W^2}{Q^2[M_W^2+Q^2]}M_3^{(0)}(1,Q^2).\label{eq:boxNachtmann}
\eeqn

In the old result by MS this connection was not written explicitly, 
\beqn
\Box^{VA}_{\gamma W}&=&\frac{\alpha}{8\pi }\int_0^\infty \frac{dQ^2 M_W^2}{M_W^2+Q^2} F(Q^2),\label{eq:BoxMS}
\eeqn
and we simply note the correspondence,
\beqn
F(Q^2)=\frac{12}{Q^2}M_3^{(0)}(1,Q^2).
\eeqn
This is the first essentially new result of our work. 

\begin{figure}[h]
\begin{center}
\includegraphics[width=8.0cm]{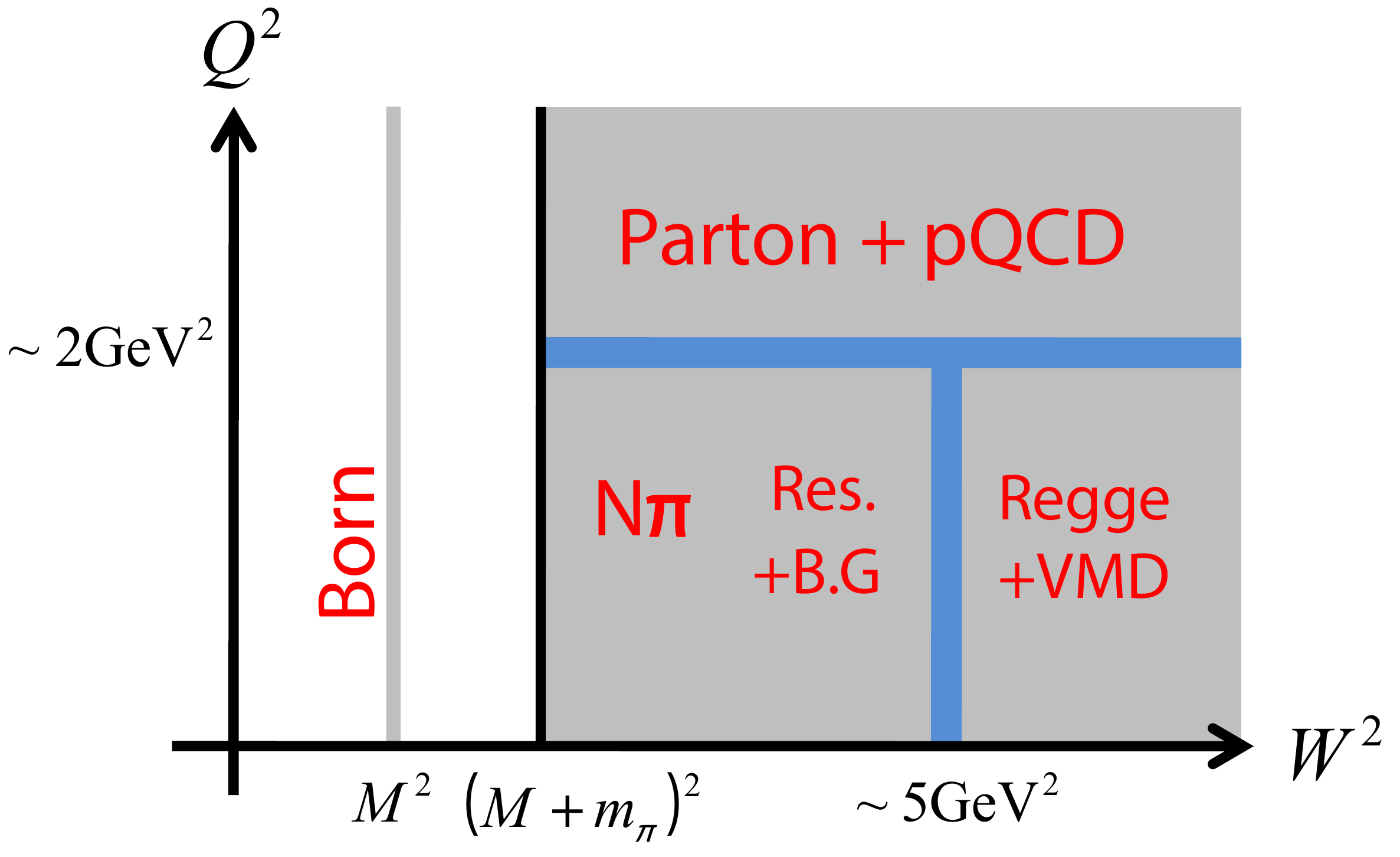}
\caption{The $W^2-Q^2$ diagram showing approximate kinematical regions which are dominated by various physical mechanisms, as indicated on the plot.}
\label{fig:W-Q2diag}
\end{center}
\end{figure}
We let the data guide us to evaluate the integral in Eq. (\ref{eq:BoxDR}): for a fixed value of $Q^2$ one has to integrate over the full spectrum in energy, and then sample all values of $Q^2$ from 0 to $\infty$. The strength is distributed differently among different energy regimes depending on $Q^2$. For low $Q^2$ the spectrum is heavily weighted towards lower part (elastic peak and resonances). As $Q^2$ grows, these contributions are however suppressed by the respective form factors. High-energy spectrum for slightly virtual and high-energy photons extends to asymptotically high energies and is well-represented by Regge exchanges. Already at moderate $Q^2\sim1.5-2.5$ GeV$^2$ this picture fades away and smoothly joins onto the partonic description which dominates the DIS regime. The regions corresponding to various physics pictures are displayed on a plane $\{W^2, Q^2\}$ with $W^2=M^2+2M\nu-Q^2$ in Fig. \ref{fig:W-Q2diag}. 
Accordingly,  our parameterization of $F_3^{(0)}$ is as follows:
\begin{equation}\label{eq:ourparam}
F_3^{(0)}\!\!=F^{(0)}_{3,\,\text{Born}}+
\begin{cases}
F^{(0)}_{3,\,\text{pQCD}}, & \!\! Q^2 \gtrsim 2\text{ GeV}^2 \\[2mm] 
\!F^{(0)}_{3,\,\pi N}\!+\!F^{(0)}_{3,\,\text{res}}\!+\!F^{(0)}_{3,\,\mathbb{R}}, & \!\!Q^2 \lesssim 2\text{ GeV}^2\,,
\end{cases}
\end{equation}
where each component supplies the dominant contribution to $F_3^{(0)}$ in various regions (elastic or Born; pQCD at high $Q^2$; $\pi N$, resonance and Regge at low $Q^2$ , respectively). 
In the following two Sections we report the procedure of relating data on inclusive charged current neutrino and neutral current electron scattering to the $\gamma-W$ interference needed to evaluate the $\gamma W$-box correction, and collect available information on the entire $\{W^2, Q^2\}$ plane.\\

Our approach can be compared with that of MS where the energy dependence is integrated over, and only regions in $Q^2$ are considered to identify $F(Q^2)$ with a certain contribution assumed to be dominant in that particular region,
\beqn
F(Q^2)=
\left\{\begin{array}{l}
F^{\rm Born}(Q^2),\;\;{Q^2\leq(0.823\,{\rm GeV})^2}\,,\\
\\
F^{\rm INT}(Q^2),\;\;{(0.823\,{\rm GeV})^2\leq Q^2\leq(1.5\,{\rm GeV})^2 },\\
\\
F^{\rm DIS}(Q^2),\;\;{Q^2\geq(1.5\,{\rm GeV})^2 }\,.
\end{array}\right.\nn
\eeqn
The Born contribution $F^{\rm Born}$ depends on the isovector axial and isoscalar magnetic nucleon form factors, and is assumed to exhaust all contributions at low $Q^2$; upon inserting it in Eq.~(\ref{eq:BoxMS}) it gives the term $\sim C_B$ in Eq.~(\ref{eq:deltaRdef}). 

The DIS contribution $F^{\rm DIS}$ dominates at high $Q^2$ and contains the parton model expectation and perturbative QCD corrections thereto,
\beqn
F^{\rm DIS}(Q^2)=\frac{1}{Q^2}\left[1-\sum_{i=1}^3C_i\left(\frac{\bar\alpha_s(Q^2)}{\pi}\right)^i\right],
\eeqn
with $\bar\alpha_s$ the strong coupling constant evaluated in the $\overline{\rm MS}$ scheme at the scale $Q^2$. Further details to this contribution are discussed in the following two Sections. 
When inserted in the integral in Eq.~(\ref{eq:BoxMS}) and integrated from $\Lambda^2$ to $\infty$, this contribution gives rise to the terms $\sim \ln(M_Z/\Lambda)+A_g$
in Eq.~(\ref{eq:deltaRdef}). Note that the mass of the $Z$-boson appears in that formula upon combining the $\gamma W$- and $ZW$-boxes together. 

The phenomenological interpolating function $F^{\rm INT}$ connects the two regions. 
Ref. \cite{Marciano:2005ec} proposed to take it in a vector dominance model (VDM)-motivated form, 
\beqn
F^{\rm INT}(Q^2)=-\frac{1.490}{Q^2+m_\rho^2}+\frac{6.855}{Q^2+m_A^2}-\frac{4.414}{Q^2+m_{\rho'}^2},
\eeqn
with $m_\rho=0.776$ GeV, $m_A=1.230$ GeV and $m_{\rho'}=1.465$ GeV, and numerical coefficients were obtained by imposing three constraints: 
\beqn
{\rm I}&&\int\limits_{\Lambda^2}^\infty \frac{dQ^2M_W^2}{M_W^2+Q^2}\left[F^{\rm INT}(Q^2)-F^{\rm DIS}Q^2)\right]=0\nn\\
{\rm II}&&Q^2F^\mathrm{INT}(Q^2)-\lim_{Q^2\rightarrow\infty}
	\left(Q^2F^\mathrm{INT}(Q^2)\right)=O\left(\frac{1}{Q^4}\right)\nn\\
{\rm III}&&F^{\rm INT}(0)=0. \label{eq:Fint}
\eeqn
Finally, the matching point $Q=0.823$ GeV is determined by requiring that $F^\mathrm{Born}(Q^2)=F^\mathrm{INT}(Q^2)$ at that point.
Upon integrating $F^{\rm INT}$ over the respective range in $Q^2$ in Eq.~(\ref{eq:BoxMS}) it gives the term $\sim C^{\rm INT}$ in Eq.~(\ref{eq:deltaRdef}).

Among the three conditions in Eq.~(\ref{eq:Fint}), the condition III requires that the following superconvergence relation is satisfied exactly,
\beqn
\int_{\nu_\pi}^\infty (d\nu/\nu^2)F_3^{(0)}(\nu,Q^2=0)=0. \label{eq:SCR}
\eeqn
\indent
To the validity of this conjecture, Ref. \cite{Marciano:2005ec} asserts that this is required by chiral perturbation theory (ChPT), and a more detailed proof will be reported in an upcoming work. Unfortunately, this proof has never been published. In Appendix \ref{app:piN} we perform an explicit calculation in relativistic ChPT and demonstrate that the integral in Eq. (\ref{eq:SCR}) does not vanish.

\section{Physics input to $F_3^{(0)}$ and $F_3^{WW}$ 
}
\label{sec:input}
It is informative to take a look at the general structure of the virtual photoabsorption spectrum displayed in Fig. \ref{fig:spectrum}. For a fixed value of $Q^2$ one clearly sees three major structures as one goes from low to high energy $\nu$: elastic peak at $Q^2/(2M)$ (broadened by radiative corrections); nucleon resonances and non-resonant pion production starting from the pion threshold $[Q^2+(M+m_\pi)^2-M^2]/(2M)$ and up to roughly 2.5 GeV above the threshold; high-energy continuum corresponding to multi-particle production that, depending on the value of $Q^2$, can be economically described by $t$-channel Regge exchanges (low $Q^2$) or quasi-free quark knock-out in the deep-inelastic regime (high $Q^2$). Exactly the same structure is expected in neutrino scattering associated with the absorption of a virtual $W$-boson.  
\begin{figure}[h]
\begin{center}
\includegraphics[width=8.0cm]{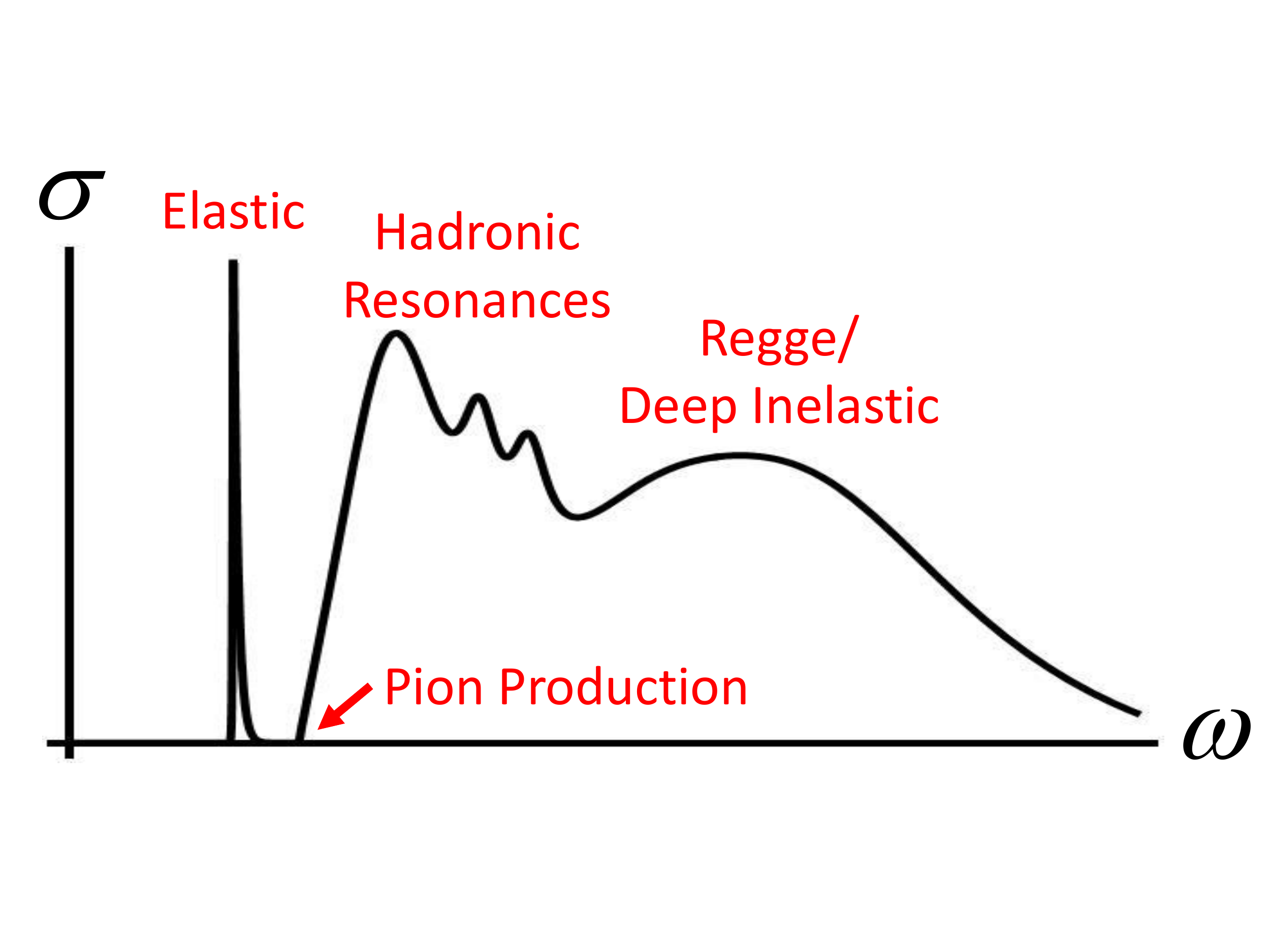}
\caption{Idealized structure of virtual photoabsorption on the nucleon.}
\label{fig:spectrum}
\end{center}
\end{figure}

While $F_3^{(0)}(\nu,Q^2)$ itself is not observable, its weak isospin partner  $F_3^{WW}(\nu,Q^2)$  is directly accessible in neutrino and antineutrino deep inelastic scattering. The two-fold differential cross section at fixed $Q^2$ as function of $x=Q^2/(2M\nu)$ and $y=\nu/E$, with $E$ the initial neutrino energy and $\nu$ the virtual $W$ laboratory frame energy, reads \cite{Onengut:2005kv}
\beqn
&&\frac{d^2\sigma^{\nu(\bar\nu)}}{dxdy}=\frac{G_F^2 ME}{\pi\left(1+Q^2/M_W^2\right)^2}\\
&&\times\left[xy^2F_1+\left(1-y-\frac{Mxy}{2E}\right)F_2\pm
x\left(y-\frac{y^2}{2}\right)F_3\right].\nn
\eeqn
The P-odd structure functions $F_3^{\nu p(\bar{\nu}p)}$ of our interest follow standard definitions:
\begin{widetext}
\begin{eqnarray}
\frac{1}{4\pi}\sum_X(2\pi)^4\delta^4(p+q-p_X)\left\langle p\right|\left(J_W^\mu\right)^\dagger\left|X\right\rangle\left\langle X\right|J_W^\nu\left|p\right\rangle&=&\frac{i\epsilon^{\mu\nu\alpha\beta}p_\alpha q_\beta}{2(p\cdot q)}F_3^{\nu p}+...\nonumber\\
\frac{1}{4\pi}\sum_X(2\pi)^4\delta^4(p+q-p_X)\left\langle p\right|J_W^\mu\left|X\right\rangle\left\langle X\right|\left(J_W^\nu\right)^\dagger\left|p\right\rangle&=&\frac{i\epsilon^{\mu\nu\alpha\beta}p_\alpha q_\beta}{2(p\cdot q)}F_3^{\bar{\nu} p}+...\label{eq:F3nupdef}
\end{eqnarray}
\end{widetext}
and their average, $F_3^{\nu p+\bar\nu p}=\frac{1}{2}[F_3^{\nu p}+F_3^{\bar\nu p}]$ can be obtained from the difference of the neutrino and antineutrino cross sections.

We follow the general structure of the parametrization of $F_3^{(0)}$ specified in Eq. (\ref{eq:ourparam}), and describe $F_3^{\nu p+\bar\nu p}$ at $Q^2\leq2$ GeV$^2$ as a sum of elastic (Born) contribution, non-resonant $\pi N$ continuum, several low-lying $\Delta$ and $N^*$-resonances, and the high-energy Regge contribution,
\beqn
F_{3,\,{\rm low}-Q^2}^{\nu p+\bar\nu p}=F_{3,\,\rm Born}^{\nu p+\bar\nu p}+F_{3,\,\pi N}^{\nu p+\bar\nu p}+F_{3,\,\rm res}^{\nu p+\bar\nu p}+F_{3,\,{\rm \mathbb{R}}}^{\nu p+\bar\nu p}.\nn\\
\label{eq:F3lowQ2}
\eeqn
Details to the elastic, $\pi N$ and resonance contributions are given in the Appendix. Since the Regge contribution plays a central role in our model, we give its explicit form here. We assume that it completely dominates at high energies, for $W\geq2.5$ GeV. At lower energies, we assume that above the two-pion production threshold $W_{th}^2=(M+2m_\pi)^2$ the Regge amplitude with an appropriate smooth threshold factor 
$f_{th}(W)=\Theta(W^{2}-W_{th}^{2})\left(1-\exp\left\{ \frac{W_{th}^{2}-W^{2}}{\Lambda_{th}^{2}}\right\} \right)$
represents on average the contribution of multi-pion and higher energy channels, 
\begin{equation}
F_{3,\,\mathbb{R}}^{\nu p+\bar\nu p}(\nu,Q^{2})=
\frac{C(Q^2)f_{th}(W)}{\left[1+{Q^2}/{m_\rho^2}\right]\left[1+{Q^2}/{m_{a_1}^2}\right]}\left(\frac{\nu}{\nu_{0}}\right)^{\alpha_{0}}
\label{eq:Regge}
\end{equation}
The Reggeized $\omega$-exchange is well described by the Regge intercept $\alpha_0\approx0.477$ \cite{Kashevarov:2017vyl}, and we choose the parameters $\nu_0=\Lambda_{th}=1$ GeV. To continue the Regge amplitude to finite $Q^2$ we assume vector (axial) meson dominance which is reflected in the usual VDM form factors above. We found however, that the pure VDM does not describe the data, so we added a phenomenological $Q^2$-dependent function $C(Q^2)$ which is obtained from a fit. That the pure VDM drops short of the virtual photoabsorption data is well-known. This fact has motivated various generalizations of the VDM which also feature phenomenological ingredients that are needed to account for this missing strength. Given the quality of the data, a simple linear form of $C(Q^2)$ was enough to describe the combined BEBC and Gargamelle data 
n the range  $Q^2\in(0.15,2.0)$ GeV$^2$ \cite{Bolognese:1982zd},
\beqn
C(Q^2) = A(1+B Q^2),
\eeqn
with $A=5.2\pm1.5$ and $B=1.08^{+0.48}_{-0.28}$. The two parameters are strongly anti-correlated. 

Above $Q^2=2$ GeV$^2$ we use the pQCD result for the Mellin moment with N$^3$LO corrections calculated in Ref. \cite{Larin:1991tj}.  
\beqn
M_3^{\nu p+\bar\nu p}(1,Q^2)=3
\left[1-\sum_{i=1}^3 C_i\left(\frac{\bar\alpha_s}{\pi}\right)^i
\right],
\eeqn
with $C_1=1$, $C_2=4.583-0.333 N_f$ and $C_3=41.440-8.020N_F+0.177N_F^2$, $N_f=3$ standing for the number of effective quark flavors, 
and $\bar\alpha_s (Q^2)$ denotes the running strong coupling constant in the modified minimal subtraction scheme with $\Lambda_{QCD}=0.2$ GeV. Note that for $Q^2\geq2$ GeV$^2$ the difference between the Nachtmann and Mellin moments is negligible. 

\begin{figure}[h]
\begin{center}
\includegraphics[width=8.0cm]{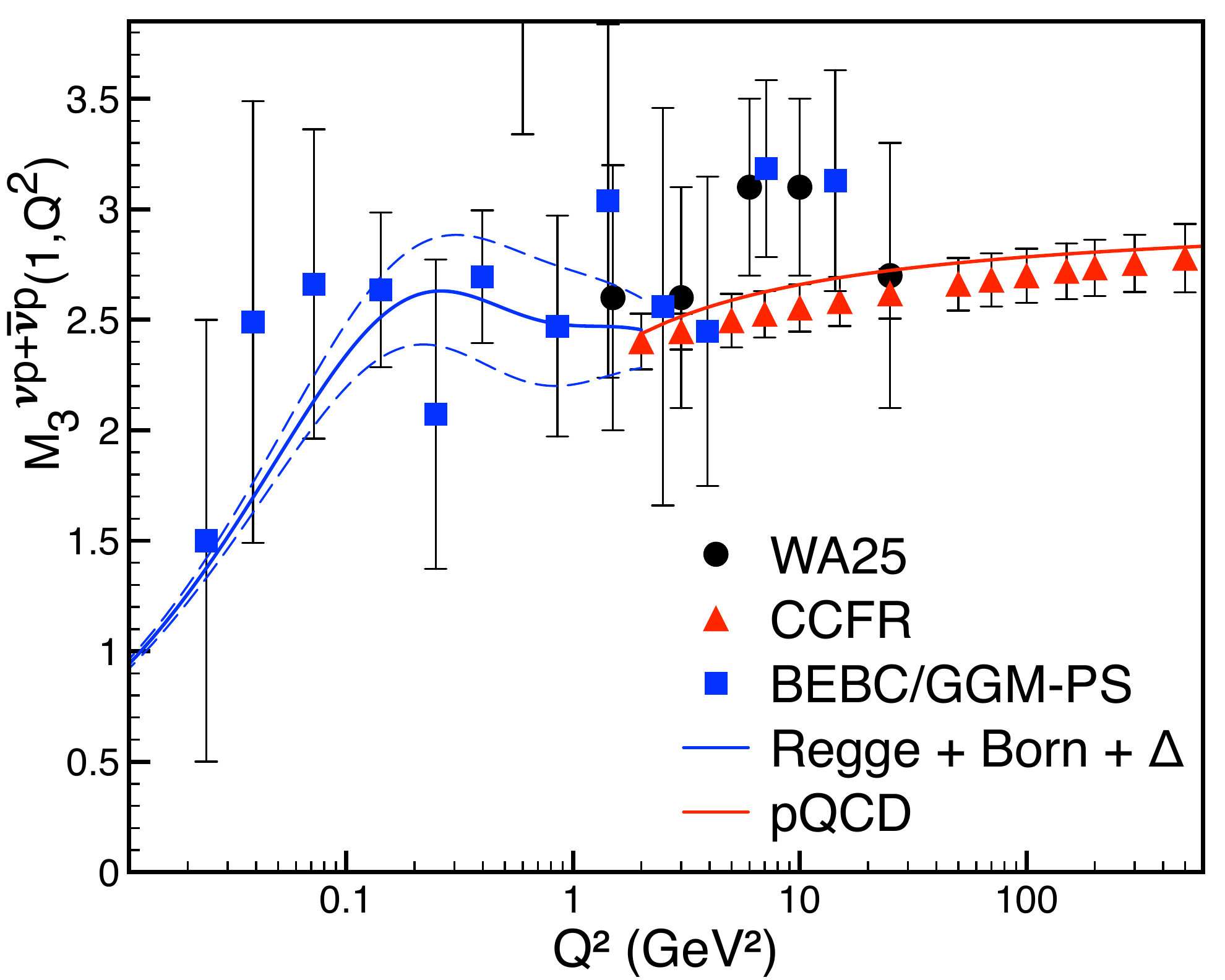}
\caption{Data on the first Nachtmann moment of $F_3^{\nu p+\bar\nu p}$ from CCFR \cite{Kataev:1994ty,Kim:1998kia}, BEBC/Gargamelle \cite{Bolognese:1982zd} and WA25 \cite{Allasia:1985hw} vs. theory. Figure adopted from Ref. \cite{Seng:2018yzq}}
\label{fig:GLSSR}
\end{center}
\end{figure}

In Fig. \ref{fig:GLSSR} we display the world data on the Nachtmann moment of $F_3^{\nu p+\bar\nu p}$ for $Q^2\in(0.01,600)$ GeV$^2$. The solid red curve shows the pQCD result of Ref. \cite{Larin:1991tj} which can be seen to nicely agree with the CCFR data \cite{Kataev:1994ty,Kim:1998kia} at $Q^2\geq2$ GeV$^2$. The solid blue curve at lower $Q^2$ shows the result of our low-$Q^2$ model as described in Eq. (\ref{eq:F3lowQ2}), and the uncertainty is represented by the dashed blue curves around it. We do not use the three left-most data points in the fit because we expect $M_3^{\nu p+\bar{\nu}p}(1,Q^2)$ to be saturated at $Q^2<0.1$ GeV$^2$ by the elastic and $\Delta$-resonance contribution \cite{Bolognese:1982zd} which are determined using more precise lower energy modern data. In our formalism, the theoretical uncertainty in the intermediate-$Q^2$ region is determined by that of the $\nu p/\bar{\nu}p$-scattering data which can be systematically improved when future, more precise data become available. This represents an advantage over the MS formalism where the physics at intermediate distances had to be  assigned a 100\% uncertainty.

We note here that while our use of a Regge-VDM parametrization of the contributions at low $Q^2$ and high energy is model-dependent, no other model describes inclusive electron scattering data in that kinematical range. Moreover, our parametrization of $F_3^{\nu p+\bar\nu p}$ can be tested explicitly by confronting it to high-energy electron spectra in inclusive CC neutrino scattering, rather than to the Nachtmann moment as we do here. Also the key ingredient of our parametrization, the effective $a_1-\rho-\omega$ vertex can be tested in exclusive neutrinoproduction of $\omega$ mesons, and in exclusive $a_1$-electroproduction. We will address the exact formulation of these tests with the existing and future data in an upcoming work.

\section{Relating Nachtmann moments of $F_3^{\nu p+\bar\nu p}$ and $F_3^{(0)}$ by isospin symmetry}
\label{sec:isospin}

After having modeled the pure CC structure function $F_3^{\nu p+\bar\nu p}$ as a sum of elastic, resonances, non-resonant $\pi N$ and Regge, we proceed to obtain $F_3^{(0)}$ via isospin rotation. This is done for each contribution separately. 
For the elastic contribution, since the intermediate is fixed at $I=1/2$ the correspondence between the two processes is simple:
\beqn
F_{3,\,\rm Born}^{\nu p+\bar\nu p}&=&-G_A(Q^2)G_M^V(Q^2)\delta(1-x),\nn\\
F_{3,\,\rm Born}^{(0)}&=&-\frac{1}{4}G_A(Q^2)G_M^S(Q^2)\delta(1-x),\label{eq:Bornisospin}
\eeqn
with the axial form factor normalized as $G_A(0)=-1.2715$, and magnetic isovector and isoscalar form factors $G_M^{V,S}(0)=\mu_p\pm\mu_n$, with the proton (netron) magnetic moment $\mu_p=2.792847356$ ($\mu_n=-1.9130427$). So the difference is simply between the isoscalar and the isovector component of the electromagnetic matrix element and an extra constant factor. 

For resonance contributions, a correspondence similar to Eq. \eqref{eq:Bornisospin} may also be stated, but with a caveat. 
The purely isovector structure function $F_3^{\nu p+\bar{\nu}p}$ receives contributions from both $I=1/2$ and $I=3/2$ resonances, with the contributions of the latter, most notably the $\Delta(1232)$, dominating over the contributions of the $I=1/2$ resonances.
Instead, only $I=1/2$ resonances contribute to $F_3^{(0)}$. The details of the calculation are given in Appendix \ref{app:resonances}. 

The Regge contribution is depicted in Fig. \ref{fig:Regge}. 
\begin{figure}[h]
	\begin{center}
		\includegraphics[width=5.0cm]{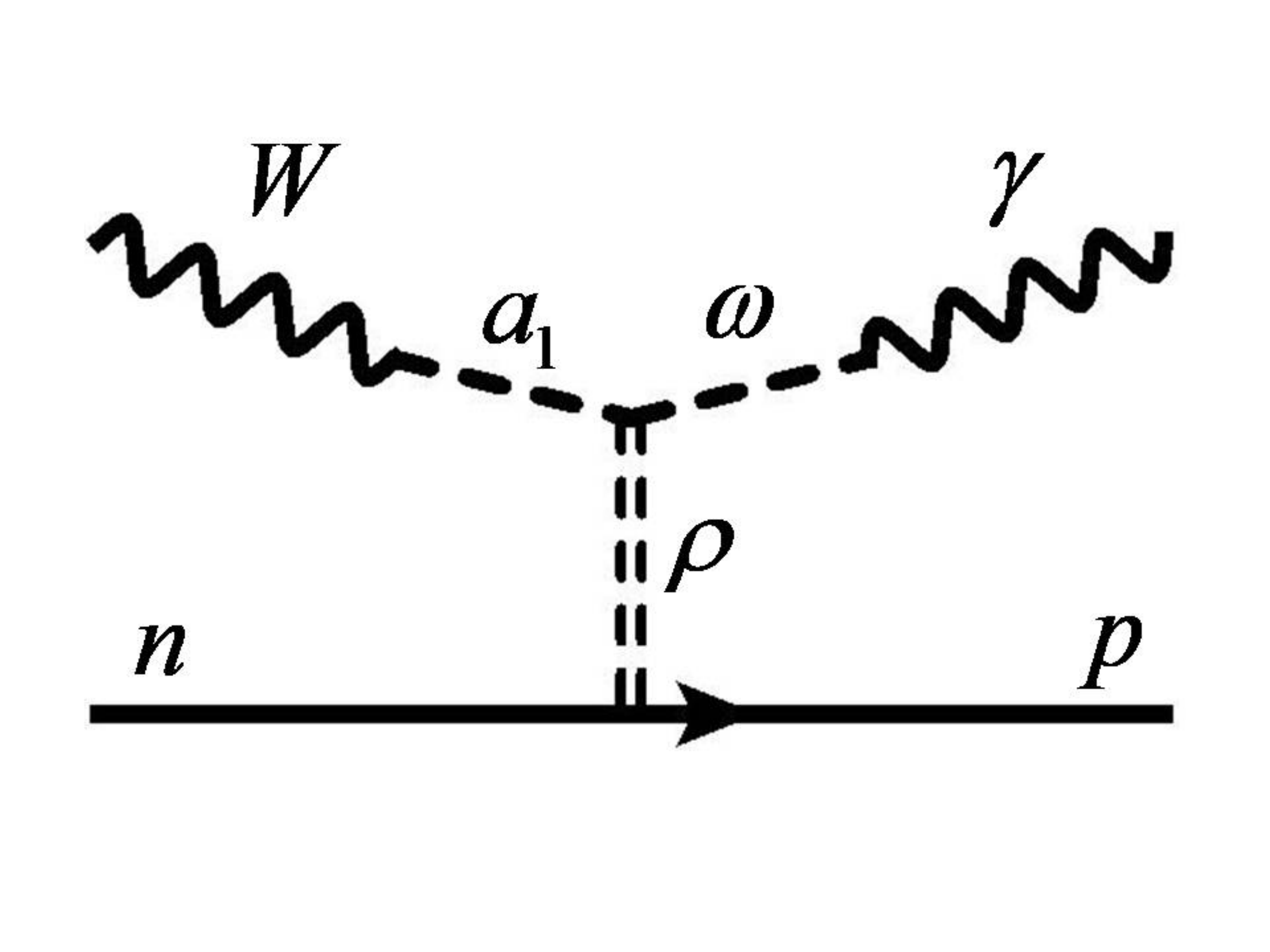}
		\includegraphics[width=5.0cm]{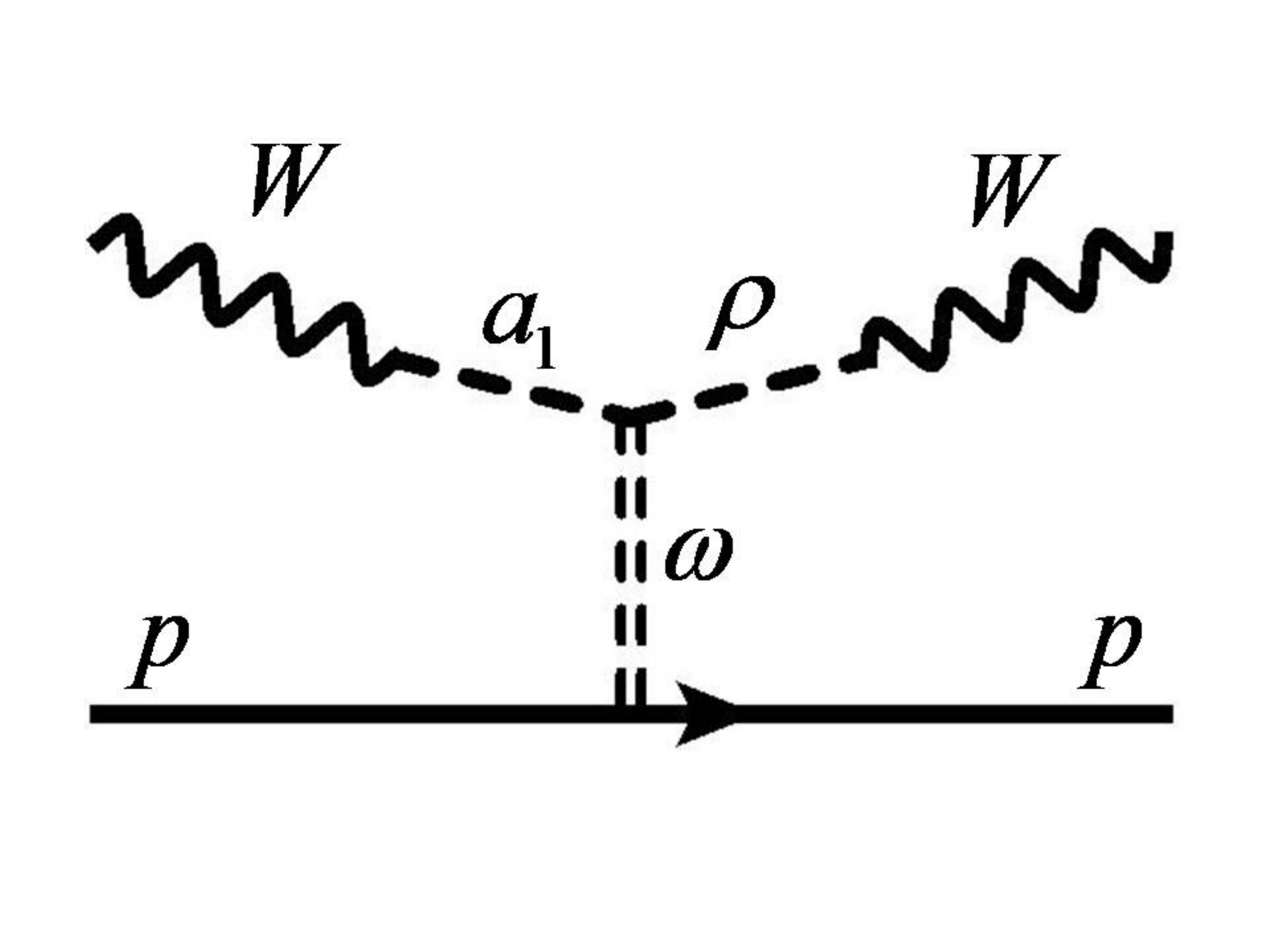}
		\caption{Regge-model description of $F_3^{(0)}$ and $F_3^{\nu p+\bar{\nu}p}$.}\label{fig:Regge}
	\end{center}
\end{figure}
It is seen that the central ingredient in this picture, the effective vertex $a_1-\rho-\omega$ is the same in both cases. Since the parameters of the $\rho$ and $\omega$ Regge trajectories and VDM propagators are nearly exactly the same, the only change would regard the respective coupling constants. As we discuss in detail in Appendix \ref{app:regge}, this entails relating the $\gamma-\omega$ and $\rho NN$ couplings entering the $\gamma W$ interference to $W-\rho$ and $\omega NN$ couplings entering the purely charge current structure function. Taking into account the correct normalization of various pieces, the isospin symmetry implies a rescaling of the Regge contribution to $F_3$ by a factor 1/36 at low $Q^2$,
\beqn
F_{3,\,\mathbb{R}}^{\nu p+\bar\nu p}
&=&
\frac{C(Q^2)f_{th}(W)}{\left[1+{Q^2}/{m_\rho^2}\right]\left[1+{Q^2}/{m_{a_1}^2}\right]}\left(\frac{\nu}{\nu_{0}}\right)^{\alpha_{0}}
\!\!\!,\\
&\downarrow&\nn\\
F_{3,\,\mathbb{R}}^{(0)}
&=&\frac{1}{36}
\frac{C_{\gamma W}(Q^2)f_{th}(W)}{\left[1+{Q^2}/{m_\rho^2}\right]\left[1+{Q^2}/{m_{a_1}^2}\right]}\left(\frac{\nu}{\nu_{0}}\right)^{\alpha_{0}}\!\!\!,\nn
\eeqn
and this rescaling straightforwardly translates in the respective change of the Nachtmann moment. 
Above, we explicitly indicate that the phenomenological $Q^2$-dependent functions $C_{WW}$ and $C_{\gamma W}$ do not have to be the same. Since we introduced  $C(Q^2)$ to correct for shortcomings (or incompleteness) of the minimal vector dominance model, it is not guaranteed that $C$ and $C_{\gamma W}$ are related anywhere except for the point $Q^2=0$. To address the shape of $C_{\gamma W}(Q^2)$, we consider the relation between the Nachtmann moments of the two structure functions at the upper limit of the applicability of our Regge parametrization, $Q^2=2$ GeV$^2$ where we can use the information from the DIS regime. 

In the parton model the relative normalization of $F_{3}^{(0)}$ with respect to $F_{3}^{\nu p+\bar\nu p}$ turns out to be $1/36$, as well. However, the running of the respective first moment has to be taken into account to extend the DIS description to $Q^2=2$ GeV$^2$. 
One of the central findings of Ref.~\cite{Marciano:2005ec} was that while the running of the first moment of $F_{3}^{\nu p+\bar\nu p}$ is fixed by the running of the GLS sum rule, that of the first moment of $F_{3}^{(0)}$ is fixed by the running of the Bjorken sum rule. Both sum rules were studied in Ref.~\cite{Larin:1991tj} in perturbative QCD at N$^3$LO, and we use their results, 
\beqn
M_3^{\nu p+\bar\nu p}(1,Q^2)&=&3
\left[1-\sum_{i=1}^3 C_i\left(\frac{\bar\alpha_s}{\pi}\right)^i\right],
\nn\\
&\downarrow&\nn\\
M_3^{(0)}(1,Q^2)&=&\frac{1}{12}
\left[1-\sum_{i=1}^3 \tilde C_i\left(\frac{\bar\alpha_s}{\pi}\right)^i\right].
\eeqn
The first two coefficients in the GLS and Bjorken sum rules are the same $\tilde C_{1,2}=C_{1,2}$, and only at N$^3$LO the difference appears: $\tilde{C}_3=41.440-7.607N_f+0.177N_F^2$ as compared to  ${C}_3=41.440-8.020N_f+0.177N_F^2$. Numerically, the change due to a 6\% shift in the value of the coefficient at $(\bar\alpha_s/\pi)^3$ is very small, and to a very good approximation the rescaling $1/36$ is thus valid for the full DIS contribution. 

Since at $Q^2=2$ GeV$^2$ our Regge contribution is matched onto the DIS one, the observed agreement of the $1/36$ rescaling rule at low and high $Q^2$ implies that $C_{\gamma W}(Q^2)=C(Q^2)$ and no additional phenomenological ingredients are necessary. We refer the reader to Appendix \ref{app:regge} for a more detailed demonstration of this equality. 

We emphasize here that relating $F_3^{\nu p+\bar{\nu}p}$ and $F_3^{(0)}$ by means of isospin symmetry introduces no additional uncertainty, up to isospin breaking corrections $\lesssim 2\%$. This is so because the axial vector charge current is a pure isovector, and the electromagnetic current is a pure isoscalar. This situation is quite different from the calculation of the energy-dependent $\gamma Z$-box correction to PV electron scattering. There, the isospin rotation was employed to obtain the NC $\gamma Z$ interference structure functions from purely electromagnetic data \cite{Gorchtein:2011mz}: the electromagnetic probe is the sum of the isoscalar and the isovector channels, and the weak NC probe additionally contains the contribution of the strange flavor channel. As a result, the isospin decomposition of the inclusive electromagnetic data together with the flavor rotation to obtain the NC $\gamma Z$ interference structure functions is the main source of the uncertainty and has been subject to an active research recently \cite{Gorchtein:2008px,Sibirtsev:2010zg,Rislow:2010vi,Gorchtein:2011mz,Blunden:2011rd,Carlson:2012yi,Blunden:2012ty,Hall:2013hta,Rislow:2013vta,Hall:2013loa,Gorchtein:2015qha,Hall:2015loa,Gorchtein:2015naa}.

\section{Results for $\Box_{\gamma W}^{VA}$,  $\Delta_R$ and $\Delta_R^V$}
\label{sec:resultsn}

We are now in the position to combine the results for the $\gamma W$-box and $\Delta_R$. 
We follow the definition 
\beqn
\Box_{\gamma W}^{VA}=\frac{\alpha}{2\pi}[C_{DIS}+C_B+C^{Regge}+C^{\pi N}+C^{Res}],
\eeqn
and give the new results for the $C$'s. 
The DIS part changes only slightly due to lowering the low $Q^2$ cut off from $(1.5\,{\rm GeV})^2$ to 2 GeV$^2$,
\beqn
C_{DIS}^{MS}=1.84&\rightarrow&C_{DIS}^{new}=1.87
\eeqn

The Born is increased because it is integrated up to infinity, rather than to the matching point $Q^2=(0.823\,{\rm GeV})^2$, but due to accounting for more recent data (see Appendix \ref{app:born}) the uncertainty is reduced,
\beqn
C_B^{MS}=0.829(83)&\rightarrow&C_B^{new}=0.91(5).
\eeqn

The biggest change affects the interpolating function introduced by MS. It is replaced by the sum of $\pi N$, resonance and Regge contributions. The central value increases considerably, yet the uncertainty is reduced, 
\beqn
C_{INT}^{MS}=0.14(14)\,\rightarrow\,C^{Regge}+C^{\pi N}+C^{Res}=0.48(7).\nn\\
\eeqn

Putting the numbers together, the result for the $\gamma W$-box increases with a significantly smaller uncertainty,
\beqn
&&\left(\Box_{\gamma W}^{VA}\right)^{MS}=2.81(16)\frac{\alpha}{2\pi}=3.26(19)\times10^{-3}\nn\\
&&\left(\Box_{\gamma W}^{VA}\right)^{\rm new}=3.26(9)\frac{\alpha}{2\pi}=3.79(10)  \times10^{-3}
\eeqn
Finally, when translating everything into $\Delta_R$ one should also take into account the uncertainty due to all neglected higher order effects; MS quotes a value of $\pm 0.0001$ as its contribution to the $\Delta_R$ uncertainty, and we have done no improvement beyond that so this number should be retained.
Thereby, the shift of the radiative correction $\Delta_R$ to the neutron decay rate reads \cite{Czarnecki:2018okw}:
\beqn
\Delta_R^{\rm old}=0.03886(38)\,\rightarrow\,\Delta_R^{\rm new}=0.03992(22),
\eeqn
or, in terms of the nucleus-independent radiative correction $\Delta_R^V$ \cite{Hardy:2014qxa,Seng:2018yzq},
\beqn
\Delta_R^{V,\rm old}=0.02361(38)\,\rightarrow\,\Delta_R^{V,\rm new}=0.02467(22).\label{eq:DeltaRV}
\eeqn

\begin{figure}[h]
	\begin{center}
		\includegraphics[width=8.0cm]{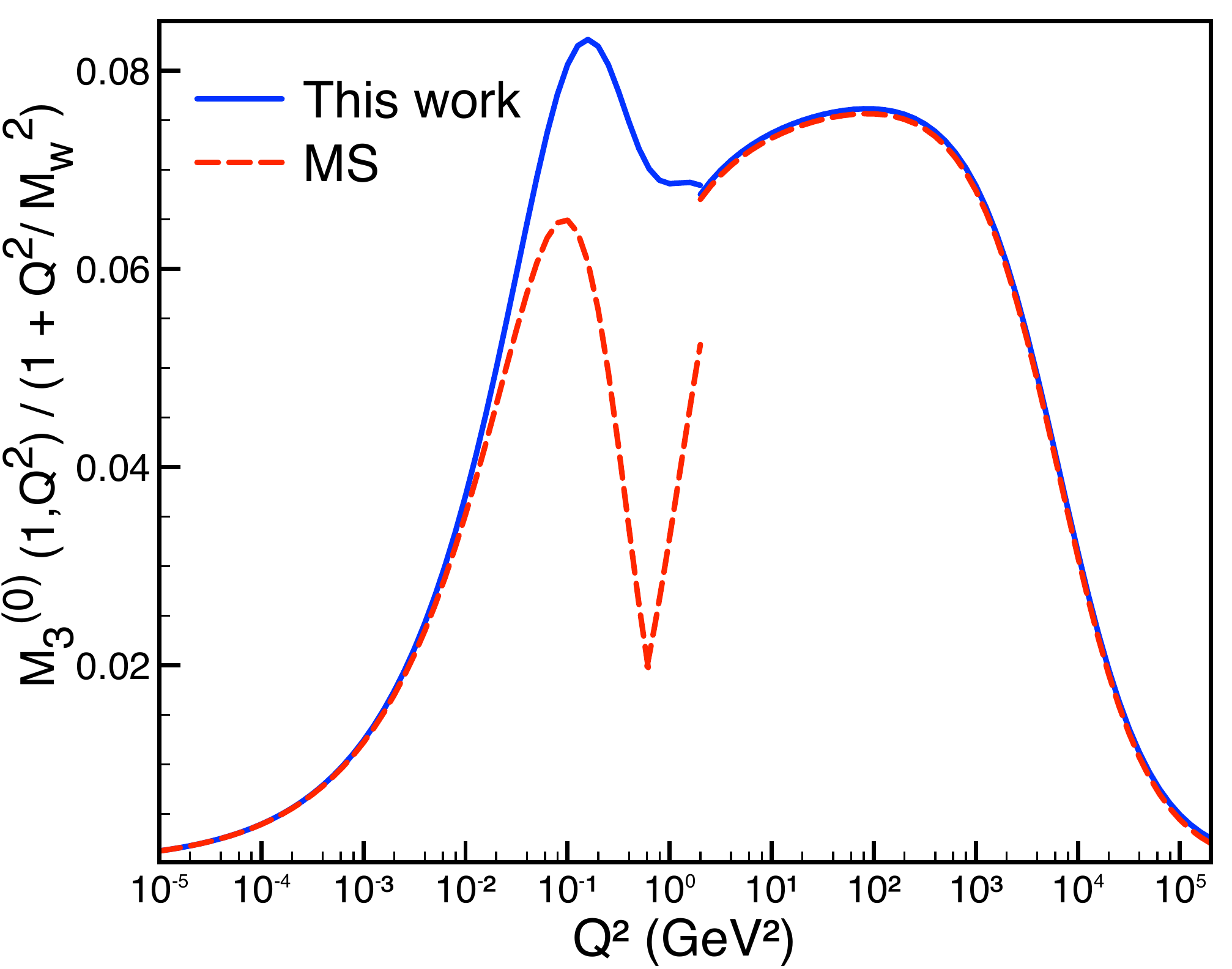}
		\caption{Our prediction of $\frac{M_W^2}{M_W^2+Q^2}M_3^{(0)}(1,Q^2)$ vs the MS's prediction. Notice that the peak around $Q^2=0.1$ GeV$^2$ is due to the Born contribution. Figure adopted from Ref. \cite{Seng:2018yzq}}
		\label{fig:M3plot}
	\end{center}
\end{figure}
The comparison between our new result and the MS result is most easily visualized through a plot of $\left(M_W^2/(M_W^2+Q^2)\right)M_3^{(0)}(1,Q^2)$ versus $Q^2$ in log scale, as shown in Fig. \ref{fig:M3plot}. 
Since $dQ^2/Q^2=d\ln Q^2$ in Eq. \eqref{eq:boxNachtmann}, the area under the curve provides a direct measure of $\Box_{\gamma W}^{VA}$. While mutually agreeing at large $Q^2$, we find three main differences between our approach and MS: (1) MS assume no physics other than Born at low $Q^2$, which is not consistent with the $W^2-Q^2$ diagram in Fig. \ref{fig:W-Q2diag}. In fact, our result shows that inelastic channels start contributing significantly already from $Q^2\approx0.1$ GeV$^2$ onwards; (2) MS require their interpolating function to vanish when $Q^2\rightarrow0$ (which turns out not to be true by explicit ChPT calculation), which causes the function to drop too fast with decreasing $Q^2$ and meet $F_{el}(Q^2)$ at relatively large matching point $Q^2=(0.823\mathrm{GeV})^2$; (3) MS require the integral of their interpolating function, instead of the function itself, to match pQCD result in the asymptotic region. This causes a discontinuity of their $F(Q^2)$ at the UV-matching point. All in all, the MS treatment of the interpolating function results in an underestimation of $\Delta_R^V$.

Our study leads to a new, more precise extraction of $V_{ud}$ from superallowed decays, as reported in Ref. \cite{Seng:2018yzq},
\beqn
&&|V_{ud}^{\rm old}| =0.97420(18)_{\rm RC}(10)_{\mathcal{F}t}\nn\\
&&\rightarrow|V_{ud}^{\rm new}|=0.97370(10)_{\rm RC}(10)_{\mathcal{F}t}.\label{eq:Vudnew}
\eeqn
It is worth noting that the uncertainty in $V_{ud}$ associated with $\Delta_R^V$ is now comparable to that due to $\mathcal{F}t$. 
This new result reflects in the first row CKM unitarity constraint,
\beqn
|V_{ud}|^2+|V_{us}|^2+|V_{ub}|^2=0.9984\pm 0.0004,
\eeqn
where 2018 PDG averages \cite{PDG2018} $|V_{us}|=0.2243(5)$ and $|V_{ub}|=0.00394(36)$ were used. The previous PDG constraint on the first row unitarity was $|V_{ud}|^2+|V_{us}|^2+|V_{ub}|^2=0.9994\pm0.0005$, roughly consistent with unitarity. Our new result suggests that, if all other SM corrections are correct, first row unitarity is violated by $(1.6\pm 0.4)\times 10^{-3}$, at the level of 4$\sigma$; the deviation reaches 5$\sigma$ if the updated determination of $V_{us}$ from the $K_{l3}$ decay in Ref. \cite{Bazavov:2018kjg} is adopted. 

One may also extract $V_{ud}$ from free neutron beta decay:
\beqn
|V_{ud}^{\rm old}|_{\rm free\:n} =0.9763(16)\,\to\,|V_{ud}^{\rm new}|_{\rm free\:n} =0.9758(16),\nn\\
\eeqn
where we have taken $\tau_n=879.3(9)$s and $\lambda=-1.2723(23)$ as quoted in Section 79 of PDG 2018 \cite{PDG2018}. Our new evaluation of $\Delta_R^V$ does not impact the total uncertainty because the latter is dominated by the experimental uncertainties due to $\lambda$ and $\tau_n$.} Note however that recently, the uncertainty of $\lambda$ was significantly reduced by the PERKEO--III experiment \cite{Markisch:2018ndu}, delivering $\lambda=-1.27641(56)$, with the statistical and systematical uncertainties added in quadrature. That reference provides an updated extraction from the free neutron decay, $|V_{ud}|=0.97351(60)$ which used an average of the three most recent lifetime measurements \cite{Serebrov:2018yxq,Pattie:2017vsj,Ezhov:2018lmo} $\tau_n=879.7(8)$s in place of the PDG average and the old evaluation of the RC. Applying our analysis of $\Delta_R$ to these new measurements we obtain 
\beqn
|V_{ud}^{\rm new}|_{\rm free\:n}^{\rm PERKEO-III} =0.97302(57),
\eeqn
in good agreement with the extraction from superallowed nuclear decays, and with the uncertainty that is now only four times larger than in the latter. This uncertainty is currently dominated by that in the lifetime, and future lifetime measurements aim at further reducing it by a factor 3-4 (see Ref.~\cite{Gonzalez-Alonso:2018omy} for a comprehensive review of experimental activities), closing the gap between the two methods. Importantly, the free neutron decay is free from nuclear uncertainties.

As mentioned already in the Introduction, the value of $V_{ud}$  extracted from the superallowed nuclear decays relies on the nuclear structure corrections $\delta_{NS}$ which are purely theoretical. There persists a discussion on the uncertainty and model dependence of those calculations, see e.g. the recent Ref.~\cite{Xayavong:2017kim} and references therein. The shell model approach with the Wood-Saxon potential advocated by Hardy and Towner is at variance with Hartree-Fock evaluations which may signal a systematic effect that has not yet been fully understood. In view of this we plan reassessing the nuclear corrections from the dispersion relation perspective in detail in the upcoming work. In the next Section we demonstrate the potential of the dispersion treatment on the example of the quasielastic contribution to the $\gamma W$-box calculation on nuclei.

{\section{Nuclear contributions to the $\Box_{\gamma W}^{VA}$ for nuclear decays}}
\label{sec:nuclear}

When extracting $V_{ud}$ from superallowed Fermi transitions, one must consider modifications of the free nucleon matrix elements due the presence of the nuclear environment. The standard approach to organizing the radiative corrections to nuclear $\beta$ decay followed in Refs. \cite{Marciano:2005ec,Hardy:2014qxa,Towner:1994mw} is summarized in Eq. (\ref{eq:superallowed}). The quantity appearing in the denominator is universal, nucleus-independent, and related to the measured $ft$ values as
\begin{equation}
\mathcal{F}t(1+\Delta_R^V)=ft(1+\delta'_R)(1-\delta_C+\delta_{NS})(1+\Delta_R^V)\ \ \ .
\end{equation}
Here, $\delta'_R$ is the nuclear charge-dependent outer correction; $\delta_C$ corrects the matrix element of the Fermi operator for the nucleus-dependent isospin symmetry breaking effects; $\Delta_R^V$ stands for the universal part that stems from the $\gamma W$-box on a free nucleon; and $\delta_{NS}$ accounts for nuclear structure corrections within the $\gamma W$-box. The latter two corrections combined together should be understood as the $\gamma W$-box evaluated on a nucleus, with the inclusive nuclear and hadronic intermediate states taken into account.  \black
\begin{figure}[h]
\begin{center}
\includegraphics[width=8.0cm]{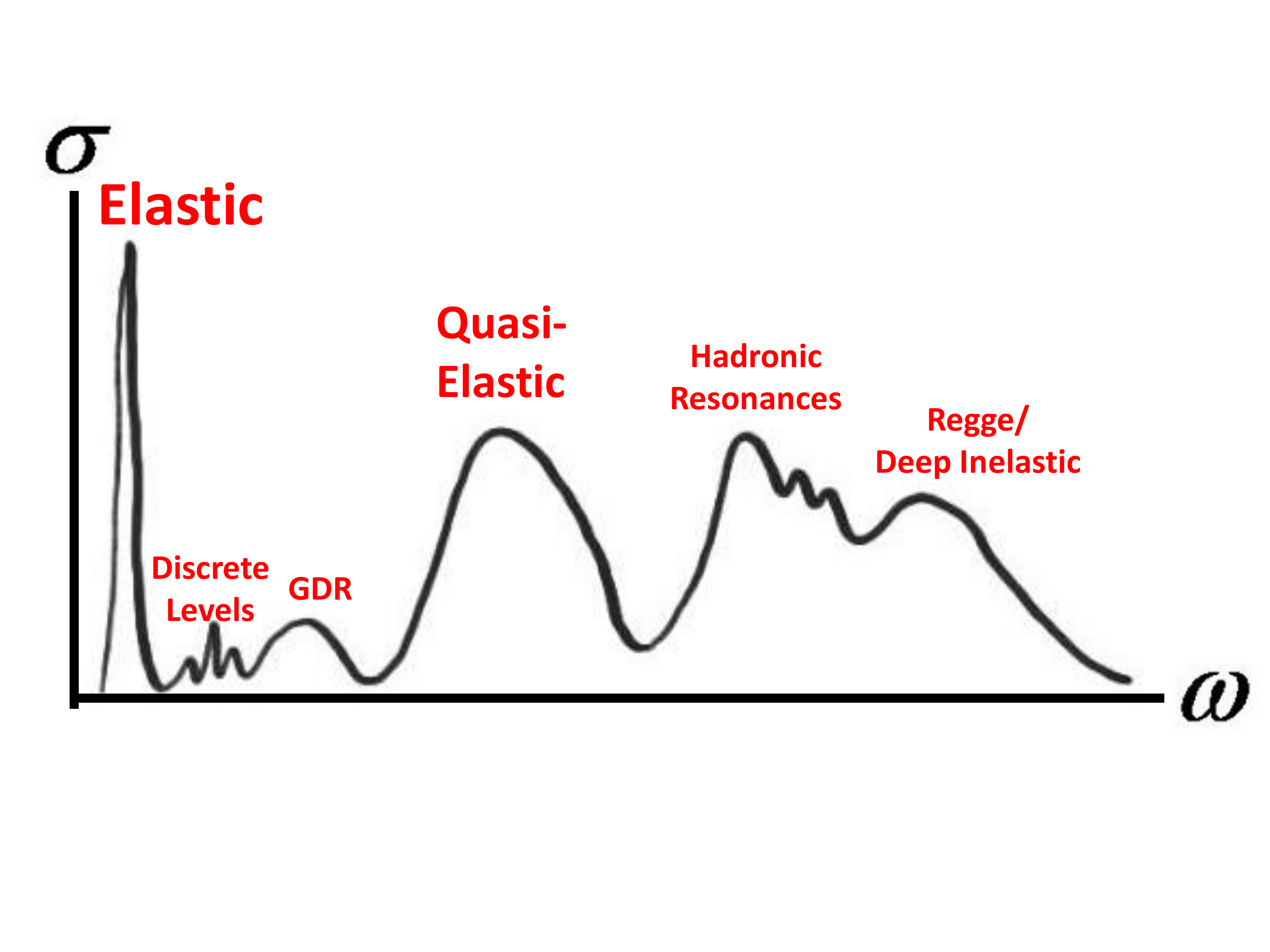}
\caption{Idealized structure of virtual photoabsorption on a nucleus.}
\label{fig:spectrum_nucl}
\end{center}
\end{figure}

In the context of dispersion relations, it is useful to visualize these contributions in terms of the nuclear response to an external lepton in a manner analogous to what is shown in Fig. \ref{fig:spectrum}. To that end, we show in Fig. \ref{fig:spectrum_nucl} an idealized structure of the nuclear electroabsorption spectrum. While the shape in the hadronic regime is similar to that for a free nucleon in Fig. \ref{fig:spectrum}, the lower part of the nuclear spectrum contains nuclear resonances and the quasielastic (QE) peak. The latter includes {the} one-nucleon knock-out as well as {the} knock-out of two or more nucleons in a single scattering process. 
The nuclear structure correction $\delta_{NS}$ thus accounts for the additional features of the electroabsportion spectrum on nuclei as compared to that on a free nucleon.

The $\gamma W$-box on a nucleus should in principle be calculated in using the full nuclear Greens function. Doing so is challenging, however, since the latter should be known in the full kinematical range to describe all the effects from lowest-lying nuclear excitations to shadowing at high energies. In practice, the nuclear modifications of the $\gamma W$-box have been calculated using the nuclear shell model with a semi-empirical Woods-Saxon potential (WSSH) \cite{Hardy:2014qxa} and nuclear density functional theory\cite{Satula:2012zz}. Attempts to address the calculation of $\delta_C$ in nuclear approaches other than WSSH suggest that the understanding of the nuclear structure corrections may not be at the level needed to warrant the current $\sim2\times10^{-4}$ relative precision of the ${\cal F}t$ values \cite{Miller:2008my,Miller:2009cg}. We refer the reader to a detailed discussion in Ref. \cite{Hardy:2014qxa} which contains the list of relevant calculations, and the critique to those from the standpoint of semiempirical Woods-Saxon potential shell model advocated by the authors of that reference. \black

In what follows, we focus on the modification of the free nucleon Born correction $(\alpha/2\pi) C_{B}$ due to the presence of the QE response. We defer a treatment of the other features of the low-lying nuclear spectrum to future work. To proceed, we recall that the procedure for dividing the full $\gamma W$-box on a nucleus into a universal and nucleus-dependent corresponds to rewriting identically,
\beqn
\Box_{\gamma W}^{\rm VA,\;Nucl.}=\Box_{\gamma W}^{\rm VA,\;free\;n}+
\Big[\Box_{\gamma W}^{\rm VA,\;Nucl.}-\Box_{\gamma W}^{\rm VA,\;free\;n}\Big].\label{eq:deltaNS}
\eeqn
The first term is then absorbed in $\Delta_R^V$, while the second term makes part of $\delta_{NS}$:
\begin{eqnarray}
&&\frac{\alpha}{2\pi} C_B^\mathrm{free\;n}  \subset   \Box_{\gamma W}^{\rm VA,\;free\;n} \subset \Delta_R^V\ \ \ , \nonumber
\\
&&2\Big[\Box_{\gamma W}^{\rm VA,\;Nucl.}-\Box_{\gamma W}^{\rm VA,\;free\;n}\Big]  \equiv  \delta_{NS} \ \ \ .
\end{eqnarray}
Note that no approximation has been made at this step. 


As a matter of self-consistency, 
one should compute the two terms entering $\delta_{NS}$ in a common framework. In practice, different approaches have been utilized to date. 
The free nucleon term has been evaluated using phenomenological input from intermediate and high-energy data as described in the previous sections. The second (nuclear) term is at present calculated in non-relativistic nuclear models. The procedure of subtracting the former from the latter may introduce additional model dependence, raising concerns about additional as of yet unquantified theoretical uncertainty. We observe that such uncertainty would have to be primarily of a systematic, nucleus-independent nature so as not to spoil the present agreement with the CVC property of the charged current weak interaction. In this Section we argue that with the use of dispersion relations one may evaluate both the free nucleon term and the nuclear $\gamma W$-box correction on an equal footing. In doing so, we will show that the previous treatment of the latter has, indeed, omitted an important, universal nuclear correction.

Working with the nucleons as the relevant degrees of freedom for describing the nuclear structure, the $\gamma W$-box calculation has two generic contributions: one arising from the one-body current operator and a second involving two-body currents. For a given nuclear model, the latter are required for consistency with the nuclear continuity equation (current conservation). Considering now the one-body current contribution, we write the nuclear $\gamma W$ Compton amplitude schematically as  
\beqn
T_{\mu\nu}^{\gamma W\, \mathrm{nuc}} \sim \langle f\vert  J^W_\mu\,  G_\mathrm{nuc}\,   J^\mathrm{EM}_\nu  \vert i\rangle\label{eq:tmununuc}
\eeqn
where $\vert i\rangle$ and $\vert f\rangle$ are the intitial and final nuclear states;  $J^W_\mu $ and  $J^\mathrm{EM}_\nu $ are the weak charged current and electromagnetic current, respectively; and 
\beqn
G_\mathrm{nuc} = \sum_n \frac{\vert n\rangle\langle n\vert}{E_n-E_0}
\eeqn
is the nuclear Green's function (we have omitted spacetime arguments for simplicity). Considering first fully relativistic nucleons described by Dirac spinors $N$, the one-body weak current in momentum space is
\begin{eqnarray}
\nonumber
J^W_\mu & = & \sum_k {\bar N_k} \left[ g_A(Q^2)\tau_3(k)  \gamma_\mu +\cdots\right]N_k\\
& \equiv&  \sum_k J_\mu^W(k)\label{eq:jweak}
\end{eqnarray}
where the \lq\lq $+\cdots$" indicate contributions from the weak magnetism and induced pseudoscalar terms and where the sum is over all nucleons $k=1,\ldots, A$.  A corresponding expression involving the charge and magnetic form factors applies to $J^\mathrm{EM}_\nu $.  

In the treatment of Ref.~\cite{Hardy:2014qxa}, the one-body contribution to the matrix element in Eq.~(\ref{eq:tmununuc}) is decomposed into two terms:  (A) a contribution singling out the same nucleon in $J^W_\mu $ and  $J^\mathrm{EM}_\nu $; (B)  a contribution involving distinct nucleons in these two operators. 
For purposes of the following discussion, it is useful to identify these two contributions using Eqs. (\ref{eq:tmununuc} - \ref{eq:jweak}):
\begin{eqnarray}
T_{\mu\nu}^{\gamma W\, \mathrm{nuc}} & \sim & \sum_{k, \ell} \langle f\vert J_\mu^W(k) \, G_\mathrm{nuc}\,   J^\mathrm{EM}_\nu(\ell)  \vert i\rangle 
\label{eq:tmunudecomp}
\\
&=& T_{\mu\nu}^A+ T_{\mu\nu}^B
\nonumber
\end{eqnarray}
where
\begin{eqnarray}
T_{\mu\nu}^A& = & \sum_{k} \langle f\vert J_\mu^W(k) \, G_\mathrm{nuc}\,   J^\mathrm{EM}_\nu(k)   \vert i\rangle  
\label{eq:comptA}\\
T_{\mu\nu}^B& = &  \sum_{k \not=\ell} \langle f\vert _\mu^W(k) \, G_\mathrm{nuc}\,   J^\mathrm{EM}_\nu(\ell)  \vert i\rangle 
\label{eq:comptB}
\end{eqnarray}
Here, $T_{\mu\nu}^A$ and $T_{\mu\nu}^B$ correspond, respectively, to contributions (A) and (B) mentioned above. The authors of Ref.~\cite{Hardy:2014qxa} refer to a part of contribution (A) as the nuclear Born term, while contribution (B) is included as a separate part of $\delta_{NS}$.

As first articulated in the earlier work of Ref. \cite{Towner:1994mw}, the nuclear Born term is evaluated by  replacing the free nucleon isovector  axial form factor $g_A(Q^2)$ and isoscalar magnetic form factor $G_M(Q^2)$ by \lq\lq quenched" values. This procedure is motivated by the observation that use of the free nucleon form factors in the one-body currents over-predicts the strength of nuclear Gamow-Teller transitions and nuclear magnetic moments \magenta \cite{Brown:1983zzc,Brown:1985zz}\black. The corresponding isoscalar magnetic moment and isovector axial coupling quenching parameters, $q_S^{(0)}$ and $q_A$, respectively, then describe the reduction of the spin-flip interaction strengths in the nuclear environment, with $q_S^{(0)},q_A\leq1$.  In evaluating the nuclear Born contribution to $\Box_{\gamma W}^{\rm VA}$, the authors of Ref.~\cite{Hardy:2014qxa} then evaluate the contribution (A) as described above but with these quenching factors applied:
\begin{eqnarray}
T_{\mu\nu}^A
\rightarrow 
\sum_{k} \langle f\vert \widetilde{ J_\mu^W} (k) \, G_\mathrm{nuc}\,   \widetilde{J^\mathrm{EM}_\nu} (k)   \vert i\rangle \\
\nonumber
\rightarrow 
\sum_{k} \langle f\vert \widetilde{J_\mu^W} (k) \,  \left[ S_F\otimes G_\mathrm{nuc}^{A^{\prime\prime}}\right] \widetilde{J^\mathrm{EM}_\nu} (k)   \vert i\rangle
\end{eqnarray}
where $\widetilde{J_\mu}$ denotes a current operator containing the quenching factor and where, in the last step, the nuclear Green's function has been replaced by the direct product of the free nucleon propagator, $S_F$, and  the Green's function for an intermediate \lq\lq spectator nucleus", $A^{\prime\prime}$. The loop integral used in obtaining $C_B$ for the free nucleon, which contains $S_F$, is then evaluated without further reference to the spectator nucleus but with the quenched form factors included. One then writes, 
\beqn
C_B^{\,\rm free\;n}\;\rightarrow\;C_B^{\,\rm Nucl.}=C_B^{\,\rm free\;n}+[q_S^{(0)}q_A-1]C_B^{\,\rm free\;n},
\label{eq:cbornnuc}
\eeqn
and includes the second term on the RHS of Eq.~(\ref{eq:cbornnuc}) in $\delta_{NS}$.

Note that this treatment relies on several assumptions:  (i) the impact of the nuclear environment is dominated by the transitions to the low-lying states $\vert n \rangle$;  (ii) the nucleon form factors entering the $\gamma W$ box graph for a single nucleon should inherit the impact of this apparent modification of the one-body currents in low-lying nuclear transitions; (iii) the quenching observed for pure Gamow-Teller and for magnetic moments and pure magnetic transitions translates directly into a mixed Gamow-Teller $\otimes$ magnetic response via the product of the corresponding quenching factors $q_S^{(0)}q_A$; and (iv) the $Q^2$-weighting inside the nucleon and nuclear box is the same. 

In effect, the foregoing assumptions amount to translating the effective quenching of the one-body operator strengths relevant to transitions involving the low-lying nuclear states $\vert n\rangle$ into a virtual free nucleon computation applicable to the $\omega=0$ and $0\leq Q^2 \leq (0.823\ \mathrm{GeV})^2$ part of the nucleon  electroabsorption spectrum (see Fig. 3).
To our knowledge, no explicit computation of the low-lying nuclear contributions to $T_{\mu\nu}^A$ has been performed. 
With these assumptions and using $C_B=0.89$, Refs. \cite{Towner:2002rg,Towner:1994mw} obtain that the quenched Born contribution for nuclei of interest  monotonically decreases from $-0.189$ for $^{10}$C to $-0.306$ for $^{74}$Rb. These results have propagated in all further evaluations of $\delta_{NS}$. Refs. \cite{Towner:2002rg,Towner:1994mw} assigned a generic 10\% uncertainty to this contribution. 

\begin{figure}[h]
\begin{center}
\includegraphics[width=8.5cm]{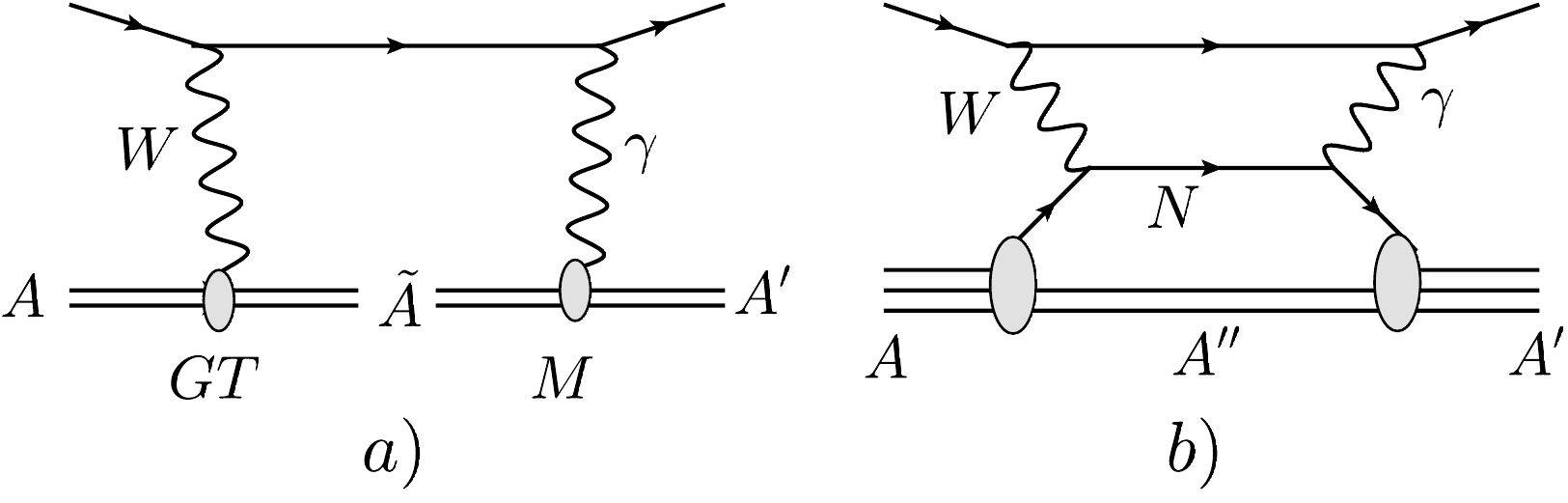}
\caption{Diagrammatic representation of the quenching mechanism of the Born contribution in the approach of Refs. \cite{Towner:2002rg,Towner:1994mw} , diagram a) with the initial (final) nucleus $A$ ($A'$), and an excited nuclear state $\tilde A$ accessed via a Gamow-Teller transition from the initial nucleus and via a magnetic transition from the final nucleus. Panel b) shows the quasielastic picture with a single-nucleon knockout.}
\label{fig:HTvsQE}
\end{center}
\end{figure}

We now argue that the assumptions underlying the approach of Ref. \cite{Towner:1994mw} are not well-justified. To that end, it is useful to refer to the generic electroabsoprtion spectrum shown in Fig. \ref{fig:spectrum_nucl}. We then observe:
\begin{itemize}
\item The strength of the nuclear response in the QE regime is significantly larger than that due to low-lying nuclear excitations, and covers a broader range of excitation energy than the latter. Thus, one might expect that the QE region generally has a more significant impact on the dispersion integral, as well. To address the nuclear modification of the free nucleon contribution in a controlled manner, the QE knock-out contribution has to be explicitly included. \black

\item The dynamics in which the {\em same} nucleon participates in the transition to a state involving a quasi-free nucleon and spectator nucleus are those of the QE response, whose peak at $\omega\sim Q^2/2M$ can lie significantly above the low-lying nuclear excitation spectrum. In the $\gamma W$-box this contribution corresponds to (i) the virtual $W^+$ knocking out one neutron from the initial nucleus, converting it to a proton and a spectator nucleus, corresponding to a subset of intermediate states $\vert n\rangle$ in the nuclear Green's function and (ii) reabsorbtion of the  quasifree proton into the final nucleus by emitting a virtual photon. 

\item The significant store of data for QE electron-nucleus scattering implies that, to a first approximation, one may obtain an adequate description of the QE response using the free-nucleon form factors without any quenching factors applied. Inclusion of subdominant effects arising from nuclear correlations and two-body currents may yield $\mathcal{O}(10-30\%)$ corrections \cite{Moniz:1971mt}. \black

\item Finally, the QE contribution to $\gamma W$-box requires a quasi-free active nucleon between the $\gamma$ and $W$ couplings rather than a bound nucleon inside an excited nuclear state;  compare  Fig. \ref{fig:HTvsQE}b) and a), respectively. The $Q^2$-dependence under the integral in the box with the low-lying excited nuclear state as in Fig. \ref{fig:HTvsQE}a), on the other hand, depends on nuclear form factors which are known to drop much faster than the free nucleon form factors, so the assumption that the integral over form factors should simply rescale as the charges is not justified. 
\end{itemize}

With these observations in mind, we propose an alternative method of addressing the modification of the free nucleon Born contribution by explicitly accounting for the QE contribution shown in Fig. \ref{fig:HTvsQE}b). This approach entails (1) employing the dispersion relation framework to evaluate the contribution from the QE component of $T_{\mu\nu}^A$ to $\delta_{NS}$, and (2)   replacing the Towner and Hardy computation of the same-nucleon contribution to $\delta_{NS}$ by our computation of the QE contribution. We defer a treatment of the contributions from low-lying nuclear excitations to a future, state-of-the-art many-body computation. We expect that such a computation will take into account the underlying many-body dynamics responsible for the quenching of spin-flip transition strengths in low-lying nuclear transitions.

We now turn to the dispersion representation of the $\gamma W$-box correction in Eq. (\ref{eq:BoxDR}) with the nuclear structure function $F_{3,\,\gamma W}^{(0),\;{\rm Nucl.}}$,
defined per active nucleon,
\beqn
\Box^{VA,\;Nucl.}_{\gamma W}&=&\frac{\alpha}{N \pi M}\int\limits_0^\infty \frac{dQ^2 M_W^2}{M_W^2+Q^2} \int\limits_0^\infty d\nu \frac{(\nu+2q)}{\nu(\nu+q)^2}\nn\\
&&\times F_{3,\,\gamma W}^{(0),\;{\rm Nucl.}}(\nu,Q^2),
\eeqn
with $N$ the number of neutrons (protons) in the $\beta^-\;(\beta^+)$ decay process, respectively.
and concentrate on the quasielastic  part only. Instead of defining the quenching via a simple rescaling of the Born we will directly calculate $C_{QE}$ from a dispersion representation,
\beqn
C_{QE}=2\int\limits_0^\infty dQ^2 \int\limits_{\nu_{min}}^{\nu_\pi} \frac{d\nu (\nu+2q)}{M\nu(\nu+q)^2}F_{3,\,\gamma W}^{(0),\;QE}(\nu,Q^2),
\label{eq:CQE}
\eeqn
with the limits of the $\nu$-integration being $\nu_{min}$, the threshold for the quasielastic breakup specified in Eq.~(\ref{eq:numin}) below 
\black and $\nu_\pi=(Q^2+(M+m_\pi)^2-M^2)/2M$ the threshold for pion production. 
Then, 
we estimate the modification of the Born contribution discussed above, as
\beqn
C_B^{\,\rm Nucl.}=C_B^{\,\rm free\;n}+[C_{QE}-C_B^{\,\rm free\;n}].
\eeqn

For purposes of this exploratory calculation, 
we describe the quasielastic peak in the $\gamma W$ box contribution to a superallowed $\beta^+$ decay process $A\rightarrow A'e^+\nu_e$ in the plane-wave impulse approximation (PWIA). In this picture, a nucleus first splits into an on-shell spectator nucleus $A''$ and an active off-shell nucleon, and the latter interacts with the gauge bosons. The effective scattering process proceeds as $AW^-\rightarrow n A''\rightarrow A'\gamma$, see Fig. \ref{fig:HTvsQE}b). The active nucleon carries an off-shell momentum $k$ before interacting with the gauge boson. To describe its distribution in the nucleus we adopt the Fermi gas model, which assumes a uniform distribution of nucleon momenta within the Fermi sphere with the Fermi momentum $k_F$.  

\black

We compute the quasielastic contribution to the structure function $F_3^{(0)}$ per proton in a nucleus. Details of the calculation are reported in Appendix \ref{app:QE}, and here we simply show the final result,
\beqn
\frac{1}{Z}F_{3,\,\gamma W}^{(0),QE}(\nu,Q^2)=-G_AG_M^S\frac{3Q^2}{32q}F_P\frac{\left((\tilde k_+)^2-(\tilde k_-)^2\right)}{k_F^3},\nn\\
\eeqn
where the $1/Z$ is  the normalization specific for $\beta^+$ process and should be replaced by $1/N$ for $\beta^-$ decay. 
The quantity $F_P(|\vec{q}|,k_F)$
\black  is a function describing the Pauli blocking effect during the interaction between the active nucleon and the gauge bosons, while $\tilde{k}_\pm=\mathrm{min}(k_F,k_\pm)$ where $k_\pm$ denote the upper and lower limits of the active nucleon three-momentum $k$. These arise due to the on-shell condition for the intermediate nucleon and are given by 
\begin{widetext}
\beqn
k_\pm=\left|\frac{q}{2}\frac{M_{A-1}+\nu-\nu_{min}}{\frac{M_A}{2}+\nu-\nu_{min}}
\pm\frac{M_A+\nu}{2}\frac{\sqrt{(\nu-\nu_{min})(2MM_{A-1}/M_A+\nu-\nu_{min})}}{\frac{M_A}{2}+\nu-\nu_{min}}\right|,\label{eq:kpm}
\eeqn
\end{widetext}
where we introduced the threshold energy for the quasielastic breakup,
\beqn
\nu_{min}=Q^2/(2M_A)+\epsilon,
\label{eq:numin}
\eeqn
with $\epsilon=M_{A-1}+M-M_A$ the nucleon removal energy. 
This nucleon removal energy is another scale that is relevant for QE scattering. Because of a mismatch between the initial and final nucleus masses for each decay (usually referred to as the $Q$-{\it value} of the decay), \black every initial-final nucleus pair involves not one, but two removal energies. Specifically, for $\beta^+$ decay these are given by
\beqn
\epsilon_1&=&M_{A''}+M_n-M_{A'},\nn\\
\epsilon_2&=&M_{A''}+M_n-M_A<\epsilon_1,
\eeqn
with $A''=A-p=A'-n$ the spectator nucleus. For $\beta^-$ decay the proton and neutron masses should be exchanged in this definition. 
We only account for bulk properties of nuclear structure at this step, and define an average removal energy for each pair, 
\beqn
\overline\epsilon&=&\sqrt{\epsilon_1\epsilon_2}
\eeqn

We consider 20 decay modes collected in the 2015 review by Hardy and Towner \cite{Hardy:2014qxa}, use the known $Q$-values of the decays and calculate relevant nucleon removal energies and summarize the 
results in Table \ref{tab1}.
\begin{table}[h]
  \begin{tabular}{c|c|c|c}
\hline
Decay & $\epsilon_1$ (MeV) &  $\epsilon_2$ (MeV) & $\overline\epsilon$ (MeV) \\
\hline
\hline
$^{10}C\to ^{10}B$ &  6.70  &  4.79 & 5.67 \\
$^{14}O\to ^{14}N$ &  8.24  &  5.41 & 6.68 \\
$^{18}Ne\to ^{18}F$ &  8.11  &  4.71 & 6.18 \\
$^{22}Mg\to ^{22}Na$ &  10.41  &  6.28 & 8.09 \\
$^{26}Si\to ^{26}Al$ &  11.14  &  6.30 & 8.38 \\
$^{30}S\to ^{30}P$ &  10.64  &  5.18 & 7.42 \\
$^{34}Ar\to ^{34}Cl$ &  11.51  &  5.44 & 7.91 \\
$^{38}Ca\to ^{38}K$ &  11.94  &  5.33 & 7.98 \\
$^{42}Ti\to ^{42}Sc$ &  11.57  &  4.55 & 7.25 \\
$^{26m}Al\to ^{26}Mg$ &  11.09  &  6.86 & 8.72 \\
$^{34}Cl\to ^{34}S$ &  11.42  &  5.92 & 8.22 \\
$^{38m}K\to ^{38}Ar$ &  11.84  &  5.79 & 8.28 \\
$^{42}Sc\to ^{42}Ca$ &  11.48  &  5.05 & 7.61 \\
$^{46}V\to ^{46}Ti$ &  13.19  &  6.14 & 9.00 \\
$^{50}Mn\to ^{50}Cr$ &  13.00  &  5.37 & 8.35 \\
$^{54}Co\to ^{54}Fe$ &  13.38  &  5.13 & 8.28 \\
$^{62}Ga\to ^{62}Zn$ &  12.90  &  3.72 & 6.94 \\
$^{66}As\to ^{66}Ge$ &  12.74  &  3.16 & 6.34 \\
$^{70}Br\to ^{70}Se$ &  13.17  &  3.20 & 6.49 \\
$^{74}Rb\to ^{74}Kr$ &  13.85  &  3.44 & 6.90 \\
\hline
 \end{tabular}
\caption{Effective removal energy $\overline\epsilon$ as calculated from the initial and final nucleus removal energies $\epsilon_{2,1}$ 
for all superallowed $\beta$ decays listed in Ref. \cite{Hardy:2014qxa}.}
\label{tab1}
\end{table}
We notice that while individual breakup thresholds vary significantly from isotope to isotope, the average removal energies all fall in a narrow range, $\overline\epsilon=7.5\pm1.5$ MeV. The Fermi momentum also varies in a small range, from 228 MeV to 245 MeV, from lightest to heaviest nucleus. 
We use the model with the average vaues of Fermi momentum and breakup threshold for calculating the bulk quasielastic contribution $\Box_{\gamma W}^{VA,\;\rm QE}$ universal for all nuclei, and do not attempt to address the nuclear-specific corrections at this time. 
The numerical evaluation of the QE contribution in Fermi gas model gives 
\beqn
C_{QE}=0.44\pm0.04\pm0.13.
\eeqn
The first uncertainty is obtained by varying the average removal energy and the Fermi momentum within their respective range. 
The second uncertainty is the uncertainty of the model which we assume to be $\sim30$\% for the free Fermi gas model. 
This way we obtain a new estimate of the ``quenching of the Born contribution" (note that  Refs. \cite{Towner:2002rg,Towner:1994mw} adopted an older result $C_B=0.89$, whereas our evaluation suggests a slightly higher value $C_B=0.91(5)$)
\beqn
C_{QE}-C_B=-0.47\pm0.14.
\eeqn

We observe that the nuclear environment reduces the size of the elastic box correction by about a half. This effect can be qualitatively understood by noticing the $\sim1/\nu^2$ weighting under the integral in Eq.~(\ref{eq:CQE}). In the free nucleon case, the $Q^2$-integration starts at zero, and so does the $\nu$ integration since $\nu=Q^2/(2M)$. In nuclei, binding effects shift that threshold to a finite value $\nu=Q^2/(2M_A)+\overline\epsilon$. Pauli blocking provides an additional  source of reduction. Indeed, Ref. \cite{Blunden:2012ty} observed the analogous effect of Pauli blocking upon the $\gamma Z$-box contribution to parity violation in heavy atoms. \black We checked that in the limit $\overline\epsilon,\,k_F\to0$ we recover the Born contribution on a free nucleon.

For a meaningful comparison with Refs. \cite{Towner:2002rg,Towner:1994mw}, we extract the average of their estimates for 20 decays,
$[q_S^{(0)}q_A-1]C_B=-0.25(6)$ and notice a significantly larger magnitude of the \black nuclear modification in our approach. This means that retaining all other nuclear corrections in Ref. \cite{Hardy:2014qxa}, the universal ${\cal F}t$ value should be corrected by
\beqn
\frac{\alpha}{\pi}(C_{QE}-q_S^{(0)}q_AC_B)=-(5.1\pm3.2)\times10^{-4},\nn\\
\eeqn
leading to a new estimate
\beqn
\overline{{\cal F}t}=3072.07(63)s\rightarrow[\overline{{\cal F}t}]^{\rm new}=3070.50(63)(98)s,
\eeqn
with the second uncertainty stemming from that of the QE contribution. 

This shift in the ${\cal F}t$ value partially cancels the large shift in the value of $V_{ud}$ that followed from the new dispersion evaluation of $\Delta_R^V$ in the previous Section,
\beqn
|V_{ud}^{\rm new}|=0.97370(14)\rightarrow |V_{ud}^{\rm new,\;QE}|=0.97395(14)(16).\nn\\
\eeqn
The corresponding change in the test of first-row CKM unitarity reads
\beqn
&&|V_{ud}|^2+|V_{us}|^2+|V_{ub}|^2=0.9984\pm0.0004\label{eq:ckm_nuc}\\
&&\to |V_{ud}|^2+|V_{us}|^2+|V_{ub}|^2=0.9989\pm0.0005.
\nn
\eeqn
The result in Eq.~(\ref{eq:ckm_nuc}) is 2.2 standard deviation away from exact unitarity and within one standard deviation from the current PDG value, $|V_{ud}|^2+|V_{us}|^2+|V_{ub}|^2=0.9994\pm0.0005$.

We can relate this new result for the ${\cal F}t$ value with the observation made in Ref. \cite{Czarnecki:2018okw} for the free neutron decay. 
While the lifetime and the axial charge individually are not very precisely known at present, a combination of them 
$\tau_n(1+3\lambda^2)$ forms a constant which is independent of the uncertainty in $\Delta_R,\Delta_R^V$:
\begin{equation}
\tau_n(1+3\lambda^2)\approx 1.70865\frac{1+\Delta_R^V}{1+\Delta_R}\mathcal{F}t=\mathrm{constant}.
\end{equation}
\indent
The constant depends on nuclear-structure effects via the $\mathcal{F}t$ value: while Ref. \cite{Czarnecki:2018okw} obtains $\mathrm{constant}=5172.0(1.1)s$ based on the analysis of Ref. \cite{Hardy:2014qxa}, our evaluation of the QE contribution shifts this value to a lower value $5169.7(2.0)s$.

\black

As mentioned above, we consider this new dispersion relation-based estimate of the quasielastic nuclear correction as exploratory since it is based on a simple free Fermi gas model and is not yet directly validated by experimental data. This motivated us to assign a generous 30\% model uncertainty to the quasielastic result. A future evaluation that will use a more sophisticated model of quasielastic nuclear response will certainly decrease this uncertainty while also being able to address the dependence of this correction on the final nucleus charge $Z$. We postpone this calculation to a future work. 
With these reservations, we believe that our new evaluation of the ``quenched Born contribution" is much better justified, as compared to the earlier approach of Ref. \cite{Towner:1994mw} used in computing $\delta_{NS}$. The dispersion relation approach also provides the basis for a unification of the universal correction $\Delta_R^V$ and the nuclear structure-dependent correction $\delta_{NS}$ within the same framework. To further advance the evaluation of these corrections, the following steps will be necessary: i) more advanced calculations of the QE single-nucleon knock-out contribution using up-to-date nuclear theory and validated by experimental QE data; ii) advanced calculations of the QE two-nucleon knock-out that is the main contribution to $\delta_{NS}$, which should also be confronted with the experimental data; iii) new computations of the contributions to $\delta_{NS}$ from low-lying nuclear states that directly incorporate the dynamics responsible for the observed quenching of spin-flip transitions; iv) computations that include nuclear shadowing effects which may affect the evaluation of $\Delta_R^V$ on a nucleus, and have not been considered in the literature. To set up this research program, a close cooperation between particle and nuclear theorists, and experimentalists will be crucial.

\section{Relation to $eN$-scattering data}
\label{sec:relationtodata}

Besides making use of the neutrino scattering data, one other possibility to probe the $\gamma W$ interference matrix element in experiment is to relate it to the $\gamma Z$ matrix element which can be measured in parity-violating (PV) $eN$-scattering through isospin symmetry. To illustrate this point, we first define a set of rank-one spherical tensors in the isospin space
using the axial current $A_{i}^{\mu}=\bar{q}\gamma^{\mu}\gamma_{5}\tau_{i}q$:
\begin{eqnarray}
A_{1}^{\pm1,\mu} & = & \mp\frac{1}{\sqrt{2}}\left(A_{1}^{\mu}\pm iA_{2}^{\mu}\right)\nonumber \\
A_{1}^{0,\mu} & = & A_{3}^{\mu}
\end{eqnarray}
such that the axial components of the charged and neutral weak currents are given by $\left(J_{W}^{\mu}\right)_{A}=(1/\sqrt{2})A_{1}^{1,\mu}$
and $\left(J_{Z}^{\mu}\right)_{A}=-(1/2)A_{1}^{0,\mu}$. With this, one can easily show using the Wigner-Eckart
theorem in the isospin space that 
\beqn
&&\left\langle p\right|J_{em}^{(0)\mu}\left(J_{W}^{\nu}\right)_{A}\left|n\right\rangle \\
&&=\left\langle p\right|J_{em}^{(0)\mu}\left(J_{Z}^{\nu}\right)_{A}\left|p\right\rangle -\left\langle n\right|J_{em}^{(0)\mu}\left(J_{Z}^{\nu}\right)_{A}\left|n\right\rangle,\nn
\eeqn
where $J_{em}^{(0)\mu}$ is the isosinglet component of the electromagnetic current (and a superscript ``3" will denote its isotriplet component).
Next, we can write $J_{em}^{(0)\mu}=J_{em}^{\mu}-J_{em}^{(3)\mu}$
at the right hand side of the equation above and argue that the terms
with $J_{em}^{(3)\mu}$ sum up to zero. The reason is simple: both
$J_{em}^{(3)\mu}$ and $\left(J_{Z}^{\nu}\right)_{A}$ are $(I=1,I_{3}=0)$
objects, so their product can only be $(I=0,I_{3}=0)$ or $(I=2,I_{3}=0)$.
The $I=2$ piece obviously vanishes when taking matrix element with
respect to $I=1/2$ nucleon states, while the matrix elements of the
$I=0$ piece are the same for the proton and neutron so they cancel
each other. Therefore we can simply replace $J_{em}^{(0)\mu}\rightarrow J_{em}^{\mu}$
at the right hand side. That leads to the following identity for the
parity-odd structure functions $F_{3}$: 
\begin{equation}
4F_3^{(0)}=F_{3,\gamma Z}^p-F_{3,\gamma Z}^n.
\end{equation}
The factor 4 at the left hand side is just due to the choice of normalization in $F_3^{(0)}$. The structure functions on the RHS are in principle measurable in PV electron scattering experiments. One should however be aware of the possible caveats of such correspondence: recall that the isoscalar component of the electromagnetic current is much smaller than its isovector component; so any attempt based on isospin argument to relate a small isoscalar EM matrix element to the full EM matrix element will be more exposed to unknown hadronic complications such as the nucleon anapole moment and the strange quark effects. 

\begin{figure}[h]
	\begin{center}
		\includegraphics[width=4.0cm]{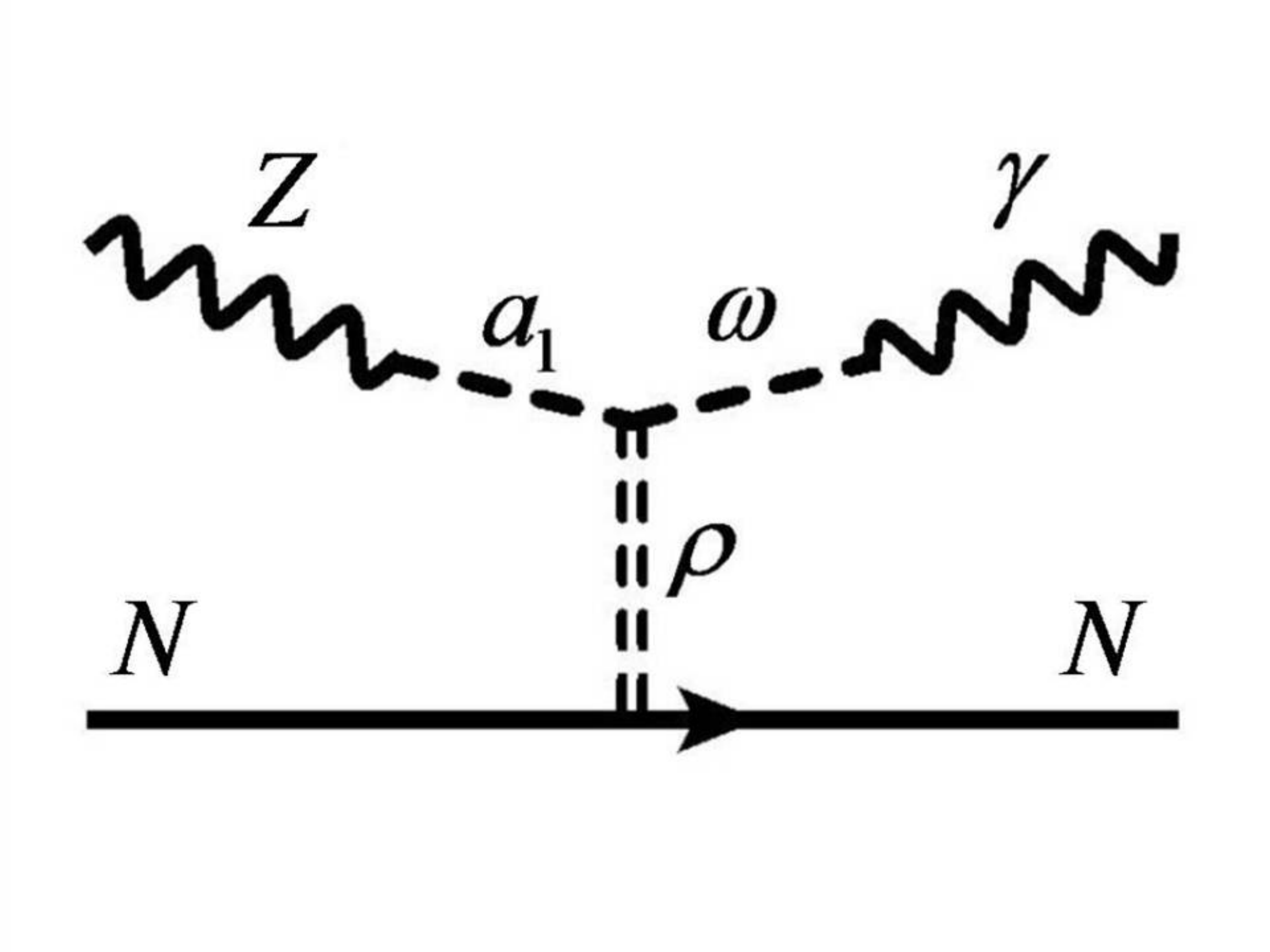}
		\includegraphics[width=4.0cm]{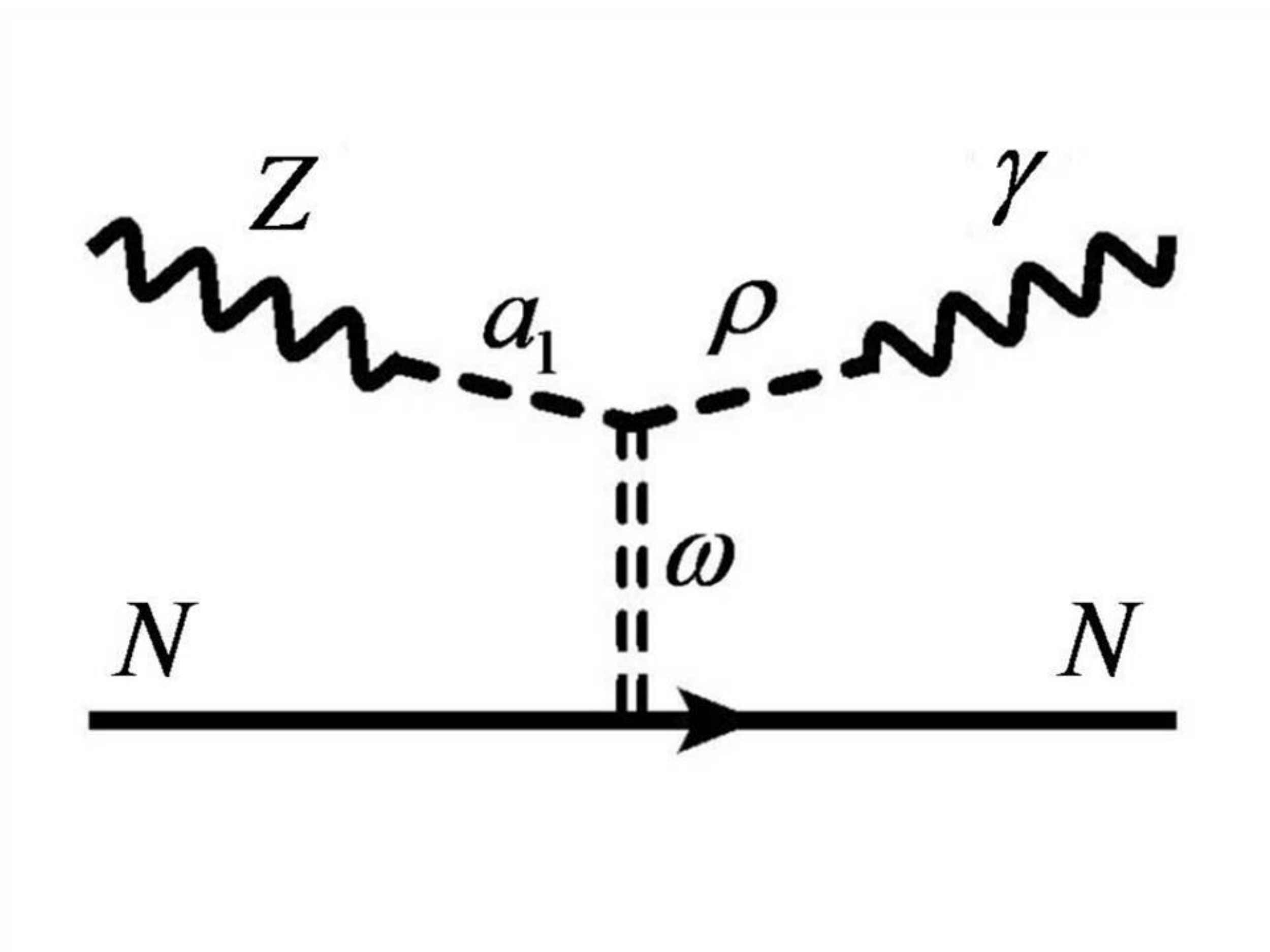}
		\caption{Regge-model description of $F_{3,\gamma Z}^{N}$.}\label{fig:ReggeZ}
	\end{center}
\end{figure}

Another significance of PV $eN$-scattering is its ability to test our current modeling of $F_3^{(0)}$ and $F_3^{\nu p+\bar{\nu}p}$ simultaneously. Recall that $F_3^{(0)}$ is probing a current product of the form isoscalar $\times$ isovector while  $F_3^{\nu p+\bar{\nu}p}$ is of isovector $\times$ isovector, they can be related to $F_{3,\gamma Z}^N$ of which the electromagnetic current contains both the isoscalar isovector components. To illustrate this point let us consider the Regge contribution to $F_{3,\gamma Z}^{N}$ in total analogy to those detailed in Appendix \ref{app:regge}. The exchanged-diagrams are depicted in Fig. \ref{fig:ReggeZ}, and one observes that the photon can fluctuate to both $\omega$ and $\rho^0$. 
The only extra ingredient needed apart from those in Appendix \ref{app:regge} is the mixing Lagrangian between $a_1$ and $Z$, also given in Ref. \cite{Lichard:1997ya}:
\begin{equation}
\mathcal{L}_{a_1Z}=-\frac{gm_{a_1}^2}{2g_\rho\cos\theta_W}w_{a_1}a_{1\mu}^0Z^\mu.
\end{equation}
With this we can write down the Regge prediction of $F_{3,\gamma Z}^{N}$ in complete analogy to Eq. \eqref{eq:F3Regge}:
\begin{widetext}
\begin{eqnarray}
F^{\gamma Z,\,N}_{3,\,\mathbb{R}}(\nu,Q^2)&=&2\left(\frac{eg}{2\cos\theta_W}\right)^{-1}\left(\frac{e}{g_\omega}\frac{m_\omega^2}{m_\omega^2+Q^2}\right)\left(-\frac{gw_{a_1}}{2g_\rho\cos\theta_W}\frac{m_{a_1}^2}{m_{a_1}^2+Q^2}\right)\left(\frac{g_\rho}{2}\tau^3_N\right)g_1H_\rho(\nu,Q^2)\nonumber\\
&+&2\left(\frac{eg}{2\cos\theta_W}\right)^{-1}\left(\frac{e}{g_\rho}\frac{m_\rho^2}{m_\rho^2+Q^2}\right)\left(-\frac{gw_{a_1}}{2g_\rho\cos\theta_W}\frac{m_{a_1}^2}{m_{a_1}^2+Q^2}\right)\left(\frac{g_\omega}{2}\right)g_2H_\omega(\nu,Q^2),
\end{eqnarray}
\end{widetext}
with $\tau^3_{p,n}=\pm1$ the nucleon isospin. 
From here one immediately observes the relations: 
\begin{eqnarray}
F^{\gamma Z,\,p}_{3,\,\mathbb{R}}-F^{\gamma Z,\,n}_{3,\,\mathbb{R}}&=&4F_{3,\,\mathbb{R}}^{(0)}\nonumber\\
F^{\gamma Z,\,p}_{3,\,\mathbb{R}}+F^{\gamma Z,\,n}_{3,\,\mathbb{R}}&=&F_{3,\,\mathbb{R}}^{\nu p+\bar{\nu}p}.\label{eq:twoisospin}
\end{eqnarray}
which are nothing but direct consequences of isospin symmetry; the first line has already been proven above and the second line works the same way. 

There are several benefits of this analysis. Firstly, according to the second line in Eq. \eqref{eq:twoisospin}, PV electron scattering (PVES) experiments on deuteron (which is essentially $p+n$) plays the same role as neutrino scattering in terms of probing the Regge contribution, thus the two different experiments may complement each other in providing input data to the dispersion relation at wider regions of $\nu$ and $Q^2$. Secondly, one major conclusion of our Regge analysis in Appendix \ref{app:regge} is that $F_{3,\mathbb{R}}^{(0)}$ and $F_{3,\mathbb{R}}^{\nu p+\bar{\nu}p}$ should have approximately the same $Q^2$-dependence; this can now be checked by comparing the $Q^2$-dependence of $F_{3,\gamma Z}$ between proton and deuteron (after subtracting out the elastic, $N\pi$ and resonance contribution by hand). The similarity in the $Q^2$-dependence of the two will be an indication of the correctness of our model prediction.  

To conclude this section, we notice a great similarity of the $\gamma W$-box correction discussed here at length, and the $\gamma Z$-box correction to parity violating interaction of the electron with the nucleon and nuclei. Its value at zero energy which is relevant for interpreting hadronic weak charges in terms of running of the weak mixing angle $\sin^2\theta_W(\mu)$, is given by our master formula for the $\gamma W$-box of Eq. (\ref{eq:boxNachtmann}) with just a few changes: we take the Nachtmann moment of the interference structure function $F_{3,\gamma Z}^p$, replace the $W$ mass by the $Z$ mass and include the scale dependence due to the running weak mixing angle in the electron's weak charge $v_e(Q^2)=-1+4\hat s^2_W(Q^2)$ under the integral to obtain:
\beqn
\Box^{A}_{\gamma Z}(0)=\frac{3\alpha}{2\pi}\int\limits_0^\infty \frac{dQ^2 M_Z^2v_e(Q^2)}{Q^2[M_Z^2+Q^2]}M_{3,\gamma Z}^{p}(1,Q^2).
\eeqn
Compare the above formula to Refs. \cite{Blunden:2011rd,Blunden:2012ty} where a similar result was obtained in terms of a series of Mellin moments which is equivalent to the Nachtmann moment. Asymptotically, the first Nachtmann moment of $F_{3,\gamma Z}^{p}$ is $M_{3,\gamma Z}^{p}(1,Q^2\to\infty)=\int_0^1dx(\frac{2}{3}u^p_V(x)+\frac{1}{3}d^p_V(x))=\frac{5}{3}$, which gives the well-known large logarithm contribution \cite{Marciano:1982mm,Marciano:1983ss}, 
$(5\alpha/2\pi)v_e(M_Z^2)\log(M_Z^2/\Lambda^2)$,
with $\Lambda$ a scale above which parton picture sets in. As to the terms not enhanced by this large logarithm and sensitive to hadronic structure, the most recent estimate stems from Refs. \cite{Blunden:2011rd,Blunden:2012ty} and includes Born and resonance contributions. To apply the free nucleon estimates of RC to the nuclear weak charges in parity violation in atoms, Ref. \cite{Blunden:2012ty} included Pauli blocking in the free Fermi gas model, but disregarded binding energy effects. In view of our new analysis of the $\gamma W$-box that showed the importance of the high-energy continuum at low $Q^2$ (our Regge-VDM parametrization), a similar contribution may be of importance for the $\gamma Z$-box, as well, and we will investigate this in an upcoming work.

\section{Conclusions}
\label{sec:conclude}
In summary, we proposed a new, dispersion relation-based formulation of the $\gamma W$-box correction to neutron and nuclear beta decays. We expressed this correction as an integral over the first Nachtmann moment of the P-odd structure function $F_{3,\,\gamma W}^{(0)}$ that arises from the interference of the isovector axial current and the isoscalar electromagnetic current. Dispersing the structure function $F_{3,\,\gamma W}^{(0)}$ in energy allows for obtaining a more detailed input needed for evaluating the correction and the uncertainty thereof. Utilizing the available neutrino and antineutrino scattering data we were able to obtain a new data-driven determination of the universal correction $\Delta_R^V$, which should replace the previous evaluation in Ref.~\cite{Marciano:2005ec}. While the uncertainties of the data are quite large, this approach allowed for a reduction of the uncertainty of $\Delta_R^V$ which is central to extracting the value of $V_{ud}$ and testing the first-row CKM unitarity. The new evaluation of $\Delta_R^V$, assuming all other corrections remain unchanged, resulted in a significant reduction of the value of $V_{ud}$ and to a violation of CKM unitarity by four standard deviations. 

While this disagreement opens up possibilities for BSM contributions which were essentially excluded by the previous CKM unitarity constraint, it prompted us to address nuclear structure corrections from the dispersion relation perspective. One particular contribution, that due to a one-nucleon knock-out, was evaluated in the free Fermi gas model. It can be identified with the ``quenched Born contribution" proposed and calculated by Towner in 1994 and included in all analyses of nuclear structure corrections by Hardy and Towner ever since. We argued that the assumptions entering that calculation are not well-justified, and that a proper inclusion of the quasielastic contribution is likely to increase that effect considerably. A free Fermi gas model estimate showed that this effect partially cancels the increase of $\Delta_R^V$ and brings the first row CKM unitarity closer to agreement. Moreover, our work showed that it is possible to unify the universal and nuclear structure-dependent corrections within the same framework.

\begin{acknowledgments}
We  are grateful to Hiren H. Patel for his participation at the early stages of this calculation. 
We acknowledge helpful discussions with Bill Marciano, John Hardy, Vincenzo Cirigliano, Peter Blunden, Emmanuel Paschos, Cheng-Pang Liu, Chung-Wen Kao, Hubert Spiesberger and Jens Erler.  
Significant progress was made during the scientific program ``Bridging the Standard Model to New Physics with the Parity Violation Program at MESA" hosted by MITP Mainz. 
M.G.'s acknowledges support by the Deutsche Forschungsgemeinschaft under the personal Grant No. GO 2604/2-1, and by the German--Mexican research collaboration grant SP 778/4--1 (DFG) and 278017 (CONACyT).
CYS's work is supported in part by the National Natural Science Foundation of China (NSFC) under Grant Nos.11575110, 11655002, 11735010, Natural Science Foundation of Shanghai under Grant No.~15DZ2272100 and No.~15ZR1423100, by Shanghai Key Laboratory for Particle Physics and Cosmology, by Key Laboratory for Particle Physics, Astrophysics and Cosmology, Ministry of Education, and by the DFG (Grant No. TRR110)
and the NSFC (Grant No. 11621131001) through the funds provided
to the Sino-German CRC 110 ``Symmetries and the Emergence of
Structure in QCD", and also appreciates the support through the Recruitment Program of Foreign Young Talents from the State Administration of Foreign Expert Affairs, China.  
MJRM was supported in part by US Department of Energy Contract DE-SC0011095.
\end{acknowledgments}

\begin{appendix}

\section{Crossing symmetry}
\label{app:crossing}
In this appendix we demonstrate the crossing symmetry of the $\gamma W$ P-odd invariant function: $T_3^{(I)}(-\nu,Q^2)=\xi^I T_3^{(I)}(\nu,Q^2)$ where $\xi^{(0)}=-1$ and $\xi^{(3)}=1$. This relation is crucial in obtaining their dispersive representation \eqref{eq:DR}. Our derivation makes use of the following time-reversal identity:
\begin{equation}
\bigl\langle\beta\bigl|\hat{O}\bigr|\alpha\bigr\rangle=\bigl\langle\tilde{\alpha}\bigl|\mathbb{T}\hat{O}^{\dagger}\mathbb{T}^{-1}\bigr|\tilde{\beta}\bigr\rangle\label{eq:TReversal}
\end{equation}
where $\hat{O}$ is a linear operator, $\mathbb{T}$ is the time-reversal operator and $\bigl|\tilde{\alpha}\bigr\rangle=\mathbb{T}\bigl|\alpha\bigr\rangle$, $\bigl|\tilde{\beta}\bigr\rangle=\mathbb{T}\bigl|\beta\bigr\rangle$ are the time-reversed state. It is convenient to also define the time-reversed four-momentum: $\tilde{p}^\mu=p_\mu$ and recall the time-reversal operation to a four-current: 
\begin{equation}
\mathbb{T}J^\mu(\vec{x},t)\mathbb{T}^{-1}=J_\mu(\vec{x},-t).\label{eq:TJmu}
\end{equation} 

Now we may decompose the electromagnetic current into isoscalar and isovector components $J_{em}^\mu=J_{em}^{(0)\mu}+J_{em}^{(3)\mu}$ and compute their corresponding $V\times A$ component of the forward Compton tensor \eqref{eq:Compton} with $q\rightarrow -q$:
\begin{widetext}
\begin{eqnarray}
T_{\gamma W,VA}^{(I)\mu\nu}(p,-q)=\frac{1}{2}\int dxe^{iq\cdot x}\langle p(p)|T[\left(J^{\nu}_W(x)\right)_AJ^{(I)\mu}_{em}(0)]|n(p)\rangle
=\frac{1}{2}\int dxe^{i\tilde{q}\cdot x}\langle n(\tilde{p})|T[J^{em(I)}_{\mu}(x)\left(J^{W}_{\nu}(0)\right)^\dagger_A]|p(\tilde{p})\rangle\nn\\\label{eq:Tpminusq}
\end{eqnarray}
\end{widetext}
where the first equality is due to translational invariance and the second equality makes use of Eqs. \eqref{eq:TReversal}, \eqref{eq:TJmu} and the translational invariance. Next, we can prove the following identity:
\beqn
&&\langle n(\tilde{p})|T[J^{em(I)}_{\mu}(x)\left(J^{W}_{\nu}(0)\right)^\dagger_A]|p(\tilde{p})\rangle\nn\\
&&=-\xi^I \langle p(\tilde{p})|T[J^{em(I)}_{\mu}(x)\left(J^{W}_{\nu}(0)\right)_A]|n(\tilde{p})\rangle
\eeqn
straightforwardly using the Wigner-Eckart theorem in the isospin space. With these, we can now express both the LHS and RHS of Eq. \eqref{eq:Tpminusq} in terms of invariant function $T_3^{(I)}$ to get:
\beqn
&&\frac{i\epsilon^{\mu\nu\alpha\beta}p_\alpha q_\beta}{2(p\cdot q)}T_3^{(I)}\left(-\frac{p\cdot q}{M},Q^2\right)\nn\\
&&=-\xi^I\frac{i\epsilon_{\mu\nu\alpha\beta}\tilde{p}^\alpha\tilde{q}^\beta}{2(\tilde{p}\cdot \tilde{q})}T_3^{(I)}\left(\frac{\tilde{p}\cdot\tilde{q}}{M},Q^2\right).
\eeqn
Finally, using $\epsilon_{\mu\nu\alpha\beta}\tilde{p}^\alpha\tilde{q}^\beta=-\epsilon^{\mu\nu\alpha\beta}p_\alpha q_\beta$ and $\tilde{p}\cdot\tilde{q}=p\cdot q$, we immediately obtain the desired crossing symmetry relation $T_3^{(I)}(-\nu,Q^2)=\xi^I T_3^{(I)}(\nu,Q^2)$.

\section{Elastic (Born) contribution}
\label{app:born}
This appendix serves as an updated evaluation of the elastic contribution to the $\gamma W$ box diagram. As clearly seen from Fig. \ref{fig:spectrum} the elastic contribution is separated from inelastic one by a finite gap. This picture remains intact for any value of $Q^2$, so it is natural to separate this piece out of the integral. To evaluate it, we need electromagnetic and weak vertices. 
The electromagnetic vertex is given by
\beqn
\Gamma_{em}^{\mu}(q)&=&\left[F_1^S(Q^2)\frac{\mathbbm{1}}{2}+F_1^V(Q^2)\frac{\tau^3}{2}\right]\gamma^\mu\\
&+&\left[F_2^S(Q^2)\frac{\mathbbm{1}}{2}+F_2^V(Q^2)\frac{\tau^3}{2}\right]i\sigma^{\mu\alpha}\frac{q_\alpha}{2M},\nn
\eeqn
with $F_{1,2}^{S,V}=F_{1,2}^p\pm F_{1,2}^n$ and $q$ the incoming momentum. The weak charged current vertex is given by
\beqn
\Gamma_W^{a,\mu}(q)=\left[F_1^W\gamma^\mu+F_2^Wi\sigma^{\mu\alpha}\frac{q_\alpha}{2M}+G_A\gamma^\mu\gamma_5\right]\tau^a;
\eeqn
here we do not display the pseudoscalar form factor $G_P$ that does not contribute to the box diagram. 

A straightforward calculation leads to the following expression for the elastic contribution to the structure function,
\beqn
F_{3,\,\rm Born}^{(0)}=-\frac{1}{4}G_A(Q^2)G_M^S(Q^2)\delta(1-x)\label{eq:F3Born}
\eeqn
where $G_M^S=F_1^S+F_2^S$ is the isoscalar magnetic Sachs form factor. The resulting contribution to the box correction reads
\begin{eqnarray}
\Box_{\gamma W}^{VA,\mathrm{Born}} &=&  -\frac{\alpha}{\pi }\int\limits_{0}^{\infty}dQ\frac{2\sqrt{4M^{2}+Q^{2}}+Q}{\left(\sqrt{4M^{2}+Q^{2}}+Q\right)^{2}}\nn\\
&&\times G_{A}(Q^{2})G_{M}^{S}(Q^{2})
\end{eqnarray}
Above, we neglected the $Q^2$-dependence of the $W$-propagator since the integral converges way below $Q^2\sim M_W^2$ due to nucleon form factors. Notice that unlike Marciano and Sirlin who only account for the elastic contribution in the low-$Q^2$ part of the integral, in the dispersive approach it extends to all $Q^2$.

Numeric evaluation with modern data on electromagnetic and weak form factors leads to
\beqn
\Box_{\gamma W}^{VA,\rm Born}=\frac{\alpha}{2\pi}0.908(49)=1.05(6)\times 10^{-3},\label{eq:Bornvalue}
\eeqn
slightly above the MS value \cite{Marciano:2005ec},
\beqn
\Box_{\gamma W}^{VA,\rm Born}\Bigr|_{\mathrm{MS}}=\frac{\alpha}{2\pi}0.829(83)=9.63(96)\times 10^{-4}.
\eeqn
The two calculations agree within the errors, but the uncertainty in the MS calculation is rather arbitrarily assigned as $\pm 10\%$, whereas ours is derived from the most recent information on nucleon form factors and is half of that in MS. This result is essentially model-independent: form factors are fixed by data on electron and neutrino scattering. If future data will further constrain the form factors, the uncertainty can be further reduced. 

In the following we present the details in obtaining the Born contribution \eqref{eq:Bornvalue} to $\Box_{\gamma W}$. First, we notice that according to our definitions, $G_{M}^{p}(Q^{2})>0$,
$G_{M}^{n}(Q^{2})<0$, $G_{A}(Q^{2})<0$ and $|G_{M}^{p}(Q^{2})|>|G_{M}^{n}(Q^{2})|$
for all relevant values of $Q^{2}$. So, we can write $\Box_{\gamma W}^{VA,\mathrm{Born}}$ 
as:
\beqn
\Box_{\gamma W}^{VA,\mathrm{Born}}&=&\frac{\alpha}{\pi}\int_0^\infty dQ\frac{2\sqrt{4m_{N}^{2}+Q^{2}}+Q}{\left(\sqrt{4m_{N}^{2}+Q^{2}}+Q\right)^{2}}\\
&\times&|G_{A}(Q^{2})|\left(|G_{M}^{p}(Q^{2})|-|G_{M}^{n}(Q^{2})|\right),\nn
\eeqn
so that every single multiplicative term in the integrand is positive
definite.

We acquire the data of the nucleon magnetic Sachs form factors from Fig. 1
and Fig. 4 of Ref. \cite{Ye:2017gyb}.
In each figure, we take the upper and lower boundary of the red shaded
band (multiplied by appropriate factors) as our $|G_{M}^{N}(Q^{2})|_{\mathrm{max}}$
and $|G_{M}^{N}(Q^{2})|_{\mathrm{min}}$ respectively ($N=p,n$),
and define:
\begin{eqnarray}
|G_{M}^{N}(Q^{2})| & = & \frac{1}{2}\left(|G_{M}^{N}(Q^{2})|_{\mathrm{max}}+|G_{M}^{N}(Q^{2})|_{\mathrm{min}}\right)\nonumber \\
\Delta|G_{M}^{N}(Q^{2})| & = & \frac{1}{2}\left(|G_{M}^{N}(Q^{2})|_{\mathrm{max}}-|G_{M}^{N}(Q^{2})|_{\mathrm{min}}\right).\nn\\
\end{eqnarray}

For the axial form factor at low $Q^{2}$, we utilize the result from
the model-independent $z$-expansion analysis in Ref. \cite{Bhattacharya:2011ah}. We take the upper and lower
end of the green error bars in their Fig. 3 as our $|G_{A}(Q^{2})|_{\mathrm{max}}$
and $|G_{A}(Q^{2})|_{\mathrm{min}}$, and define $|G_{A}(Q^{2})|$
and $\Delta|G_{A}(Q^{2})|$ accordingly. For $Q^{2}>1$ GeV$^{2}$,
we extrapolate each of the upper and lower curve using a dipole function
$g_{A}(1+Q^{2}/m_{A}^{2})^{-2}$ with $g_A\approx 1.27$. We find that, for a smooth exctrapolation
of $|G_{A}(Q^{2})|_{\mathrm{max}}$ we need $m_{A}=1.27$ GeV while
for $|G_{A}(Q^{2})|_{\mathrm{min}}$ we need $m_{A}=1.09$ GeV. 

For the uncertainty estimation, we define:
\begin{widetext}
\begin{eqnarray}
\delta_1 & = & \frac{\alpha}{\pi}\int_0^\infty dQ\frac{2\sqrt{4m_{N}^{2}+Q^{2}}+Q}{\left(\sqrt{4m_{N}^{2}+Q^{2}}+Q\right)^{2}}\Delta|G_{A}(Q^{2})|\left(|G_{M}^{p}(Q^{2})|-|G_{M}^{n}(Q^{2})|\right)\nonumber \\
\delta_2 & = & \frac{\alpha}{\pi}\int_0^\infty dQ\frac{2\sqrt{4m_{N}^{2}+Q^{2}}+Q}{\left(\sqrt{4m_{N}^{2}+Q^{2}}+Q\right)^{2}}|G_{A}(Q^{2})|\left(\Delta|G_{M}^{p}(Q^{2})|\right)\nonumber \\
\delta_3 & = & \frac{\alpha}{\pi}\int_0^\infty dQ\frac{2\sqrt{4m_{N}^{2}+Q^{2}}+Q}{\left(\sqrt{4m_{N}^{2}+Q^{2}}+Q\right)^{2}}|G_{A}(Q^{2})|\left(\Delta|G_{M}^{n}(Q^{2})|\right)
\end{eqnarray}
\end{widetext}
to account for the uncertainty due to $G_{A}(Q^{2})$, $G_{M}^{p}(Q^{2})$
and $G_{M}^{n}(Q^{2})$ respectively. These three pieces are added
up in quadrature to obtain a total uncertainty:
\begin{equation}
\delta=\sqrt{\delta_1^2+\delta_2^2+\delta_3^2}.
\end{equation}

Due to the limitation of available data, in the numerical evaluation we integrate $Q^2$ up to $10$ GeV$^{2}$, which is practically equivalent to setting $Q_{\mathrm{max}}\rightarrow\infty$. The resulting central value is $1.05\times 10^{-3}$, with $\delta_1=4.9\times 10^{-5}$,
$\delta_2=1.5\times 10^{-5}$ and $\delta_3=2.7\times 10^{-5}$,
so the full result is $(1.05\pm 0.06)\times 10^{-3}$ as shown in Eq. \eqref{eq:Bornvalue}.

\section{$\pi N$ intermediate state contribution to $F_3^{\gamma W}$}
\label{app:piN}
\begin{figure}
\includegraphics[width=3.5cm]{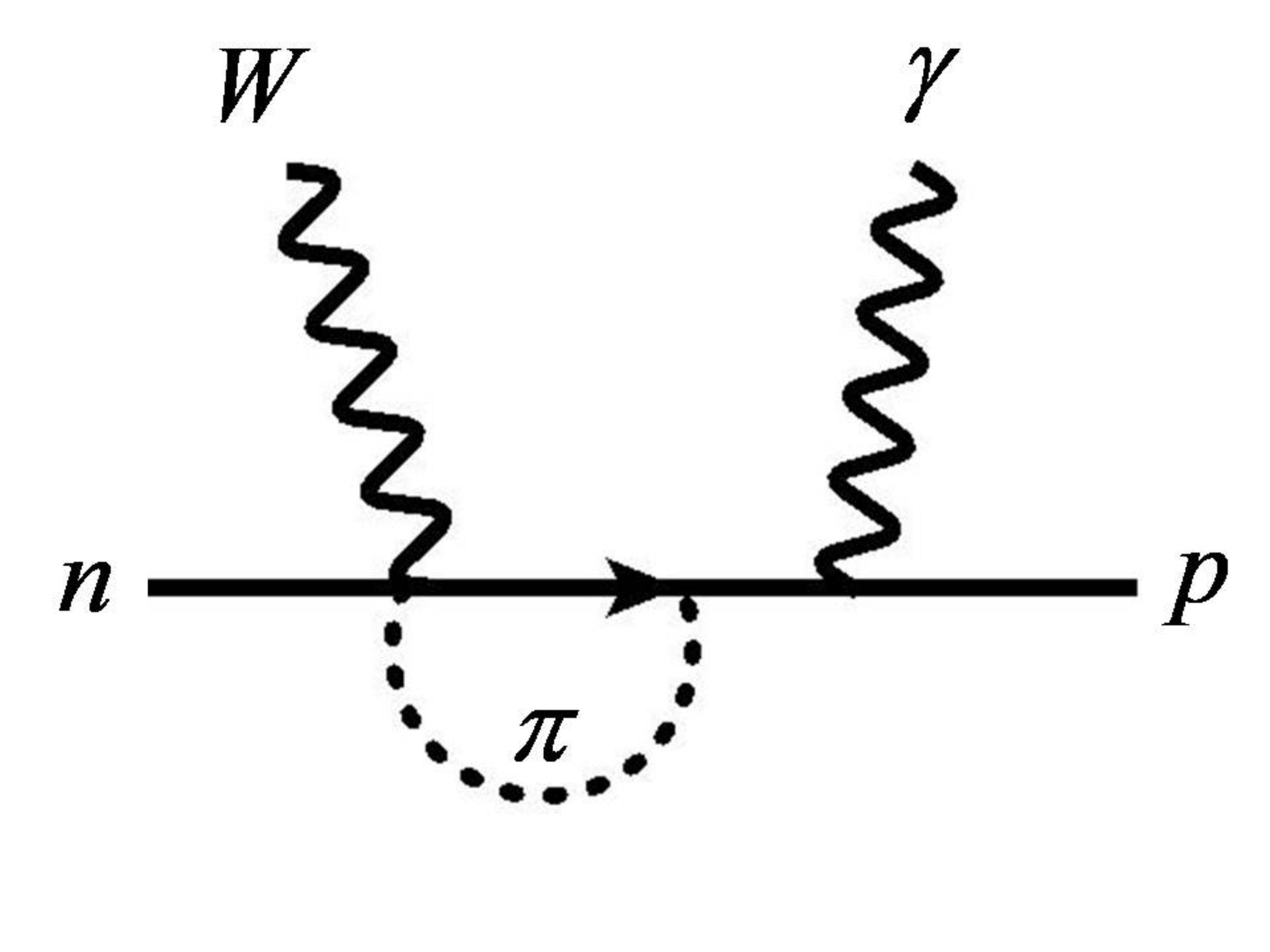}
\includegraphics[width=3.5cm]{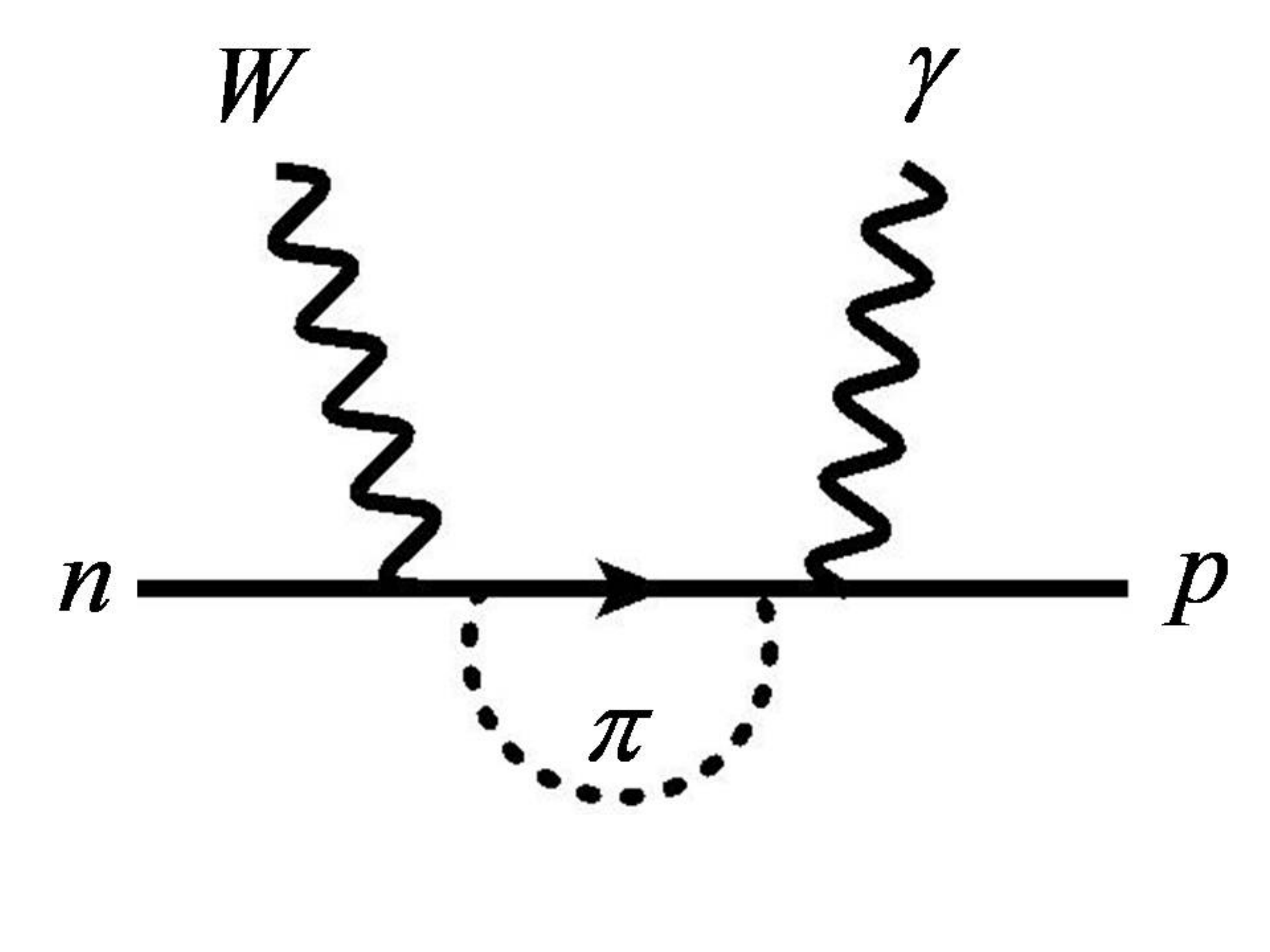}
\includegraphics[width=3.5cm]{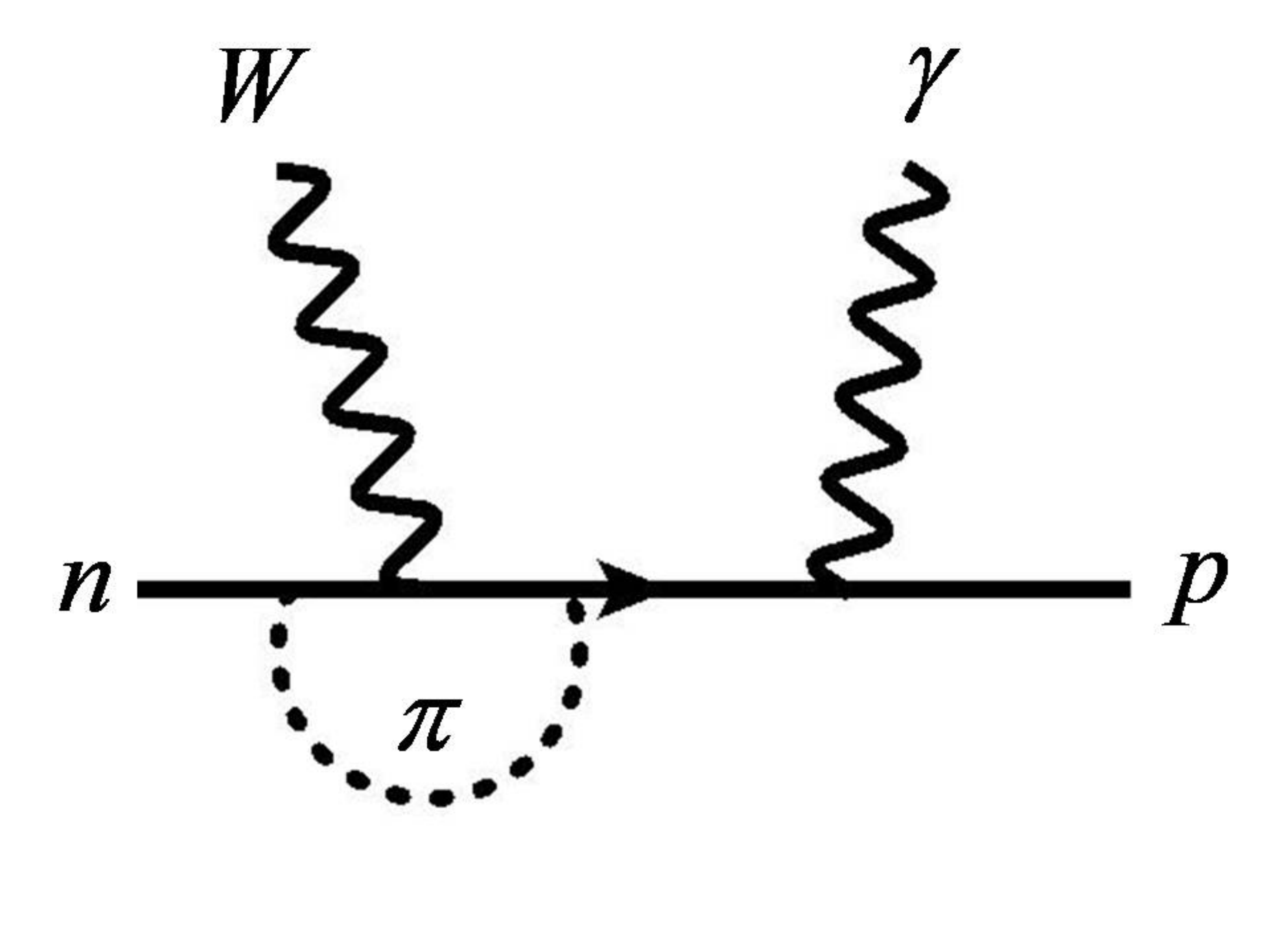}
\includegraphics[width=3.5cm]{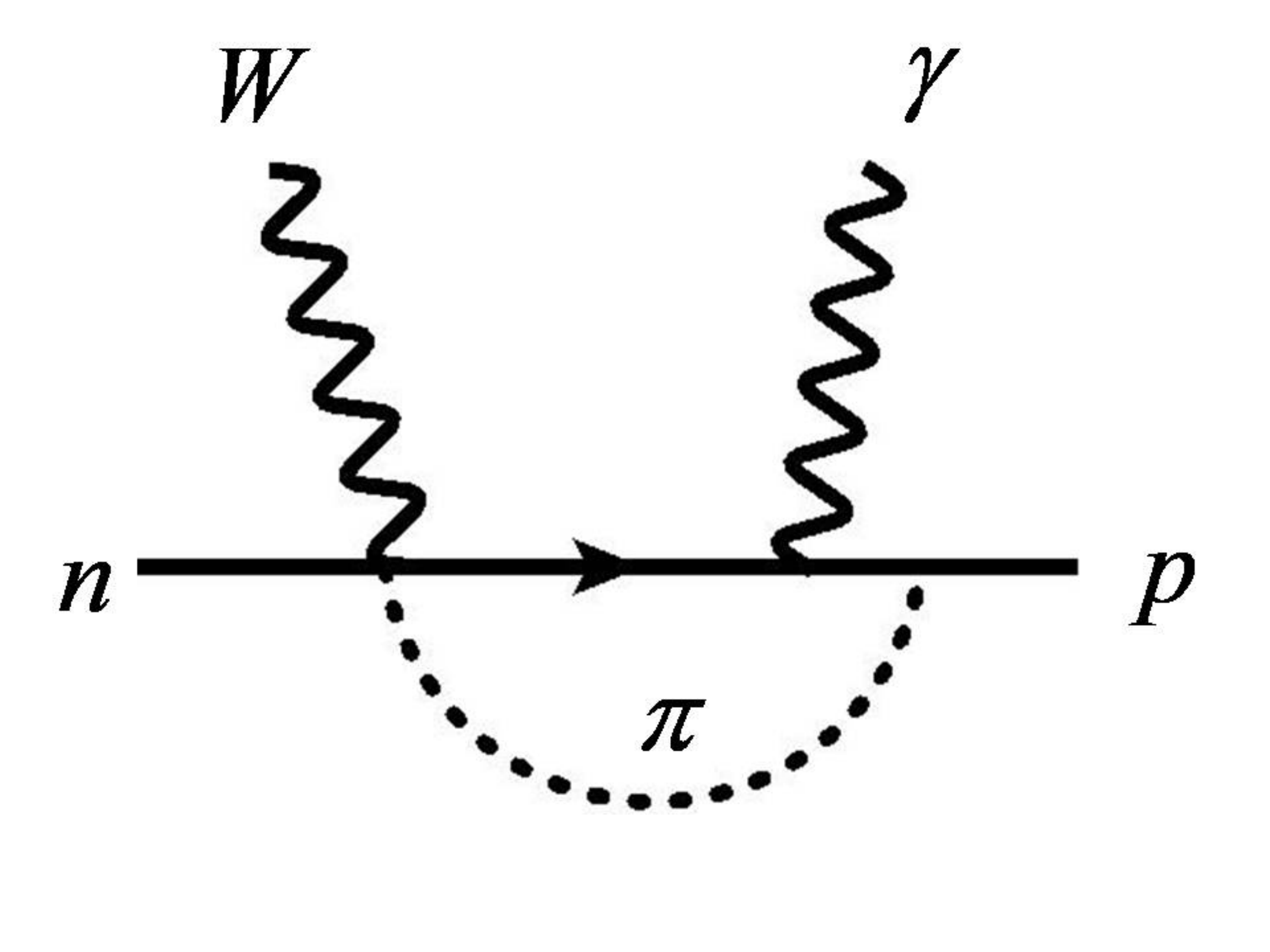}
\includegraphics[width=3.5cm]{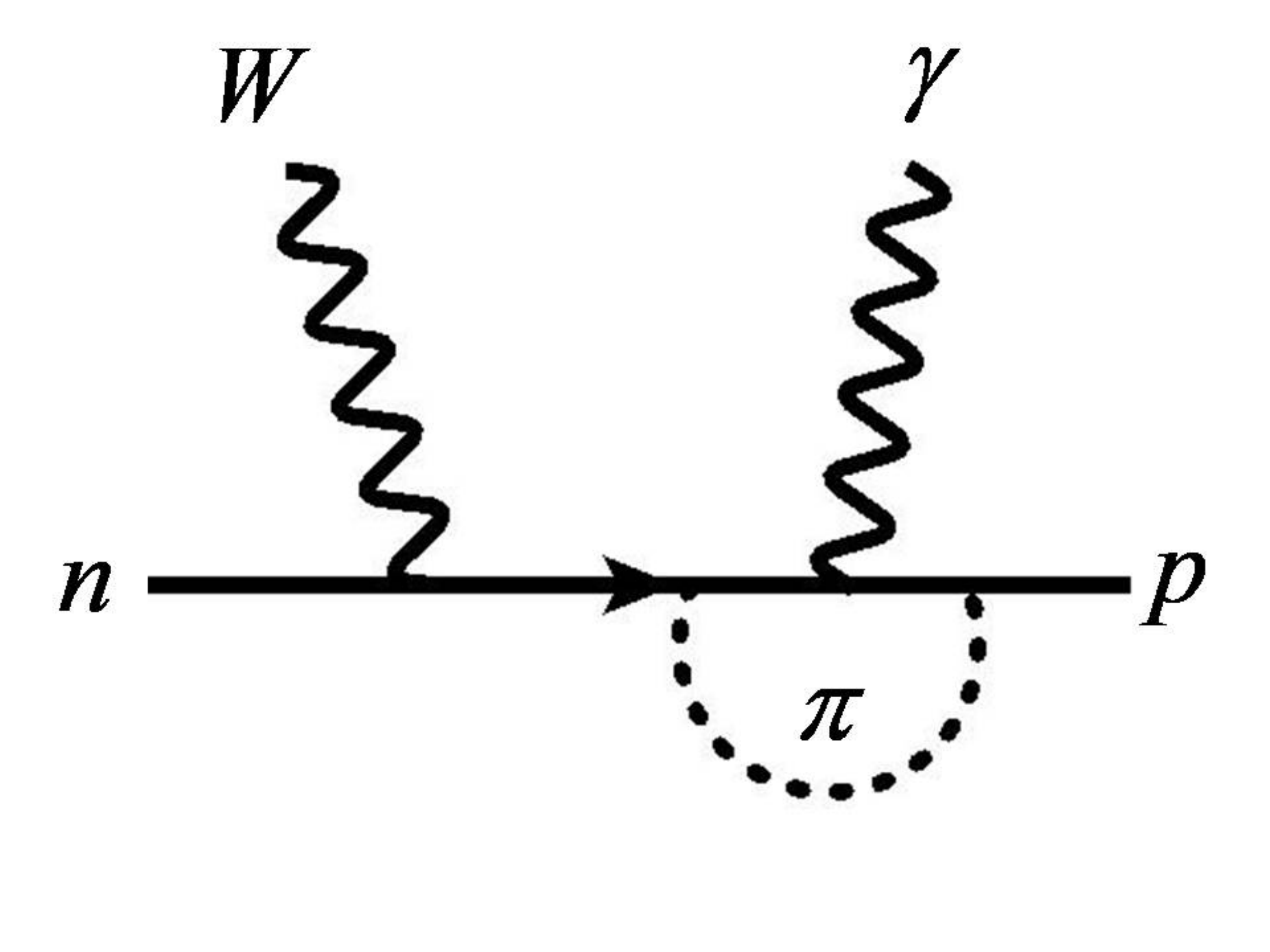}
\includegraphics[width=3.5cm]{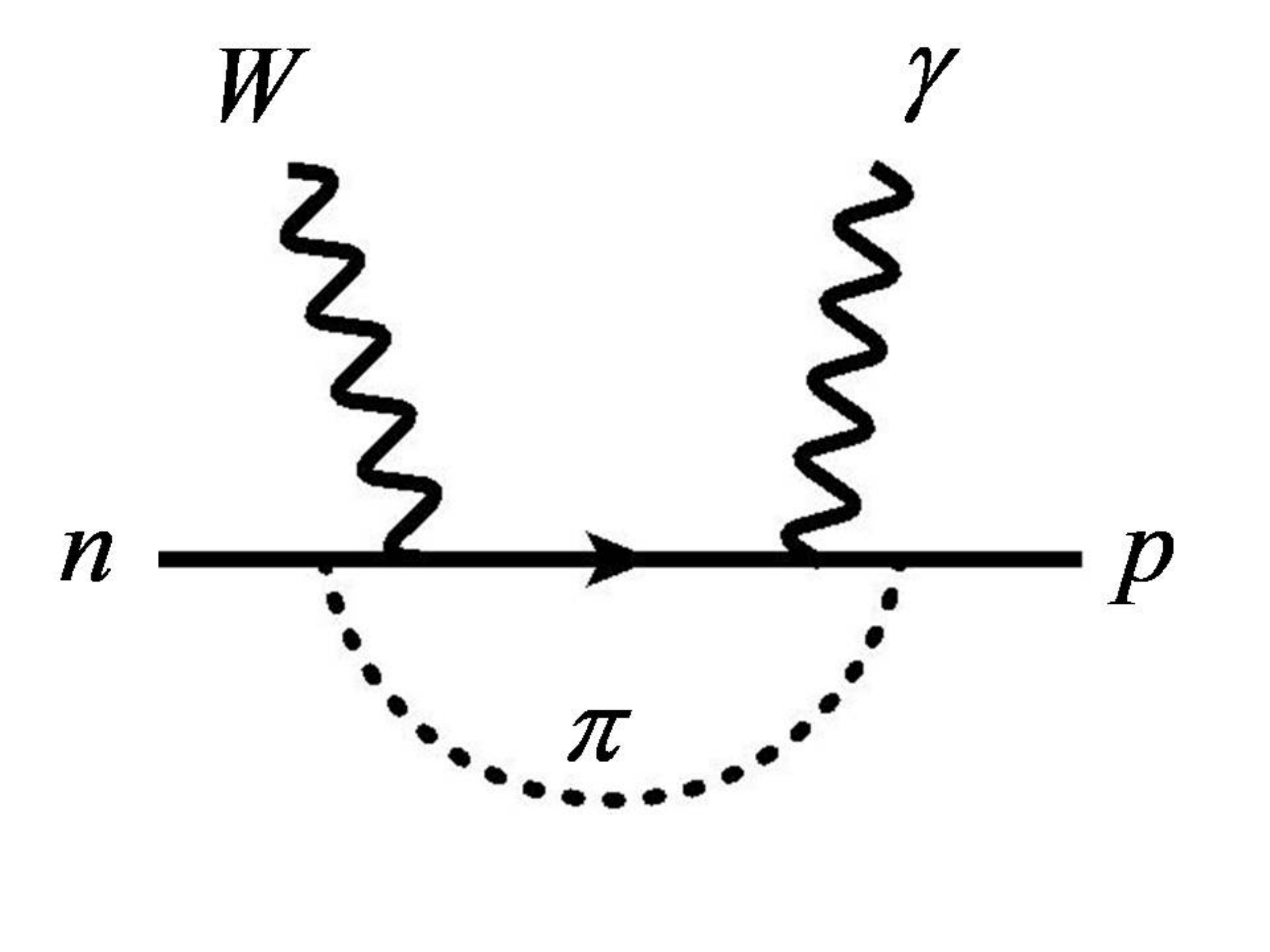}
\caption{Diagrams for $\pi N$ contribution to $F_3^{(0)}$.}
\label{fig:piN}
\end{figure}

Here we present the explicit result for the $\pi N$ contribution to the structure function $F_3^{(0)}(\nu,Q^2)$. It may be written as a sum of six terms coming from the discontinuity of the six one-loop diagrams depicted in Fig. \ref{fig:piN} respectively (which is also equivalent to calculating $M(nW\rightarrow N\pi)M^*(N\pi\rightarrow p\gamma)$ at tree level and integrating over the phase space):
\begin{equation}
F_{3,\, \pi N}^{(0)}=\sum_{i=1}^6 F_{3,i}^{{(0)}\pi N}.
\end{equation}
They are calculated using relativistic baryon chiral perturbation theory (ChPT) at leading order, keeping in mind that the photon is coupled to the isosinglet electromagnetic current, which leads to slight modifications of the corresponding Feynman rules; alternatively one could also calculate $F_{3,\gamma Z}^p-F_{3,\gamma Z}^n$ with the full Feynman rules, as shown in Section \ref{sec:relationtodata}.

The results are:
\begin{widetext}
\begin{eqnarray}
F_{3,1}^{{(0)}\pi N}&=&-\frac{g_AM\nu(M^4-M^2 m_\pi^2-2M^2W^2-m_\pi^2W^2+W^4)(E_+-E_-)}{64\pi^2 F_\pi^2W^2(W^2-M^2)q}\\
F_{3,2}^{{(0)}\pi N}&=&\frac{3g_A^3M\nu(M^6-M^4 m_\pi^2-(M^4+6M^2m_\pi^2)W^2-(M^2+m_\pi^2)W^4+W^6)(E_+-E_-)}{256\pi^2 F_\pi^2W^2(W^2-M^2)^2q}\\
F_{3,3}^{{(0)}\pi N} & = & -\frac{g_{A}^{3}\nu}{512\pi^{2}F_{\pi}^{2}q^{3}(W^{2}-M^{2})}\left\{ \left[4M^{2}W^{2}\left(2M^{4}+(2M^{2}+Q^{2})(W^{2}+Q^{2}-m_{\pi}^{2})+Q^{2}W^{2}\right)\right.\right.\nonumber\\
&  & \left.-(M^{2}+W^{2})(M^{2}+Q^{2}+W^{2})^{2}(W^{2}+M^{2}-m_{\pi}^{2})\right]\frac{E_{+}-E_{-}}{2MW^{2}}-\left[M^{6}\right.\nonumber\\
&  & +M^{4}(-2m_{\pi}^{2}+Q^{2}-W^{2})+M^{2}W^{2}(4m_{\pi}^{2}-2Q^{2}-W^{2})-2m_{\pi}^{4}Q^{2}-2m_{\pi}^{2}W^{4}\nonumber\\
&  & \left.\left.+W^{4}(Q^{2}+W^{2})\right]\ln\Xi\right\}\\
F_{3,4}^{{(0)}\pi N} & = & -\frac{g_{A}\nu}{128\pi^{2}F_{\pi}^{2}q^{3}}\left\{ \left[4M^{2}W^{2}(2M^{2}-m_{\pi}^{2}+Q^{2})\right.\right.\nonumber\\
&  & \left.-(M^{2}+Q^{2}+W^{2})^{2}(W^{2}+M^{2}-m_{\pi}^{2})\right]\frac{E_{+}-E_{-}}{2MW^{2}}-\left[M^{4}-M^{2}(Q^{2}+2W^{2})\right.\nonumber\\
&  & \left.\left.+2m_{\pi}^{2}Q^{2}+W^{2}(Q^{2}+W^{2})\right]\ln\Xi\right\}\nn\\
F_{3,5}^{{(0)}\pi N} & = & \frac{3g_{A}^{3}\nu}{512\pi^{2}F_{\pi}^{2}q^{3}(W^{2}-M^{2})}\left\{ \left[4M^{2}W^{2}\left(2M^{4}+2M^{2}(Q^{2}+W^{2})+m_{\pi}^{2}(Q^{2}-2W^{2})\right.\right.\right.\nonumber\\
&  & \left.\left.+Q^{4}+2Q^{2}W^{2}\right)-(M^{2}+W^{2})(M^{2}+Q^{2}+W^{2})^{2}(W^{2}+M^{2}-m_{\pi}^{2})\right]\frac{E_{+}-E_{-}}{2MW^{2}}\nonumber\\
&  & -\left[M^{6}+M^{4}(Q^{2}-W^{2})-M^{2}(2m_{\pi}^{2}Q^{2}+2W^{2}Q^{2}+W^{4})+2m_{\pi}^{4}Q^{2}+2m_{\pi}^{2}Q^{2}W^{2}\right.\nonumber\\
&  & \left.\left.+W^{4}(Q^{2}+W^{2})\right]\ln\Xi\right\}\nonumber
\eeqn
\beqn
F_{3,6}^{{(0)}\pi N} & = & \frac{g_{A}^{3}\nu}{512\pi^{2}F_{\pi}^{2}q^{3}}\left\{ \left[\left(4M^{2}W^{2}-(M^{2}+Q^{2}+W^{2})^{2}\right)\left(W^{2}+M^{2}-m_{\pi}^{2}\right)\right.\right.\nonumber\\
&  & \left.-\frac{8M^{2}m_{\pi}^{2}Q^{2}W^{4}\left(-3M^{2}+2m_{\pi}^{2}-Q^{2}-W^{2}\right)}{M^{6}-2M^{4}W^{2}+M^{2}(3m_{\pi}^{2}Q^{2}+W^{4})+m_{\pi}^{2}Q^{2}(W^{2}+Q^{2}-m_{\pi}^{2})}\right]\frac{E_{+}-E_{-}}{2MW^{2}}\nonumber\\
&  & \left.+2m_{\pi}^{2}\left(M^{2}+Q^{2}-W^{2}\right)\ln\Xi\right\} \label{eq:F3piN}
\end{eqnarray}
\end{widetext}
where $\lambda(a,b,c)=a^2+b^2+c^2-2ab-2bc-2ca$ is the usual  K\"{a}ll\'{e}n function, $E_\pm$ is the maximum and minimum energy of nucleon in an $N\pi$-final state:
	\beqn
	E_\pm&=&\frac{(W^2+M^2+Q^2)(W^2+M^2-m_\pi^2)}{4MW^2}\nn\\
	&\pm&\frac{q\lambda^{\frac{1}{2}}(M^2,m_\pi^2,W^2)}{2W^2}
	\eeqn
	and 
		\begin{equation}
		\ln \Xi=\ln \left(\frac{2ME_+-M^2+m_\pi^2-Q^2-W^2}{2ME_--M^2+m_\pi^2-Q^2-W^2}\right)
		\end{equation}
is a frequently-appearing logarithmic function.

It is well-known that ChPT is an effective field theory at low energy, so the results above apply only at low $Q^2$; in particular, it clearly overestimated the size of $N\pi$ effects at large $Q^2$. Form factors may be supplemented to suppress large-$Q^2$ contributions, but they should be added in a way to preserve the vector and axial current conservation:
\begin{equation}
q_\mu\left\langle p\right|J_{em}^{(0)\mu}\left|N\pi\right\rangle=0\:\:\:,\:\:\:\lim_{m_\pi\rightarrow 0}q_\mu\left\langle N\pi\right|\left(J_{W}^{\mu}\right)_A\left|n\right\rangle=0.
\end{equation} 
Here we choose to multiply our ChPT result by the isosinglet electromagnetic form factor $F_1^S(Q^2)$ and the axial form factor $-g_A^{-1}G_A(Q^2)$, which is a natural extension of the na\"{\i}ve Feynman rules applied in the ChPT calculation. 

With these results at hand, we can check the assumption of Ref. \cite{Marciano:2005ec} that chiral symmetry requires $F(Q^2=0)=0$. 
\begin{figure}[h]
\includegraphics[width=8cm]{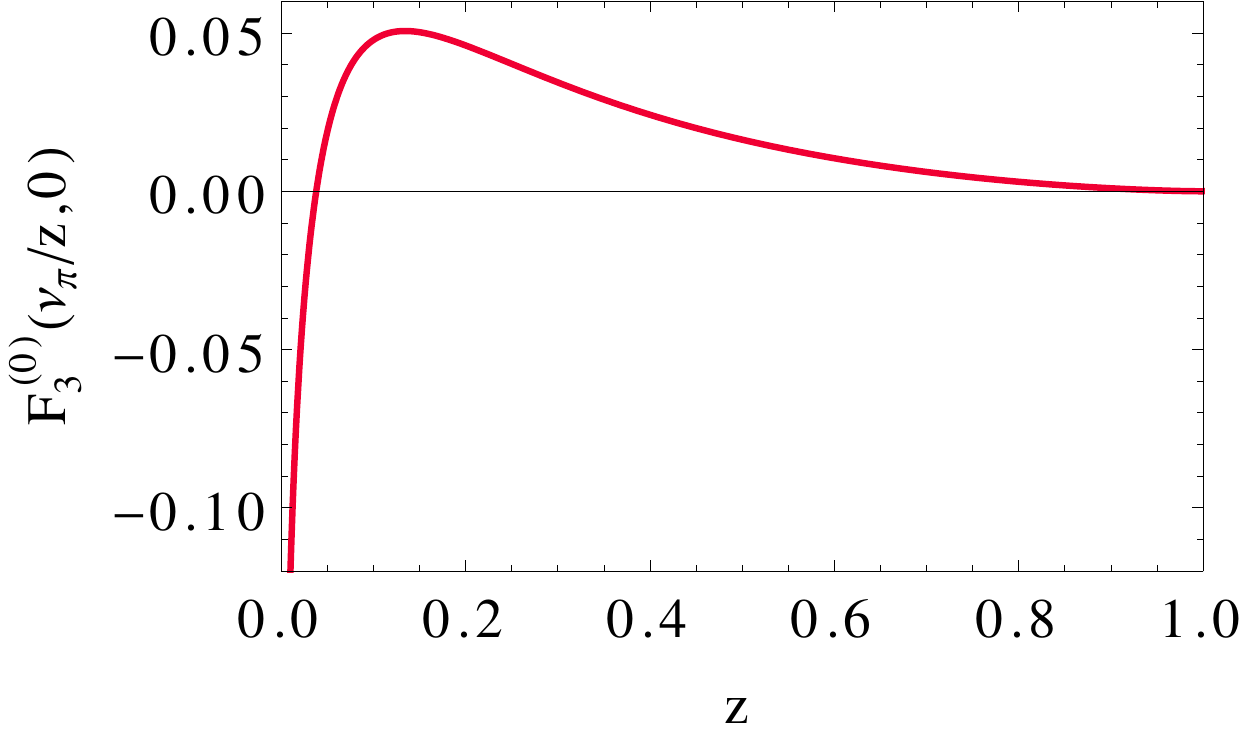}
\caption{The integrand of Eq. (\ref{eq:scrz}) as function of $z=\nu_\pi/\nu$}
\label{fig:SCR}
\end{figure}
This condition translates in the following superconvergence relation,
\beqn
\int_{\nu_\pi}^\infty\frac{d\nu}{\nu^2}F_{3,\, \pi N}^{(0)}(\nu,0)=0.
\eeqn
We find that this relation does not hold. To visualize it, we plot $F_3^{(0)\pi N}(\nu,0)$ as function of $z=\nu_\pi/\nu$ for which the superconvergence relation becomes simply 
\beqn
\int_0^1 dzF_3^{(0)\pi N}(\nu_\pi/z,0)=0,\label{eq:scrz}
\eeqn
in the full range of the variable $z\in(0,1)$ in Fig. \ref{fig:SCR}. It is evident that the surface below and above the curve are not equal, and the integral does not vanish. Numerically, using $g_A=1.27$ and $F_\pi=93$ MeV, we find 
\beqn
F(0)=6\int\limits_{\nu_\pi}^\infty\frac{d\nu}{M\nu^2}F_{3,\, \pi N}^{(0)}(\nu,0)=\frac{0.648}{{\rm GeV}^{2}}\neq0.
\eeqn
The result for $F(0)$ is finite, scales as $1/F_\pi^2$, and the contributions listed in Eq. (\ref{eq:F3piN}) exhaust all the contributions at the leading chiral order. Contributions from higher chiral orders will scale as higher powers of $1/F_\pi$ and cannot modify this result. This proves that $F(0)\neq0$.


\section{Resonance contribution}
\label{app:resonances}
Next we consider the effects of a baryon resonance $R$ to $\Box_{\gamma W}^{VA}$. We first observe that since only the isoscalar component of the electromagnetic current is involved, the contributing resonances must have isospin-1/2. This immediately gets rid of all isospin-3/2 resonances including the $\Delta$s which are usually the main contributors in the resonance region. For practical purpose we assume that the resonance $R$ takes the Breit-Wigner form with mass $m_R$ and width $\Gamma_R$. With these, the resonance contribution to $F_3^{(0)}$ is given by:
\beqn
F_{3,\,\rm res}^{(0)}&=&\sum_R\frac{\frac{\nu}{q}\Gamma_Rm_R}{(W^2-m_R^2)^2+\Gamma_R^2m_R^2}\sqrt{\frac{M(m_R^2-M^2)}{16\pi^3\alpha}}\nn\\
&\times&\sum_{J_z=1/2,3/2}(A_{em,J_z}^{R,p}+A_{em,J_z}^{R,n})^*A_{w,J_z}^R
\eeqn
where 
\begin{eqnarray}
A_{em,1/2}^{R,N} & = & -\sqrt{\frac{\pi\alpha}{M(m_{R}^{2}-M^{2})}}\left\langle R,+\frac{1}{2}\right|J_{em}^{+}\left|N,-\frac{1}{2}\right\rangle \nonumber \\
A_{em,3/2}^{R,N} & = & -\sqrt{\frac{\pi\alpha}{M(m_{R}^{2}-M^{2})}}\left\langle R,+\frac{3}{2}\right|J_{em}^{+}\left|N,+\frac{1}{2}\right\rangle \nn\\
\end{eqnarray}
are the standard electromagnetic transverse helicity amplitudes for the resonance $R$; the transverse components of the electromagnetic current are defined as $J_{em}^\pm=\mp (1/\sqrt{2})(J_{em}^x\pm iJ_{em}^y)$. In principle,
there is a sign ambiguity in the definition of the helicity amplitude,
which is related to the choice of sign for the $R\rightarrow N\pi$
coupling; throughout the paper we shall choose the same sign convention
as in Ref. \cite{Lalakulich:2006sw}. The transverse helicity amplitudes of the axial charged weak current are similarly defined as:
\beqn
A_{w,1/2}^R&=&\left\langle R,+\frac{1}{2}\right|\left(J_{w}^{+}\right)_A\left|n,-\frac{1}{2}\right\rangle,\nn\\
A_{w,3/2}^R&=&\left\langle R,+\frac{3}{2}\right|\left(J_{w}^{+}\right)_A\left|n,+\frac{1}{2}\right\rangle.
\eeqn

The values of the electromagnetic helicity amplitudes $A_{em,1/2}^{R,N},A_{em,3/2}^{R,N}$ in the vicinity of $Q^2=0$ can be obtained from partial wave analyses of $\pi,\eta$ photo- and electro-production; however for the evaluation of the dispersion integral their full $Q^2$-dependence is required and thus inputs from models are unavoidable at this point. In this work their functional form are taken from the unitary isobar model MAID2007 \cite{Drechsel:2007if,Tiator:2008kd}. In the meantime, there are much less studies of the transition matrix elements of the axial charged current. Here we make use of results of Ref. \cite{Lalakulich:2006sw} in which axial transition matrix elements of the lowest-lying spin-1/2 and 3/2 resonances at $Q^2=0$ are inferred from the $R\rightarrow N\pi$ decay width, and a modified dipole form is assumed for their $Q^2$-dependence. 

Applying the parameterizations in Ref. \cite{Drechsel:2007if,Tiator:2008kd,Lalakulich:2006sw}, we study the effects of the few lowest-lying $N^*$ resonances: $P_{11}(1440)$, $S_{11}(1535)$, $D_{13}(1520)$ and $S_{11}(1650)$, and find that their contributions to $\Box_{\gamma W}^{VA}$ are generally of the order $10^{-7}-10^{-5}$ which are negligibly small. This can be partially understood as due to the smallness of the isosinglet electromagnetic transition matrix element $A_{em}^{R,p}+A_{em}^{R,n}$ because the isosinglet component of the electromagnetic current is by itself small. Thus, we conclude that it is safe to disregard the contributions from discrete resonances and focus on the continuum. 

In the meantime, in order to extract the Regge piece in the inclusive neutrino scattering data, resonance contributions must be subtracted out by hand so we shall calculate the latter in the same manner. The main difference is, however, that in this case the resonances can have either $I=1/2$ or $3/2$; therefore the contribution from $\Delta$ is obviously the dominant piece so it is sufficient to consider only this piece. Also, according to isospin symmetry one has $F_{3,\Delta}^{\bar{\nu}p}=(1/3)F_{3,\Delta}^{\nu p}$ and therefore $F_{3,\Delta}^{\nu p+\bar{\nu}p}=(2/3)F_{3,\Delta}^{\nu p}$. Extensive model study of the $N-\Delta$ matrix element of charged weak current can be found in Ref. \cite{Lalakulich:2005cs} so here we simply quote their results relevant to our work after taking into account all differences in conventions:
\begin{equation}
F_{3,\Delta}^{\nu p+\bar{\nu}p}=-\frac{2\nu}{M}\frac{m_\Delta \Gamma_\Delta}{\pi}\frac{1}{(W^2-m_\Delta^2)^2+m_\Delta^2\Gamma_\Delta^2}\frac{V_3}{3}
\end{equation}
where 
\begin{eqnarray}
\frac{V_3}{3}&=&\frac{4}{3m_\Delta}\left[-\frac{C_3^VC_4^A}{M}(M\nu-Q^2)-C_3^VC_5^AM\right]\\
&\times&\left[2m_\Delta^2+2Mm_\Delta+Q^2-M\nu\right]\nonumber\\
&+&\frac{4}{3}\left[M\nu-Q^2\right]\left[-\frac{C_4^VC_4^A}{M^2}(M\nu-Q^2)-C_4^VC_5^A\right]\nn
\end{eqnarray}
and $m_\Delta$, $\Gamma_\Delta$ are the mass and width of the $\Delta$-resonance respectively. The $C$-functions are parameterized as:
\begin{eqnarray}
C_3^V&=&\frac{1.95}{\left(1+\frac{Q^2}{m_V^2}\right)^2}\frac{1}{1+\frac{Q^2}{4m_V^2}}\nonumber\\
C_4^V&=&-C_3^V\frac{M}{\sqrt{W^2}}\nonumber\\
C_5^A&=&\frac{1.2}{\left(1+\frac{Q^2}{m_A^2}\right)^2}\frac{1}{1+\frac{Q^2}{3m_A^2}}\nonumber\\
C_4^A&=&-\frac{C_5^A}{4}
\end{eqnarray}
where $m_V=0.84$GeV and $m_A=1.05$GeV.

\section{Regge contribution}
\label{app:regge}
In this appendix we show in detail how we obtain data input to the Regge contribution to the $\gamma W$-box from the inclusive $\nu p$ and $\bar{\nu} p$-scattering. 

The main objects of study are the P-odd structure functions $F_3^{(0)}$ and $F_3^{\nu p+\bar{\nu}p}$, and we will show that they are proportional to each other in the framework of Regge model with vector (axial vector) meson dominance. The physical picture is the following: in the VDM, the two energetic and slightly virtual gauge bosons fluctuate into a vector/axial-vector meson. The transition between the initial and final state proceeds via a $t$-channel exchange of a mesonic Regge trajectory with appropriate quantum numbers. 

The mechanism described above is depicted in Fig. \ref{fig:Regge}. The structure function $F_3^{(0)}$ involves the product between the isosinglet component of the electromagnetic current and the axial component of the charged weak current. Then, the $W$-boson should fluctuate into an isovector axial meson $a_1$ while the  isoscalar photon should fluctuate into an  isoscalar vector meson $\omega$. The exchanged meson in the $t$-channel should be the isotriplet vector meson $\rho$ in order to conserve isospin. Meanwhile, the three-meson vertex responsible for $F_3^{\nu p+\bar{\nu}p}$ is also $a_1\omega\rho$, except that this time the currents are the vector and axial components of the charged weak current, so the W-bosons should fluctuate into $\rho$ and $a_1$, respectively, whereas the exchanged-meson in the $t$-channel is $\omega$. The relevant gauge boson-vector meson mixing Lagrangian can be written as \cite{Lichard:1997ya}:
\beqn
\mathcal{L}_{\mathrm{VDM}}&=&\left(\frac{em_\rho^2}{g_\rho}\rho_\mu^0+\frac{em_\omega^2}{g_\omega}\omega_\mu\right) A^\mu\nn\\
&+&\frac{gm_\rho^2}{2g_\rho}V_{ud}\left(W_\mu^-\rho^{+\mu}+h.c.\right)\nn\\
&-&\frac{gm_{a_1}^2}{2g_{\rho}}w_{a_1}V_{ud}\left(a_{1\mu}^+W^{-\mu}+h.c.\right)
\eeqn
with the assumption that $g_{\omega}=3g_\rho$ following static SU(6) prediction \cite{deSwart:1963pdg}. Correspondingly, the vector meson coupling to nucleon is:
\begin{equation}
\mathcal{L}_{NNV}=\frac{g_\rho}{2}\bar{N}\gamma^\mu\vec{\tau}\cdot\vec{\rho}_\mu N+\frac{g_\omega}{2}\bar{N}\gamma^\mu \omega_\mu N.
\end{equation}
Note that the coupling strengths $g_\rho$ and $g_\omega$ appeared are the same as those in the mixing Lagrangian $\mathcal{L}_\mathrm{VDM}$ as a consequence of the universality relation within the VDM framework. Finally, one may write down a phenomenological $\rho-\omega-a_1$ interaction Lagrangian with two independent operators:
\begin{equation}
\mathcal{L}_{\rho\omega a_1}=g_1\epsilon^{\mu\nu\alpha\beta}(\partial_\mu\omega_\nu)\rho^i_\alpha a_{1\beta}^i+g_2\epsilon^{\mu\nu\alpha\beta}(\partial_\mu\rho^i_\nu)\omega_\alpha a_{1\beta}^i.
\end{equation}
It is obvious that the first operator only contributes to $F_3^{(0)}$ whereas the second operator only contributes to $F_3^{\nu p}$ and $F_3^{\bar{\nu}p}$ because there is no momentum exchange in the t-channel. We will further assume the exact equality of the two coupling constants $g_1,g_2$ as suggested in \cite{Harvey:2007rd} that comes from large-$N_c$ expansion, and will see that this is an important relation that leads to the near-degeneracy between the first Nachtmann moment of the Regge-induced $F_3^{(0)}$ and $F_3^{\nu p+\bar{\nu}p}$. 

With the ingredients above we may now proceed to write down the general form of the Regge contribution to $F_3^{(0)}$ and $F_3^{\nu p}$:
\begin{widetext}
\begin{eqnarray}
F_{3,\,\mathbb{R}}^{(0)}(\nu,Q^2)&=&\frac{1}{2}\left(\frac{eg}{2\sqrt{2}}V_{ud}\right)^{-1}\left(\frac{e}{g_\omega}\frac{m_\omega^2}{m_\omega^2+Q^2}\right)\left(-\frac{g}{2g_{\rho}}w_{a_1}V_{ud}\frac{m_{a_1}^2}{m_{a_1}^2+Q^2}\right)\left(\frac{g_\rho}{\sqrt{2}}\right)g_1H_\rho(\nu,Q^2)\nonumber\\
F_{3,\,\mathbb{R}}^{\nu p}(\nu,Q^2)&=&2\left(-\frac{g}{2\sqrt{2}}V_{ud}\right)^{-2}\left(\frac{g}{2g_\rho}V_{ud}\frac{m_\rho^2}{m_\rho^2+Q^2}\right)\left(-\frac{g}{2g_{\rho}}w_{a_1}V_{ud}\frac{m_{a_1}^2}{m_{a_1}^2+Q^2}\right)\left(\frac{g_\omega}{2}\right)g_2H_\omega(\nu,Q^2)\label{eq:F3Regge}
\end{eqnarray}
\end{widetext}
Notice the existence of the factor 1/2 and 2 in front of $F_{3,\,\mathbb{R}}^{(0)}$ and $F_{3,\,\mathbb{R}}^{\nu p}$ respectively; the first factor is due to our definition of $W_{\gamma W}^{\mu\nu}$ that contains a prefactor of $1/(8\pi)$ instead of $1/(4\pi)$; whereas the second factor is because $a_1$ can either couple to the incoming or outgoing $W$-boson in $F_{3,\,\mathbb{R}}^{\nu p}$. The function $H_V(\nu,Q^2)$ ($V=\rho,\omega$) encodes the information of the Regge-trajectory as well as all other universal multiplicative factors. From a pure Regge point of view one should expect $H_\rho(\nu,Q^2)\approx H_{\omega}(\nu,Q^2)$ due to the almost degenerate trajectories of $\rho$ and $\omega$; but here we may allow them to be different in order to account for other physics that could break such universality. The Regge contribution to $F_3^{\bar{\nu}p}$ can be calculated accordingly and it turns out to be identical to $F_{3,\mathbb{R}}^{\nu p}$. Upon setting $g_\omega=3g_\rho$ and $g_2=g_1$ and approximating $m_\omega\approx m_\rho$, one observes the following ratio: 
\begin{equation}
\frac{F_{3,\,\mathbb{R}}^{\nu p+\bar{\nu}p}(\nu,Q^2)}{F_{3,\,\mathbb{R}}^{(0)}(\nu,Q^2)}\approx \frac{36H_\omega(\nu,Q^2)}{H_\rho(\nu,Q^2)}.\label{eq:F3ratio}
\end{equation} 

One may now parameterize the first Nachtmann moment of both the $F_3$ functions as follows:
\begin{eqnarray}
&&M_{3,\,\mathbb{R}}^{(0)}(1,Q^2)=\frac{m_\omega^2}{m_\omega^2+Q^2}\frac{m_{a_1}^2}{m_{a_1}^2+Q^2}Q^2G_\rho(Q^2)\label{eq:M3parameter}\\
&&M_{3,\,\mathbb{R}}^{\nu p+\bar{\nu}p}(1,Q^2)=36\frac{m_\rho^2}{m_\rho^2+Q^2}\frac{m_{a_1}^2}{m_{a_1}^2+Q^2}Q^2G_\omega(Q^2).\nn
\end{eqnarray} 
A few explanations: (1) we retain the vector meson propagators as they are untouched by the integration over $\nu$; (2) we include an explicit factor of $Q^2$ in the parameterization to emphasize the fact that by definition the first Nachtmann moments vanish as $Q^2$ when $Q^2\rightarrow 0$, and (3) the explicit factor of 36 in $M_3^{\nu p+\bar{\nu}p}$ reflects the ratio in Eq. \eqref{eq:F3ratio}. 

We may now investigate the relation between the two functions $G_\rho(Q^2)$ and $G_\omega(Q^2)$ through a few matching conditions. First, it is natural to assume that $H_\omega(\nu,Q^2)$ and $H_\rho(\nu,Q^2)$ equal each other in the $Q^2\rightarrow 0$ limit, which reflects our belief that the VDM+Regge picture does appropriately describe the physics at low $Q^2$; that implies the constraint $G_\rho(0)=G_\omega(0)$. Second, we require the Regge prediction to smoothly match the perturbative QCD prediction at some matching value of $Q^2$. Recall that in the partonic description both the P-odd structure functions can be expressed in terms of valence quark PDFs of proton: $F_3^{(0)}(x)=(1/24)u_V^p(x)$ (as discussed in Appendix \ref{app:DIS}) and $F_3^{\nu p+\bar{\nu}p}=u_V^p(x)+d_V^p(x)$, which means their first Mellin's moments obey the following sum rules in the $Q^2\rightarrow \infty$ limit:
\begin{equation}
\int_0^1dx F_3^{(0)}(x)=\frac{1}{12}\:\:\:,\:\:\:\int_0^1dxF_3^{\nu p+\bar{\nu}p}(x)=3.
\end{equation}
In particular, the second line is just the GLS sum rule. At finite $Q^2$ both expressions are modified by pQCD. As pointed out in Ref. \cite{Marciano:2005ec}, the pQCD correction to the first equation follows that of the Bjorken sum rule for polarized electroproduction, which turns out to be identical to that of the GLS sum rule up to very small corrections of order $(\alpha_s/\pi)^3$ \cite{Larin:1991tj}. In other words, the ratio
\begin{equation}
\frac{\int_0^1dxF_3^{\nu p+\bar{\nu}p}(x)}{\int_0^1dx F_3^{(0)}(x)}=36\label{eq:pdfratio}
\end{equation} 
is stable with respect to pQCD corrections. As the first Nachtmann moment reduces to the first Mellin's moment at large $Q^2$, our second matching condition is to require this ratio to be satisfied by $M_3^{(0)}(Q^2)/M_3^{\nu p+\bar{\nu}p}(Q^2)$ at some matching point $Q^2=Q_0^2$.

Consider for instance a simple parameterization of $G_i(Q^2)$ with two free parameters: $G_i(Q^2)=F(a_i,b_i,Q^2)$. For $i=\omega$, the parameters can be fitted to existing experimental data of the GLS sum rule after subtracting out the elastic, $N\pi$ and resonance contributions by hand. Meanwhile, through the inspection of Eq. \eqref{eq:M3parameter} and \eqref{eq:pdfratio}, one immediately finds that upon approximating $m_\rho\approx m_\omega$, the two matching conditions simply imply $F(a_\omega,b_\omega,0)=F(a_\rho,b_\rho,0)$ and $F(a_\omega,b_\omega,Q_0^2)=F(a_\rho,b_\rho,Q_0^2)$, and the most natural solution is $a_\omega=a_\rho$, $b_\omega=b_\rho$, i.e. $G_\rho(Q^2)=G_\omega(Q^2)$. The key feature that leads to this conclusion is the existence of the same relative factor 36 in both equations: in Eq. \eqref{eq:M3parameter} this factor results from the model-predicted relations among $\{g_\rho,g_\omega\}$ as well as $\{g_1,g_2\}$ that are expected to hold at low $Q^2$, whereas in Eq. \eqref{eq:pdfratio} it is just a parton-model prediction which works at large $Q^2$. Such an agreement between two predictions at very different $Q^2$ suggests that it is reasonable to regard $M_3^{\nu p+\bar{\nu}p}(1,Q^2)$ and $M_3^{(0)}(1,Q^2)$ as being proportional to each other. Then, we extract the former from neutrino scattering data and divide it by 36 to obtain the later. Despite of lack of direct experimental data, this straightforward procedure provides the first solid and data-driven prediction for the Regge contribution to $\gamma W$ box diagram.

\section{DIS contribution}
\label{app:DIS}
Recall that we have expressed the vector-axial inference contribution to the $\gamma W$ box diagram in terms of the $Q^2$-integral over the first Nachtmann moment of the P-odd structure function $F_3^{(0)}$.
The dominant contributor to the integral lies in the large-$Q^2$ regime because the integral scales as $\ln (M_W^2/\Lambda^2)$ where $\Lambda$ is an effective infrared cutoff to the $Q^2$-integration. Fortunately at high-$Q^2$ one enters the DIS regime and the behavior of the first Nachtmann moment is quite well-understood as pQCD applies. Therefore we shall isolate the high-$Q^2$ piece:
\begin{equation}
\Box^{VA,\mathrm{DIS}}_{\gamma W}=\frac{3\alpha}{2\pi}\int_{\Lambda^2}^\infty\frac{dQ^2M_W^2}{Q^2\left[M_W^2+Q^2\right]}M_3^{(0)}(1,Q^2)
\end{equation}
and study its behavior separately.

When $\Lambda^2\gg M^2$, the first Nachtmann moment reduces effectively to the first Mellin's moment:
\begin{equation}
M_3^{(0)}(1,Q^2)\rightarrow \int_0^1dxF_3^{(0)}(x,Q^2)
\end{equation}
where $x=Q^2/(2M\nu)$ is the Bjorken variable. In the parton model the structure function $F_3^{(0)}$ depends on a combination of PDF's
\beqn
F_3^{(0)}(x)=\frac{e_u+e_d}{8}(d_n(x)-\bar u_n(x)).
\eeqn
Assuming further a symmetric sea in the neutron, $\bar u_n=\bar d_n$, the integral over $x$ simply gives the number of valence $d$-quarks inside the neutron (or equivalently, the number of valence $u$-quark inside the proton), $\int_0^1dx(d_n(x)-\bar d_n(x))=\int_0^1dxd_V(x)=2$, and we obtain the large logarithm term already obtained by MS:
\beqn
\Box_{\gamma W}^{VA,\rm DIS}&\approx&\frac{3\alpha}{2\pi}\frac{e_u+e_d}{4}\ln\frac{M_W^2}{\Lambda^2}=
\frac{\alpha}{4\pi}\ln\frac{M_W}{\Lambda},
\eeqn
An important result from Ref. \cite{Marciano:2005ec} was to realize that all pQCD corrections to this leading logarithm term are identical to those entering Bjorken sum rule for polarized electroproduction. These corrections modify the leading log (LL) result for the MS function $F(Q^2)$,
\beqn
&&F^{LL}(Q^2)=\frac{1}{Q^2}\label{eq:pQCD}\\
&&F^{\rm pQCD}=\frac{1}{Q^2}
\left[1-\frac{\bar\alpha_s}{\pi}-C_2\left(\frac{\bar\alpha_s}{\pi}\right)^2-C_3\left(\frac{\bar\alpha_s}{\pi}\right)^3
\right],\nn
\eeqn
with $C_2=4.583-0.333 N_F$ and $C_3=41.440-7.607N_F+0.177N_F^2$, $N_F$ standing for the number of effective quark flavors, 
and $\bar\alpha_s(Q^2)$ denotes the running strong coupling constant in the modified minimal subtraction scheme. 
Numerically, the pQCD corrections reduce the large logarithm $\ln(M_W/\Lambda)\approx3.98$ by roughly 8 \%.

\section{Quasielastic contribution to the $\beta^+$ decay of heavy nucleus}
\label{app:QE}
In this appendix we provide details of the calculation for the modification of the Born contribution to $F_3$ due to binding effects and Fermi motion in a nucleus.

We start by approximating the full nuclear Green's function by the subset with one active nucleon and a nuclear spectator, also known as the plane wave impulse approximation (PWIA). We write for the initial state $|A\rangle$ and final state $|A'\rangle$ wave functions involved in a $\beta^+$ decay process,
\beqn
|A\rangle&=&\sqrt{2E_A}\sum_{p\in A}\int\frac{d^3\vec k \phi^p_A(k)|p(\vec k),A-p(-\vec k)\rangle}{(2\pi)^3\sqrt{2E_{A-1}\,2E_n}}\\
|A'\rangle&=&\sqrt{2E_{A'}}\sum_{n\in A'}\int\frac{d^3\vec k \phi^n_{A'}(k)|n(\vec k),A'-n(-\vec k)\rangle}{(2\pi)^3\sqrt{2E_{A-1}\,2E_n}},\nn
\eeqn
with the on-shell condition for the intermediate nuclear state $A-p=A'-n$ but in general off-shell active nucleon with a 3-momentum $\vec k$. The momentum distribution function is normalized according to
\beqn
\int\frac{d^3\vec k}{(2\pi)^3}|\phi(k)|^2=1, 
\eeqn
while the nuclear state normalization is 
\beqn
\langle A(\vec k)|A(\vec k')\rangle=(2\pi)^32E_A\delta^3(\vec k-\vec k').
\eeqn

The $\gamma W$ interference Compton tensor in the $\beta^+$ decay of a nucleus $A$ can be defined as
\beqn
T^{\mu\nu}_{\gamma W,A}=\int dxe^{iqx}\langle A'|T[J^\mu_{em}(x)\left(J^{\nu}_W(0)\right)^\dagger]|A\rangle.
\eeqn
Using the above definitions we arrive to
the following expression in PWIA:
\beqn
T_{\gamma W,A}^{\mu\nu}&=&\sum_{p\in A}\int\frac{d^3\vec k}{(2\pi)^3}(\phi^n_{A'}(k))^*\phi^p_A(k)T_{\gamma W,p}^{\mu\nu}.
\eeqn
We aim at a universal correction that only takes into account the bulk nuclear properties, not the fine details of each initial and final nucleus. To this precision we assume that the momentum distribution of protons in the initial nucleus and neutrons in the final nucleus are the same, $\phi^p_{A'}(k)=\phi^n_A(k)=\phi(k)$. This assumption is natural, e.g. in the Fermi gas model of nucleus that we shall describe later. We then obtain a master formula in PWIA:
\beqn
T_{\gamma W,A}^{\mu\nu}(P,q)=\sum_{p\in A}\int\frac{d^3\vec k}{(2\pi)^3}|\phi(k)|^2T_{\gamma W,p}^{\mu\nu}(k,q),\label{eq:PWIA}
\eeqn
with the nuclear momentum taken at rest, $P^\mu=(M_A,\vec0)$. The Compton tensor $T^{\mu\nu}_{\gamma W,A}$ can be decomposed in terms of invariant functions as in Eq. \eqref{eq:Wmunu}, and we are interested in the P-odd invariant function $T_{3,A}^{(0)}$. 

It is informative to consider the limit of non-interacting nucleons (nucleon tensor independent of $k$) where we obtain
\beqn
T_{\gamma W,A}^{\mu\nu}&=&\sum_{p\in A}T_{\gamma W,p}^{\mu\nu}.
\eeqn
The relation Dis$T_i=4\pi F_i$ holds for each target with the respective mass. The nucleus is considered at rest, and neglecting nucleon recoil corrections $\sim\vec k^2/M^2$ we obtain for the structure function of interest,
\beqn
F_{3,A}^{(0)}&=&\sum_{p\in A}F_{3,p}^{(0)}.\label{eq:F3An}
\eeqn
In the limit of non-interacting nucleons, the nuclear structure function scales as the number of protons, just like the tree-level vector coupling of the $W$ boson to the nucleus. This confirms the result of Marciano and Sirlin for $C_B$ obtained in the free nucleon limit.

We now want to go beyond this limit by using Eq. (\ref{eq:PWIA}) to obtain: 
\beqn
F_{3,A}^{(0)}(P\cdot q,Q^2)=\sum_{p\in A}\int\frac{d^3\vec k}{(2\pi)^3}|\phi(k)|^2F_{3,p}^{(0)}(k\cdot q,Q^2).\nn
\eeqn
Here we are interested in the quasielastic contribution to $F_{3,A}^{(0)}$ which results from the smearing of the elastic term of the free proton. The latter can be inferred from Eq. \eqref{eq:F3Born}:
\beqn
&&F_{3,p}^{(0),B}(k\cdot q,Q^2)=F_{3}^{(0),B}(k\cdot q,Q^2)\\
&&=-\frac{Q^2}{4}G_A(Q^2)G_M^S(Q^2)\delta((k+q)^2-M^2).\nn
\eeqn
To perform the integral over $d^3\vec k=k^2dk d\cos\theta d\phi$ we choose the $z$-axis along the direction of the virtual photon, $q^\mu=(\nu,0,0,q)$ with $\nu=(P\cdot q)/M_A$ and $q=\sqrt{\nu^2+Q^2}$. 
The $\phi$ integration is trivial, while the $\delta$-function removes that over $d\cos\theta$ via 
\beqn
\delta((k+q)^2-M^2)=\frac{1}{2kq}\delta(\cos\theta-\cos\theta_k),
\eeqn
with
\beqn
&&\cos\theta_k\\
&&=\frac{(M_A+\nu)^2-2(M_A+\nu)p_{A-1}+M_{A-1}^2-M^2-\vec q^2}{2kq},\nn
\eeqn
with $p_{A-1}=\sqrt{M_{A-1}^2+\vec k^2}$ the energy of the on-shell spectator. One is left with an integral over $k=|\vec k|$ to obtain the quasielastic contribution to $F_{3,A}^{(0)}$:
\beqn
F_{3,A}^{(0),QE}(\nu,Q^2)=-\sum_{p\in A}G_AG_M^S\frac{Q^2}{32\pi^2q}
\int\limits_{k_-}^{k_+} kdk|\phi(k)|^2.\nn\\
\eeqn
Requiring that $-1\leq\cos\theta_k\leq1$ yields the upper and lower limits of $k$ as shown in Eq. \eqref{eq:kpm} upon neglecting terms of order $\epsilon/M_A$. 
Thus, for the initial nucleus with $Z$ protons we obtain:
\beqn
F_{3,A}^{(0),QE}(\nu,Q^2)=-ZG_AG_M^S\frac{Q^2}{16q} \langle\frac{1}{k}\rangle(\nu,Q^2),
\eeqn
with the average inverse nucleon momentum defined as
\beqn
\langle\frac{1}{k}\rangle(\nu,Q^2)=\int\limits_{k_-(\nu,Q^2)}^{k_+(\nu,Q^2)} \frac{kdk}{2\pi^2}|\phi(k)|^2.
\eeqn

As an exploratory model we consider the free Fermi gas model that corresponds to a uniform momentum distribution inside the sphere with the  radius equal to the Fermi momentum $k_F$
\beqn
\frac{1}{(2\pi)^3}|\phi(k)|^2=\frac{3}{4\pi k_F^3}\theta(k_F-|\vec k|),
\eeqn
resulting in 
\beqn
\langle\frac{1}{k}\rangle(\nu,Q^2)&=&\frac{3\left((\tilde k_+)^2-(\tilde k_-)^2\right)}{2 k_F^3},
\eeqn
with $\tilde k_\pm=\mathrm{min}(k_\pm,k_F)$.
Finally, we account for Pauli blocking by means of the function
\beqn
F_P(|\vec q|,k_F)=\frac{3|\vec q|}{4k_F}\left[1-\frac{\vec q^2}{12k_F^2}\right]\;\;\;{\rm for}\;\;|\vec q|\leq2k_F,
\eeqn
and $F_P=1$ otherwise. With these we obtain $F_{3,A}^{(0),QE}$ per proton for $\beta^+$ decay of a heavy nucleus as:
\beqn
\frac{1}{Z}F_{3,A}^{(0),QE}(\nu,Q^2)=-G_AG_M^S\frac{3Q^2}{32q}F_P\frac{\left((\tilde k_+)^2-(\tilde k_-)^2\right)}{k_F^3}.\nn\\
\eeqn

\end{appendix}


\end{document}